\documentclass[graybox, nosecnum]{svmult}

\usepackage{mathptmx}       
\usepackage{helvet}         
\usepackage{courier}        
\usepackage{type1cm}        
\usepackage{makeidx}         
\usepackage{graphicx}        
\usepackage{multicol}        
\usepackage[bottom]{footmisc}
\usepackage{hyperref}        
\usepackage{soul}            
\hypersetup{colorlinks=true,urlcolor=blue}
\usepackage{amsmath}
\usepackage{marginnote}
\usepackage[square,numbers]{natbib}
\usepackage{multirow}
\usepackage{booktabs}
\usepackage{upgreek}
\makeindex             

\newcommand{\degree}{^{\circ}}

\begin{document}
\title*{Compton Polarimetry} 
\author{Ettore Del Monte\thanks{corresponding author}, Sergio Fabiani and Mark Pearce}
\institute{Ettore Del Monte \at Istituto di Astrofisica e Planetologia Spaziali (IAPS), Via Fosso del Cavaliere 100 I-00133 Roma (Italy); INFN -- Roma Tor Vergata, Via della Ricerca Scientifica 1, I-00133 Roma, (Italy), \email{ettore.delmonte@inaf.it}, https://orcid.org/0000-0002-3013-6334
\and Sergio Fabiani \at Istituto di Astrofisica e Planetologia Spaziali (IAPS), Via Fosso del Cavaliere 100, I-00133 Roma (Italy), \email{sergio.fabiani@inaf.it}, https://orcid.org/0000-0003-1533-0283
\and Mark Pearce \at KTH Royal Institute of Technology, Department of Physics, 10691 Stockholm, Sweden. The Oskar Klein Centre for Cosmoparticle Physics, AlbaNova University Centre, 10691 Stockholm, Sweden. \email{pearce@kth.se}, https://orcid.org/0000-0001-7011-7229}

%
%
\maketitle
\abstract{Photons preferentially Compton scatter perpendicular to the plane of polarisation. This property can be exploited to design instruments to measure the linear polarisation of hard X-rays ($\sim$10 -- 100~keV). Photons may undergo two interactions in the sensitive volume of the instrument, i.e. a scattering followed by an absorption. Depending on the materials used to detect these two interactions, the Compton polarimeter can be classified as single-phase (same material for scattering and absorption detectors) or dual-phase (different materials). Different designs have been studied and adopted, and current instruments are predominantly with sensors based on scintillation- or solid-state detectors. X-ray polarimetry requires much higher statistics than e.g. spectrometry or timing, thus systematic effects must be accurately measured and accounted for. In this chapter we introduce the basic formalism of the Compton effect; we describe the design schemes developed so far for scattering polarimeters, including both the single-phase and dual-phase approaches; we overview the calibration methods to reduce the systematic effects; and we describe sources of background which affect the measurements.}

\section{Keywords}

Compton scattering, X-ray astrophysics, X-ray detector, Satellite-borne instrumentation, Balloon-borne instrumentation, Polarisation, Gamma Ray Burst, Compact Object, Scintillator, Solid-state detector

\vspace{1 cm}

\section{1 Introduction}

The hard X-ray ($\sim$10 -- 100~keV) sky is populated by variable sources such as accreting neutron stars, black holes and pulsars in our Galaxy, as well as extragalactic Gamma Ray Bursts (GRBs) and active galactic nuclei. The Sun itself is also a source of bright X-ray flares. Since the beginning of X-ray astrophysics, these sources have been studied by accumulating images, spectra and time series. The measurement of the polarisation of the emission is still lagging behind due to the difficulty in developing instrumentation with an adequate sensitivity. Polarisation is an important characteristic of the electromagnetic waves which emerge when a preferential direction in the source is present and the emission mechanism is not spherically symmetric, for example in case of synchrotron emission along a non-symmetric magnetic field or scattering on a non spherically-symmetric distribution of matter. The measurement of polarisation can thus provide new information on the emission mechanisms of astrophysical sources (see  Refs.~\cite{2011APh....34..550K,2019arXiv191010092S} and references therein).

The first pioneering measurement of the X-ray polarisation was obtained from the Crab Nebula at 2.6 keV and 5.2 keV by the satellite OSO-8 in 1976~\cite{Weisskopf1976}. Using the same instrument, the authors found an upper limit from the galactic source Sco X-1 \cite{1978ApJ...221L..13W}. PolarLight, launched at the end of 2018 aboard the CubeSat Tongchuan-1 \cite{2020NatAs...4..547F}, measured again the polarisation of the Crab Nebula \cite{2021ApJ...912L..28L} and found a significant detection of the polarisation of Sco X-1 in the 4 -- 8 keV energy range \cite{2022ApJ...924L..13L}. At the end of 2021, the NASA SMall EXplorer mission IXPE was launched to measure the polarisation of X-ray sources in the soft X-ray band, 2 -- 8 keV~\cite{2022JATIS...8b6002W}.

Initially, instruments designed for spectral and temporal studies were exploited to measure the polarisation of galactic sources and GRBs in the hard X-ray band. For some of these instruments the sensitivity to polarisation was specifically calibrated before launch, for others it was not.  After this, several groups developed polarimeters based on Compton scattering and some of them conducted observations on stratospheric balloons (e.g. GRAPE~\cite{McConnell2014}, PoGOLite and PoGO+~\cite{Chauvin.20160pr,Chauvin2016,Chauvin.2017qw}, X-Calibur~\cite{Abarr.2020p1n,Abarr.2021,2019AAS...23421508K}) or even installed aboard a space station (e.g. POLAR~\cite{2018APh...103...74X}). A satellite-borne instrument to measure the polarisation of X-rays of higher energy (i.e. in the Compton scattering regime) with a sensitivity comparable to IXPE has not yet flown.

In this chapter we discuss how to exploit Compton scattering to measure the polarisation of hard X-rays. We start with the definition of the main physical quantities related to the Compton scattering and we show how the most important performance of a polarimeter (i.e. modulation factor and minimum detectable polarisation) is related to these quantities. We then describe the design solutions and the type of detectors used to realise an actual polarimeter. We discuss the systematic effects and the methods to calibrate the polarimeter, the main sources of background and mitigation actions, and the different types of platform used to put such an instrument into orbit. We conclude with a summary of the future perspectives from the technology and science standpoints.

\section{2 Definitions and useful formulae}
\label{sec:formalism}

A beam of X-rays can be described as an electromagnetic wave comprising coupled electric and magnetic fields, which oscillate perpendicularly to each other and the direction of wave motion. The temporal and spatial development of the fields, and their coupling, is described by Maxwell's equations. The conventional description of a wave travelling in the $z$ direction at the speed of light, $c$, is $\mathbf{E}(z,t)=\mathbf{E}_{0} \cos{(\omega t - \omega z / c -\varphi)}$, where $t$ is time, $\omega$ is angular frequency, and $\varphi$ is an arbitrary phase. Consequently, the $x-y$ plane components are
\begin{equation}
E_{x}(t)=E_{x}(0)\cos{(\omega t - \varphi_{1})}; \,
E_{y}(t)=E_{y}(0)\cos{(\omega t - \varphi_{2})}.
\end{equation}
The polarisation angle, $\psi$, is defined in a counterclockwise direction between $\mathbf{E}(t)$ and the positive $x$-axis.
In the general case, the electric field vector traces an elliptical path in the $x-y$ plane, resulting in elliptical polarisation. Circular polarisation is obtained if $\lvert \varphi_{1}-\varphi_{2} \rvert=45^\circ$. For the particular case where
$\varphi_{1} = \varphi_{2}$, linear polarisation is obtained, and the electric field oscillates in a time-independent direction in the $x-y$ plane, with
$0\le\psi\le180^\circ$\footnote{Although a wave-based formalism is adopted here, through wave-particle duality, polarisation is also a property of individual photons through their spin.}. X-ray polarimeters developed thus far are only sensitive to linearly polarised radiation.

If waves are emitted from a source with constant $\psi$, the wave is said to have a polarisation fraction of 100\%. Conversely, an unpolarised wave with zero polarisation fraction is formed when a source emits waves with random values of $\psi$. To achieve emission with non-zero polarisation, an astrophysical source must therefore have a net deviation from spherical symmetry in terms of physical or magnetic field geometry.

The linear polarisation of an X-ray beam can be measured by determining a characteristic angle when it interacts in a detector. The nature of the interaction depends on the energy, $E$, of the incident X-ray.
Below a few keV coherent (Rayleigh) scattering from the atomic electron cloud is important.
At higher energies (with $E\ll m_e \mathrm{c}^2$ = 511 keV, where $m_e$ is the electron rest mass), classical Thomson scattering occurs. The incident ($E$) and scattered ($E^{\prime}$) X-ray energy are equal, i.e. the electron does not recoil. Above $\sim$20~keV, Compton scattering dominates, which is the focus of this chapter. The Compton effect is a form of incoherent scattering since the energy of the photon (and consequently also the involved electron) is changed after the interaction.

The differential cross-section for Compton scattering off a free electron in the detector material, averaged over all polarisation states for the scattered photon, is given by the Klein-Nishina relationship~\cite{Klein.1929},
\begin{equation}
\frac{\mathrm{d}\sigma}{\mathrm{d}\Omega} =
\frac{3\sigma_{\mathrm{T}}}{16\pi}
\left(
\frac{E^{\prime}}{E}
\right)^{2}
\left(
\frac{E}{E^{\prime}} +
\frac{E^{\prime}}{E} -
2\sin^2 \theta\,\cos^2 \phi
\right)
\label{eq:KN}
\end{equation}
where $\sigma_{\mathrm{T}}$ is the Thomson cross-section (6.652 $\times$ 10$^{-29}$ m$^2$) and the Compton equation states that,
\begin{equation}
E^{\prime}=\frac{E}{1+\frac{E}{m_ec^2}(1-\cos\theta)}.
\end{equation}

\begin{figure}[tb]
\begin{center}
    \includegraphics[width=0.50\linewidth]{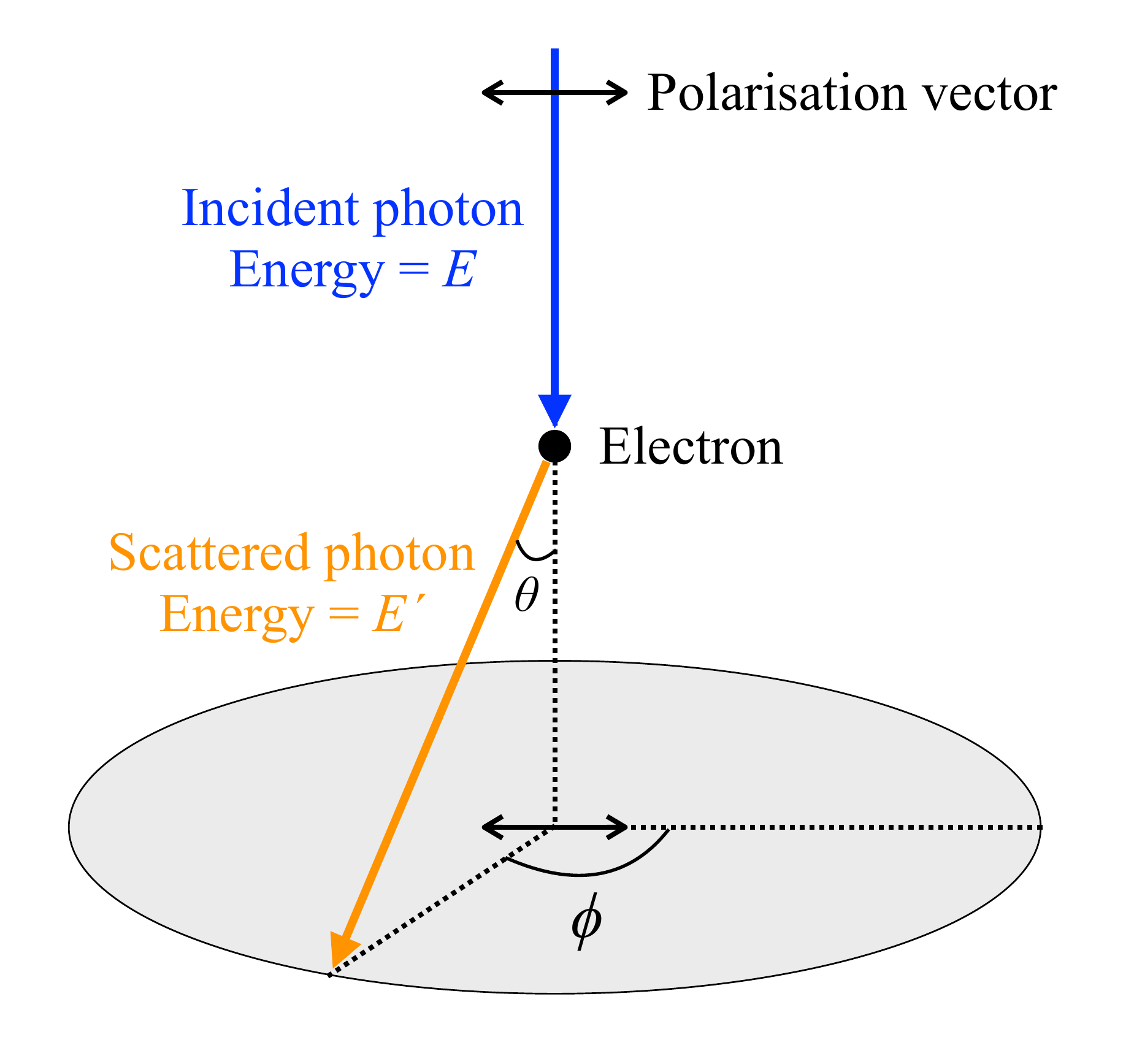}
\end{center}
\caption{A schematic representation of the Compton scattering geometry described by the Klein Nishina relationship shown in Equation~\ref{eq:KN}.}
\label{fig:KNgeom}
\end{figure}

As shown in Figure~\ref{fig:KNgeom}, $\theta$ is the photon scattering angle and $\phi$ is defined as the azimuthal angle between the scattering direction and the polarisation vector of the incident photon. A common feature of the aforementioned scattering processes is that photons scatter preferentially perpendicular to the polarisation (electric field) vector. The recoil energy of the electron after the scattering is given by
\begin{equation}\label{eq:deltaE}
\Delta E (\theta) = E - E^{\prime} = \frac{E^2 (1 - \cos \theta )}{m_e c^2 + E (1 - \cos \theta )}.
\end{equation}

\noindent For a real scatterer the theoretical Klein-Nishina cross-section is multiplied by a scattering function, which accounts for the distribution of atomic electrons and binding energies~\cite{Fabiani.2013vom}.

\begin{figure}[tb]
\begin{center}
    \includegraphics[width=0.75\linewidth]{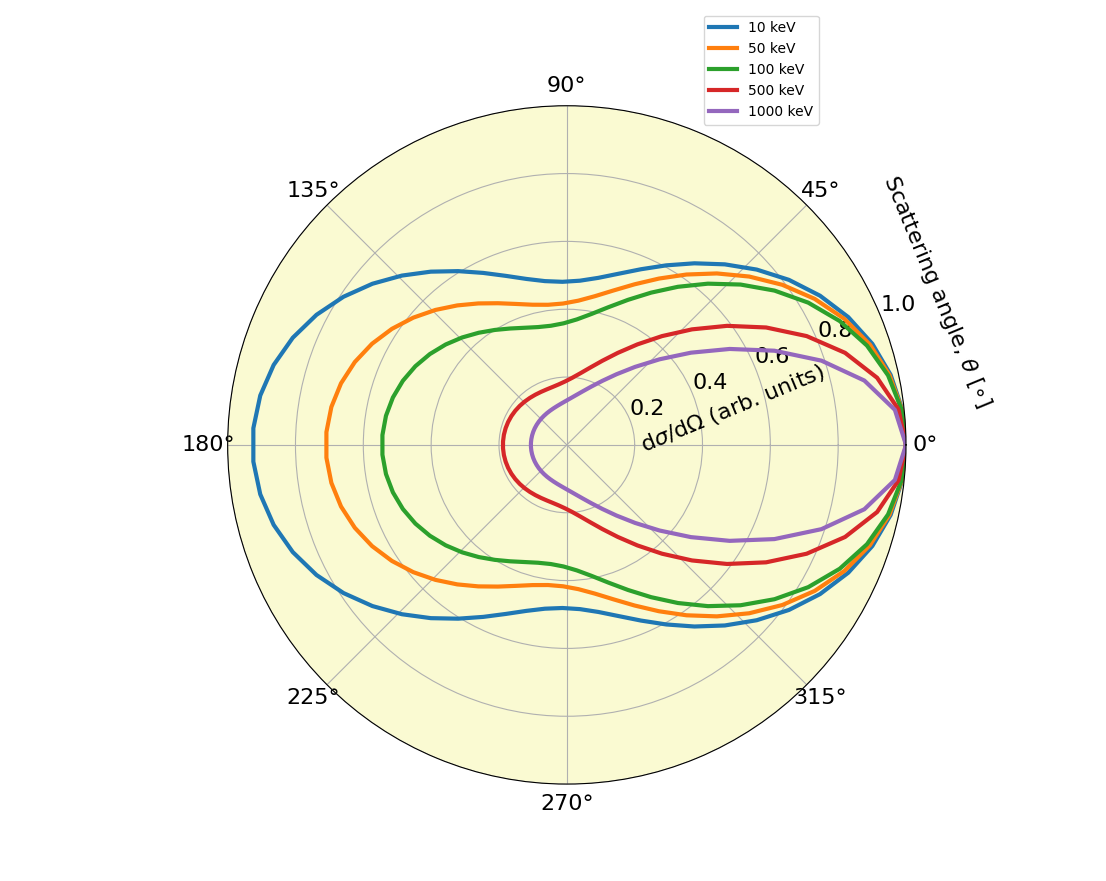}
\end{center}
\caption{The relationship between the Klein Nishina cross-section and the scattering angle, $\theta$. The photon travels along the horizontal axis from left ($180 \degree$) to right ($0 \degree$) and scatters off an electron placed at the origin. The anisotropy of the azimuthal scattering angle is seen to decrease as the photon energy increases.}
\label{fig:peanut}
\end{figure}

A graphical representation of the Klein Nishina scattering cross-section is shown in Figure~\ref{fig:peanut}.
As the incident energy of the photon increases, it is more likely to forward scatter which results in a more isotropic azimuthal scattering angle distribution. Because of this, Compton polarimetry is of limited use above the pair production threshold ({$\sim$1~MeV}). The low energy limit is defined when the photoelectric interaction cross-section becomes dominant.

The polarisation parameters of the beam are encoded in the distribution of azimuthal scattering angles.
This harmonic function is referred to as a modulation curve, and may be parameterised as,
\begin{equation}
C(\phi)=A \cos{(2(\phi-\psi+\pi/2))} + B
\end{equation}
where $C(\phi)$ is the number of scattered X-rays registered at a specified azimuthal angle, and $A$, $B$ are constants. The modulation factor, $\mu$, is derived from the maximum ($C_{\mathrm{max}}$) and minimum ($C_{\mathrm{min}}$) extent of the modulation curve as,
\begin{equation}
\mu = \frac{C_{\mathrm{max}}-C_{\mathrm{min}}}{C_{\mathrm{max}}+C_{\mathrm{min}}} = \frac{A}{B}.
\end{equation}

The phase of the modulation curve defines the polarisation angle.
In order to determine the polarisation fraction, $p$, of the beam, the modulation response to a 100\% polarised beam (preferably determined through experimental calibration, and not only using computer simulations), $\mu_{100}$, is required since,
\begin{equation}
p = \frac{\mu}{\mu_{100}}.
\end{equation}

\begin{figure}[tb]
\begin{center}
    \includegraphics[width=0.75\linewidth]{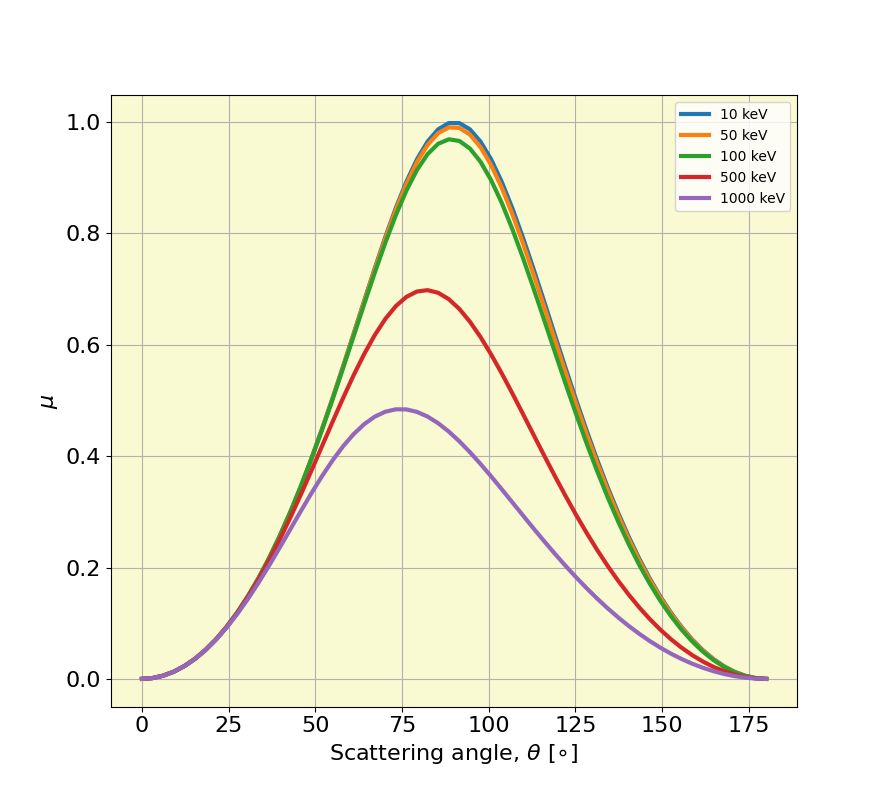}
\end{center}
\caption{The dependence of the modulation factor, $\mu$, on the scattering angle, $\theta$.
}
\label{fig:challenge}
\end{figure}

The nature of the Compton scattering process presents an experimental challenge. For an ideal polarimeter, Figure~\ref{fig:challenge} shows the scattering angle and incident photon energy dependence of $\mu$, where for $N$ scattered photons,
\begin{equation}
\mu(\theta)=
\frac{N_{\mathrm{max}}(\theta) - N_{\mathrm{min}}(\theta)}{N_{\mathrm{max}}(\theta) + N_{\mathrm{min}}(\theta)} =
\frac{
\left(
\frac{\mathrm{d}\sigma}{\mathrm{d}\Omega}
\right)_{\phi=90^\circ} -
\left(
\frac{\mathrm{d}\sigma}{\mathrm{d}\Omega}
\right)_{\phi=0^\circ}
}
{
\left(
\frac{\mathrm{d}\sigma}{\mathrm{d}\Omega}
\right)_{\phi=90^\circ} +
\left(
\frac{\mathrm{d}\sigma}{\mathrm{d}\Omega}
\right)_{\phi=0^\circ}
} =
\frac{\sin^{2}\theta}{E/E^{\prime} + E^{\prime}/E - \sin^{2}\theta}.
\label{eq:challenge}
\end{equation}
It is evident from Equation~\ref{eq:challenge} and Figure~\ref{fig:challenge} that forward- ($\theta=0^\circ$) and backward- ($\theta=180^\circ$) scattered photons convey no polarimetric information. The highest modulation response is obtained at $\theta=90^\circ$ (where $E=E^{\prime}$, the Thomson scattering limit).
A practical polarimeter will integrate the scattered counts over a range of angles in order to achieve reasonable efficiency, thereby reducing $\mu$.
Furthermore, a polarimeter should be designed to ensure that single scattering events dominate since multiple scatterings will wash out the polarisation signature.
The maximum value of $\mu$ is seen to decrease as the incident energy increases. Depending on the polarimeter design, $\mu$ may therefore be energy dependent.

Fitting harmonic functions to the modulation curve provides an intuitive method to extract polarisation parameters from a distribution of azimuthal scattering angles.
A more practical approach is to use Stokes parameters~\cite{1851TCaPS...9..399S}. Unlike $p$ and $\psi$, these are additive (e.g. allowing straight-forward background subtraction) and are statistically well-defined, being described by a bivariate normal distribution. An unbinned analysis becomes possible, which avoids possible systematic effects arising from fitting binned data. The prescription presented in Ref.~\cite{Kislat.2015} has been widely adopted, where $p$ and $\psi$ are determined by forming a weighted sum of intensity (I)-normalised Q and U Stokes parameters ($\mathcal{Q}$=Q/I and $\mathcal{U}$=U/I) for $N$ photon scattering events reconstructed in the polarimeter with polarisation angle $\psi_i=\phi_i-90^\circ$, where,
\begin{equation}
\mathrm{Q}=\sum^{N}_{i=1} w_i \cos(2\psi_i);
\mathrm{U}=\sum^{N}_{i=1} w_i \sin(2\psi_i);
\mathrm{I}=\sum^{N}_{i=1} w_i.
\end{equation}
The weights, $w_i$, can be used to compensate for systematic effects, e.g. variable angular acceptance for the scattered beam due to the detector geometry, or to account for different exposure times for source and background measurements.
For a 100\% linearly polarised beam, $\mathcal{Q}=1$ ($\mathcal{U}=1$) for a beam polarised in the north-south (northeast-southwest) direction respectively. In the absence of background\footnote{Expressions including background can be found in Volume 4, Section 4 of this handbook: M.~Kiss and M.~Pearce, Bayesian analysis of the data from PoGO+ (cross reference).}, the values of $p, \psi$ reconstructed by a polarimeter ($p_r, \psi_r$) are given by,
\begin{equation}
p_r=\frac{2}{\mu_{100}}\sqrt{\mathcal{Q}^2 + \mathcal{U}^2}; \psi_r=\frac{1}{2}\arctan\left( \frac{\mathcal{U}}{\mathcal{Q}} \right).
\end{equation}

\noindent The values of $p_r, \psi_r$ will only have Gaussian distributed uncertainties in the regime where signal photons dominate the overall counting rate (arising from both signal and background sources). This may be difficult to achieve when initially large data-sets are subdivided, as required to determine the energy dependence of polarisation parameters, or when temporal binning is needed, e.g. for phase resolved polarimetry of pulsars or studies of prompt phase GRB emission.
If $p_0$ denotes the polarisation fraction generated at a source, then even if $p_0=0$, statistical fluctuations will result in $\langle p_r \rangle > 0$, i.e. the measurement is positive definite.

The Minimum Detectable Polarisation (MDP)~\cite{Weisskopf.2010855k} is an established figure-of-merit for a polarimeter which can be used to determine if a polarisation measurement is statistically significant. There is a 1\% chance to measure $p_r \ge\!\mathrm{MDP}$ for an unpolarised beam. In Ref.~\cite{Weisskopf.2010855k} the MDP is defined at 99\% confidence level as,
\begin{equation}\label{eq:MDP_1}
    \mathrm{MDP}=\frac{4.29}{\mu_{100} \; F_s \; \epsilon A}
    \sqrt{\frac{F_s \; \epsilon A + R_{b}}{t_{\rm obs}}}
\end{equation}
where
$F_s$ is the source flux, $\epsilon$ is the polarimetric detector efficiency, $A$ is the polarimeter detection area (with $\epsilon A$ referred to as the effective area),
$R_b$ is the detected background count rate, $R_s = F_s \; \epsilon A$ is the signal count rate, and $t_{\rm obs}$ is the on-source integration time.

In Ref.~\cite{Mikhalev.2018} a method to calculate the bias on $p_r$ is presented for an analysis framework using Stokes parameters. The importance of designing a mission with an appropriate MDP performance is highlighted since for cases where a polarimeter makes a measurement such that MDP/$p_r \sim 1$, $p_r$ will be over-estimated by $\sim$20\%.

The MDP shows that values of $\mu_{100}$ and $\epsilon A$ should be maximised to provide the best polarimetric sensitivity.
The emission from astrophysical sources is generally expected to be weakly polarised.
For a well-designed polarimeter with $\mu_{100}=0.5$ and with no background, 10$^6$ counts are required to obtain MDP = 1\%.
The magnitude of such a dataset poses a significant experimental challenge, as the instrument response needs to be determined with comparable statistics.
When comparing different polarimeter designs, it is useful to introduce a quality factor, $Q$, which, following from the definition of MDP, and in the absence of background, is defined as $Q=\mu_{100}\sqrt{\epsilon}$. This quantity may be energy dependent.

\section{3 Polarimeter design}
\subsection{3.1 General concept of a Compton scattering polarimeter}\label{sec:PolDesign}

\begin{figure}[tb]
\begin{center}
    \includegraphics[width=0.75\linewidth]{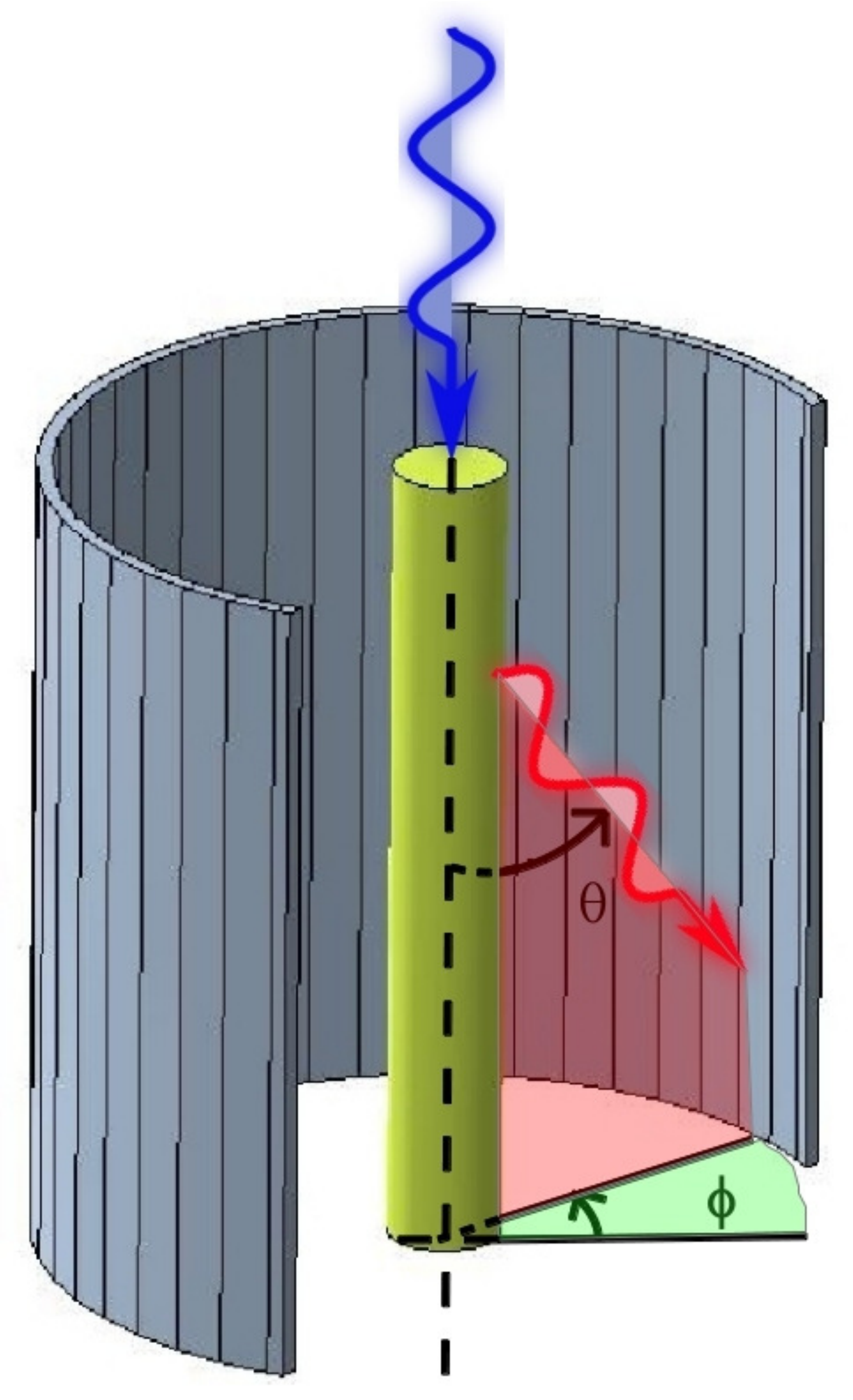}
\end{center}
\caption{Schematic of the general concept of a scattering polarimeter, reprinted from Astroparticle Physics, Vol 44, Sergio Fabiani et al., ``Characterization of scatterers for an active focal plane Compton polarimeter'', 91 – 101 (2013) \cite{Fabiani.2013vom} with permission from Elsevier. The scattering takes place in the central yellow rod and the photon is then absorbed in the external grey detector. The rod is few centimeters long and has a diameter of some millimeters. The angles $\theta$ and $\phi$ are defined in Equation~\ref{eq:KN} and shown in Figure~\ref{fig:KNgeom}.}
\label{fig:Compton_schematic}
\end{figure}

A simplified schematic of a Compton scattering polarimeter is shown in Figure~\ref{fig:Compton_schematic} from Ref.~\cite{Fabiani.2013vom}. The scattering takes place in the central yellow rod and the photon is then absorbed in the external grey detector. As shown by the Klein-Nishina cross-section in Equation~\ref{eq:KN}, the most probable azimuthal scattering direction $\phi$ is orthogonal to the polarisation vector of the incoming photon. Consequently, if we measure $\phi$ we can reconstruct the polarisation of the X-ray beam.

In the limit of a non-relativistic interaction ($E \ll m_e c^2$), the modulation factor is maximum at $\theta=90^\circ$, i.e. orthogonally to the direction of the incoming photon (see also Figure~\ref{fig:challenge}). By increasing the energy, the polar scattering directions peak in the forward-direction ($\theta < 90^\circ$, see Equation~\ref{eq:challenge} and Figure~\ref{fig:challenge}), consequently, the sensitivity to polarisation is highest in the plane orthogonal to the direction of the incoming photons.

At least two interactions occur in a scattering polarimeter and the first one is a Compton scattering. The larger the distance between the two interaction points and the more finely segmented the sensitive volume, the better the precision of the angular measurement. The most favourable condition occurs if the fraction of energy released by the incoming photon in the scatterer is detected (see Equation~\ref{eq:deltaE}) and the scattered photon is subsequently stopped via the photoelectric effect in the absorber. The choice of the materials of the scatterer and the absorber elements, as well as their geometrical shape and spatial configuration, determine the capability of the polarimeter to accurately reconstruct the polarisation.

In order to have a small MDP (in Equation~\ref{eq:MDP_1}), $\mu_{100}$ and the detection efficiency should be maximised. However, by narrowing the angle of acceptance of the polar angles $\theta$ around the peak of the modulation factor distribution, the efficiency in detecting scattering/absorption events decreases and vice versa. Therefore, a trade-off between modulation factor and detection efficiency is needed and this defines the geometrical configuration of the polarimeter assembly. Moreover, to ensure a high probability of Compton interaction one would have the scattering volume extended parallel to the incoming photon direction, while maintaining a small transverse size (compared to the attenuation length) to minimise multiple scattering events, in order to preserve the initial scattering direction needed to measure the polarisation.

The choice of materials of both scattering and absorbing elements is the other key parameter. The scattering volume should maximise the probability of a Compton interaction, while minimising the photoelectric absorption that prevails at low energy, thereby defining the low energy threshold of detection. Materials with a low atomic number $Z$ have a large scattering/absorption probability ratio, thus they are a suitable choice as scatterers. On the contrary, materials with a high atomic number are suitable as absorbers because they efficiently absorb via photoelectric effect the scattered photon.
\begin{figure}
 \begin{center}
\begin{tabular}{c}
\includegraphics[scale=0.11]{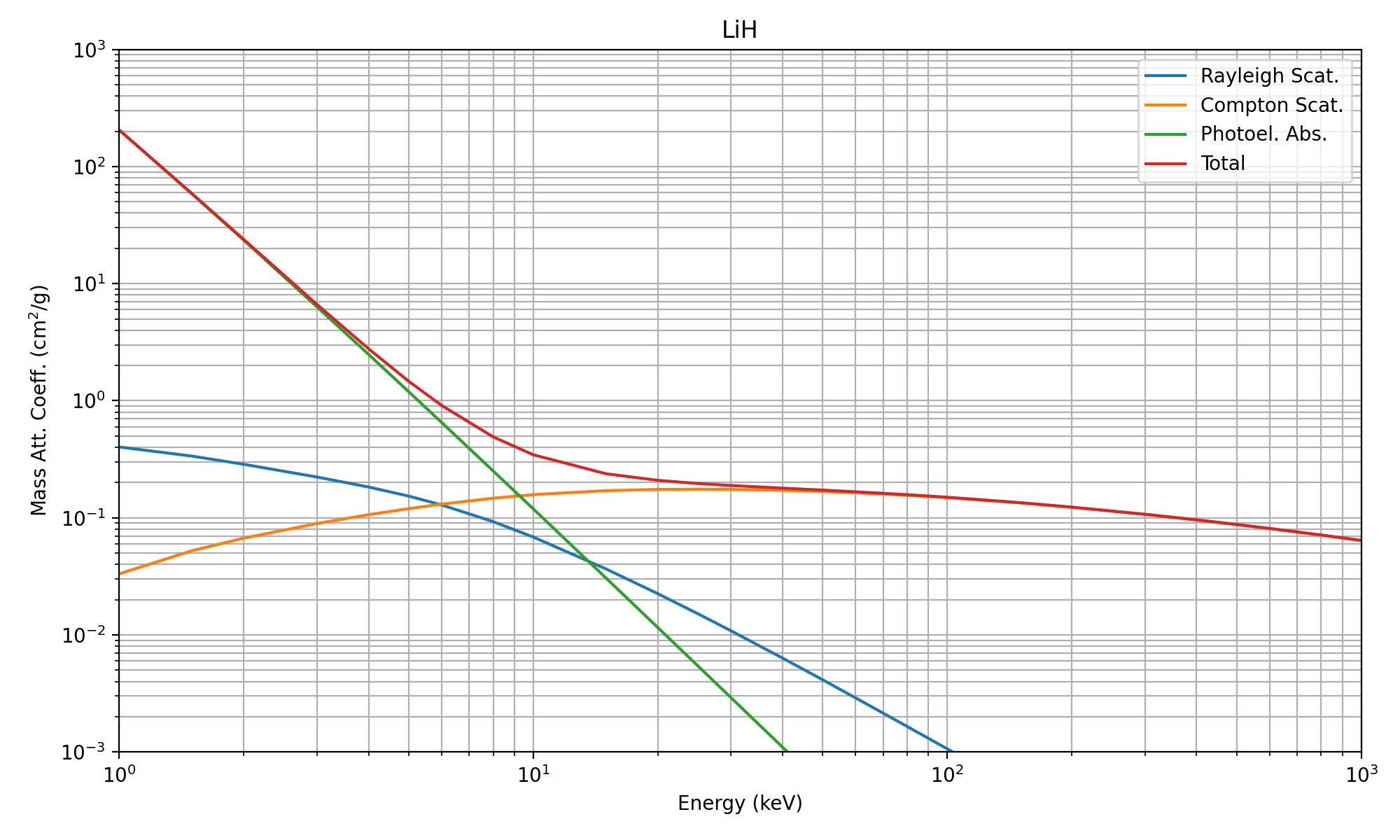}
\includegraphics[scale=0.11]{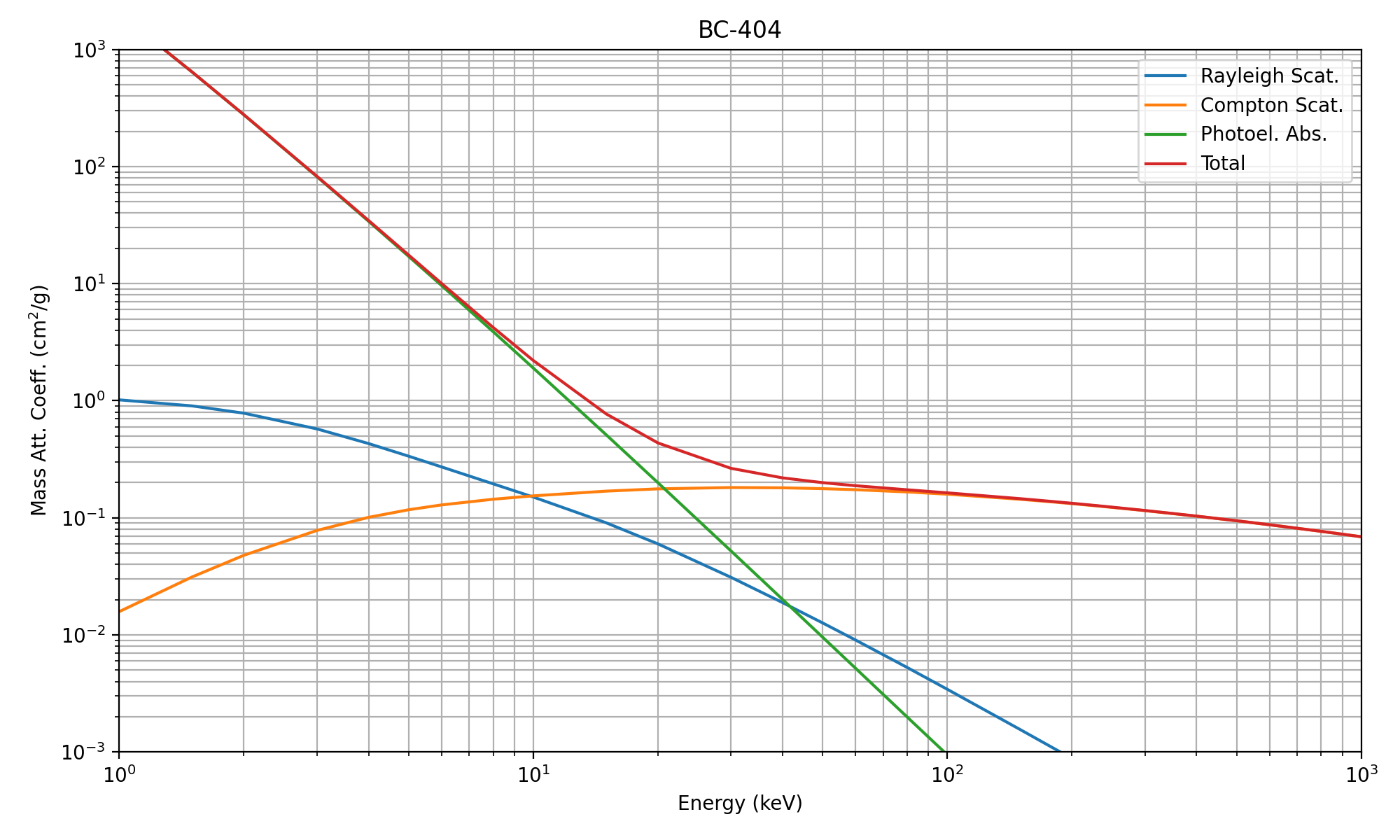}\\
\includegraphics[scale=0.11]{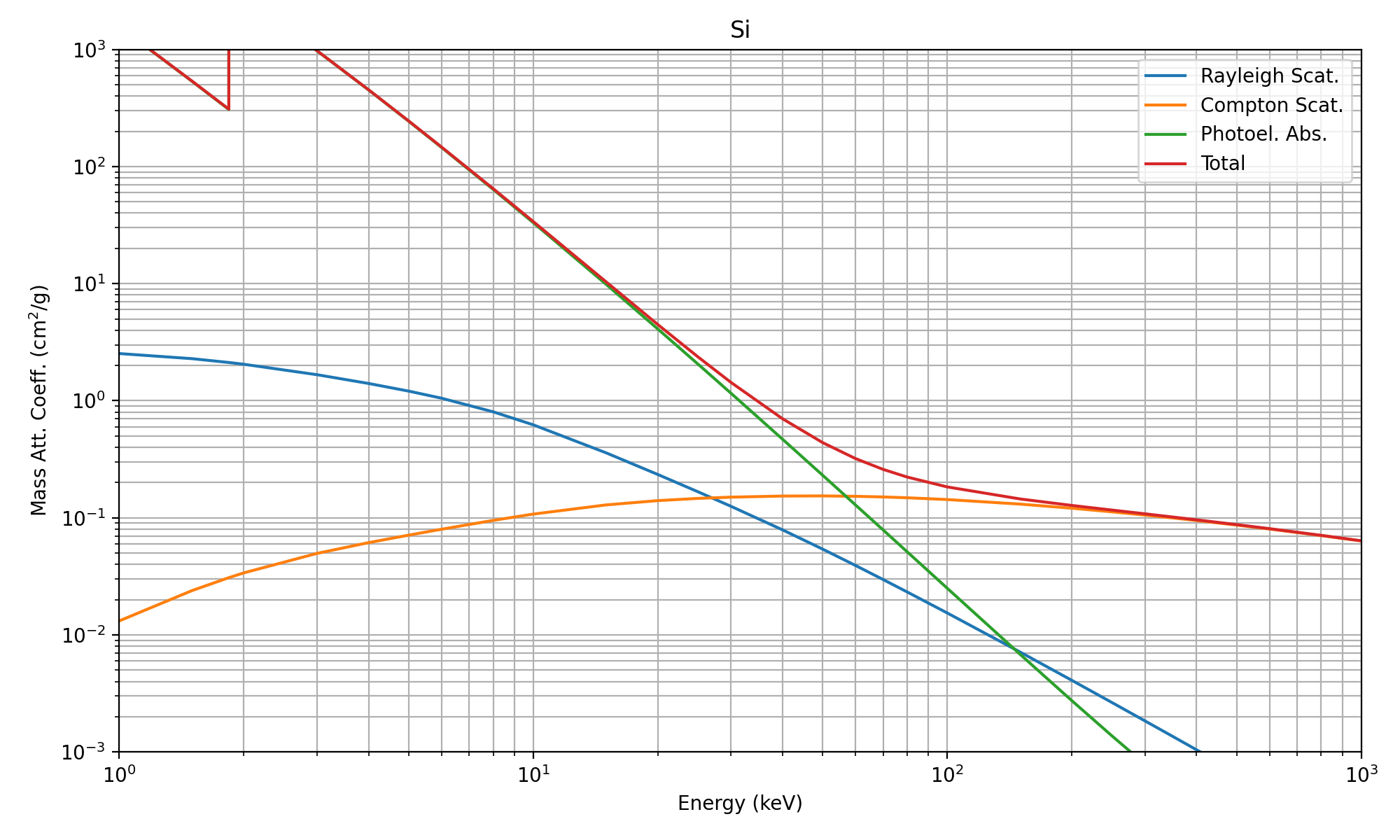}
\includegraphics[scale=0.11]{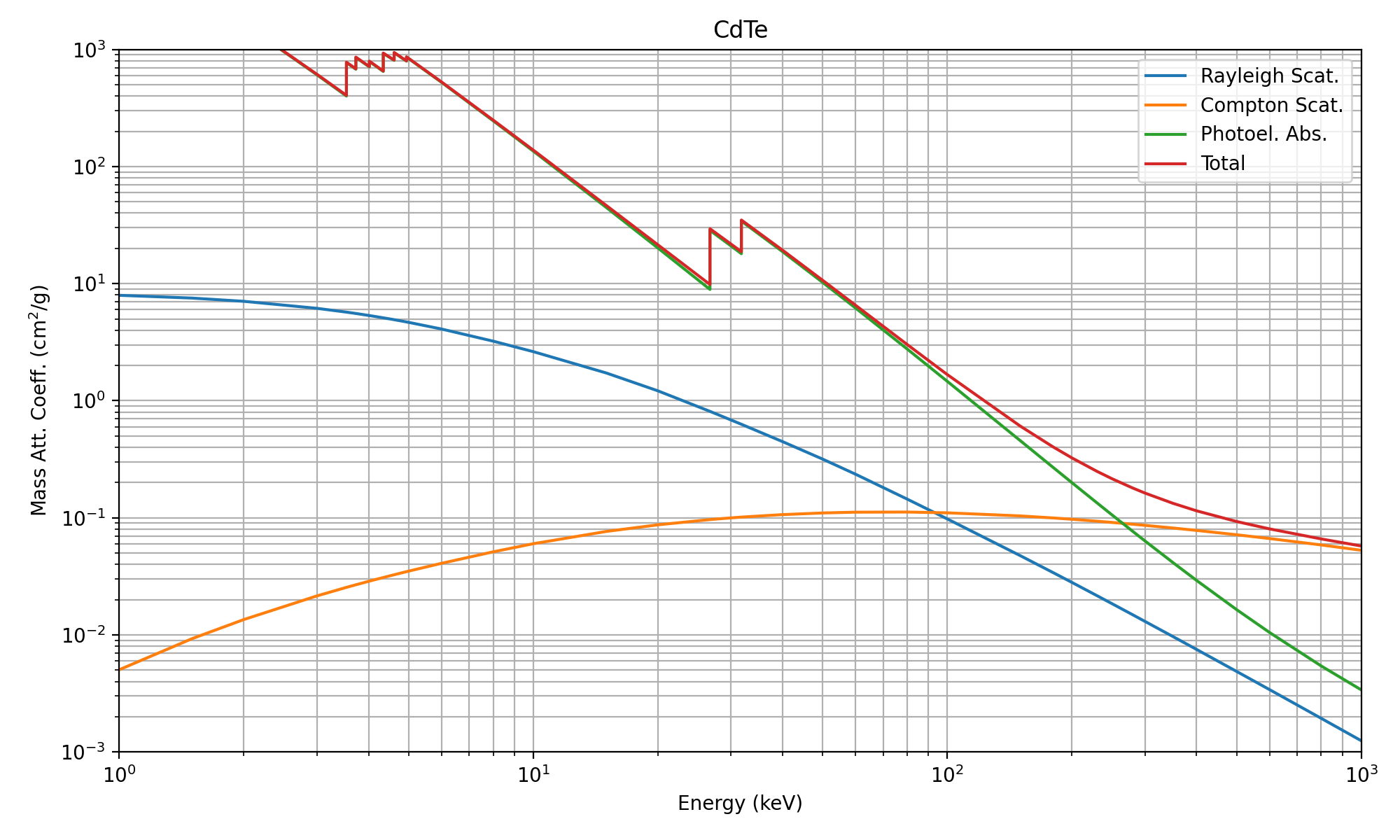}\\
\includegraphics[scale=0.11]{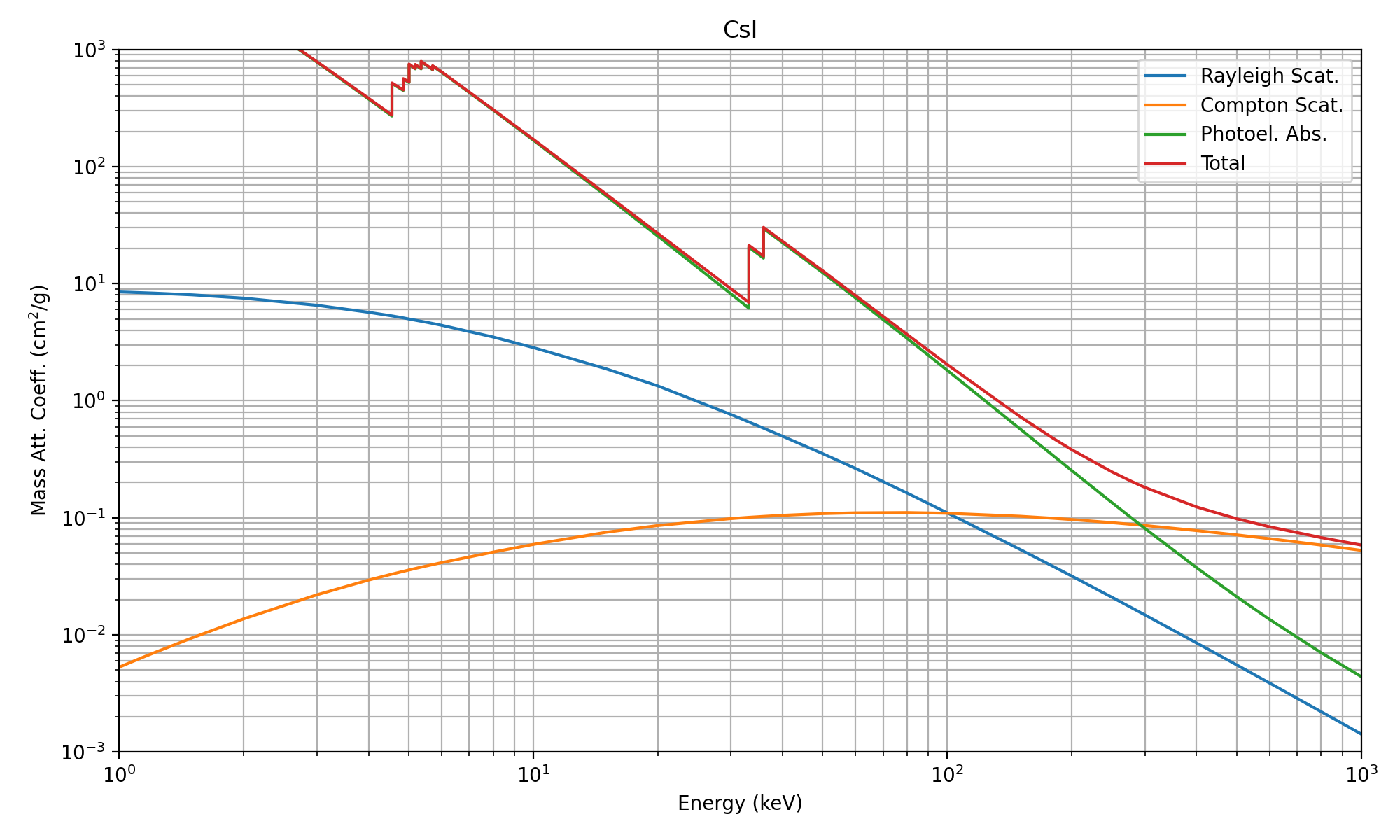}
\includegraphics[scale=0.11]{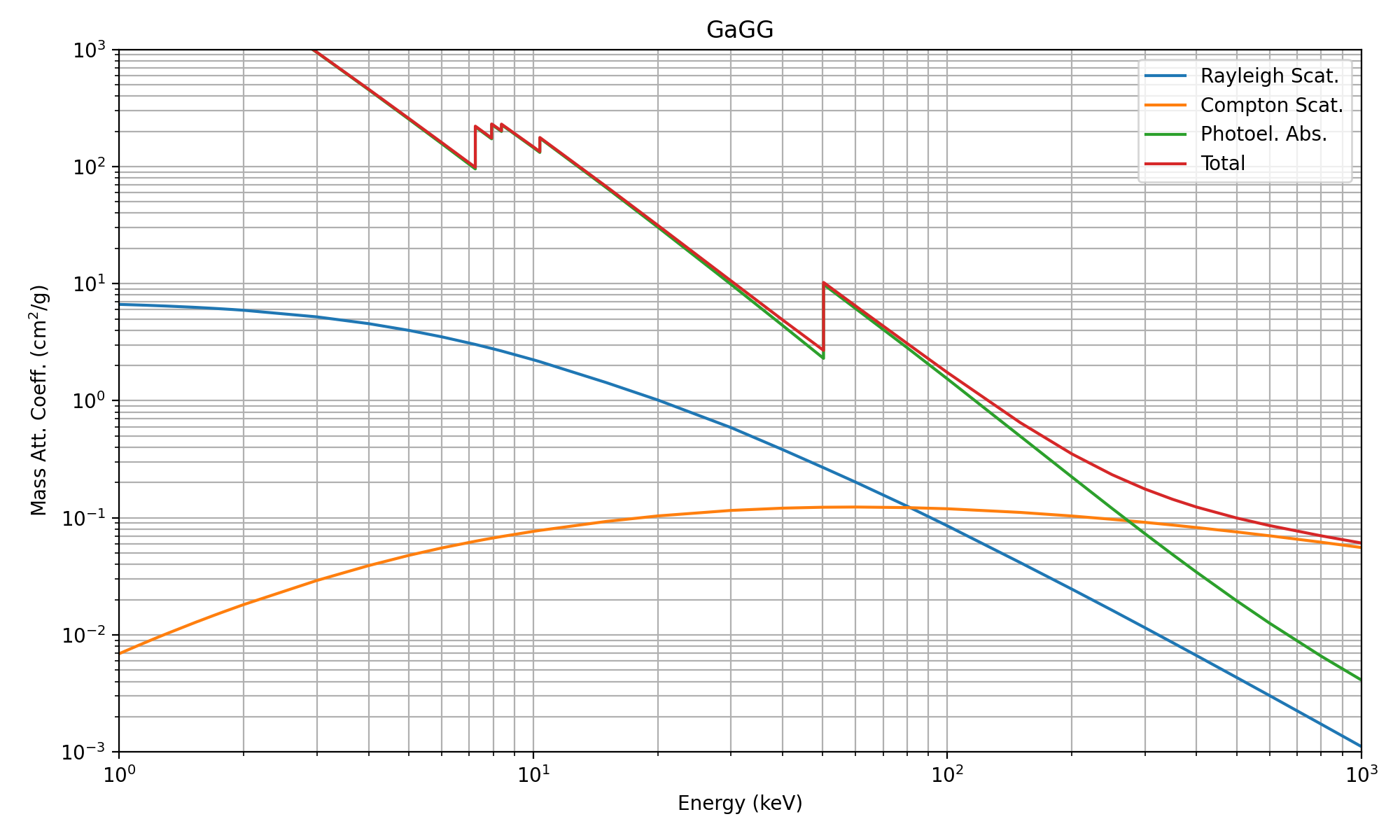}
\end{tabular}
\caption{Mass attenuation coefficients for Rayleigh and Compton scattering and for photoelectric absorption of LiH, BC-404 plastic scintillator (and the equivalent EJ-204), Si, CdTe (and the similar CdZnTe), CsI and GAGG(Ce).} \label{fig:ComptonVsPhoto}
 \end{center}
 \end{figure}

Figure~\ref{fig:ComptonVsPhoto} shows the mass attenuation coefficients for Rayleigh and Compton scattering and for photoelectric absorption of LiH, plastic scintillator (under the commercial name BC-404 or EJ-204), Si, CdTe (and the similar CdZnTe), CsI and Gd Al Ga garnet (known as GAGG). In light materials like LiH (a typical passive scatterer) and BC-404 or EJ-204 (active scatterers), the scattering process is favoured down to about 9 keV and 21 keV, respectively. In some applications Si is used as scatterer, but the scattering energy threshold rises to about 57 keV. CdZnTe and CdTe are proposed usually as absorbing materials. They are also proposed as scatterers in imager detectors with polarimetric capability as a by product of Compton scattering taking place in the detector, but efficiently only above $\sim$260 keV for both materials. Inorganic scintillators like CsI and GAGG(Ce) are suitable as absorbing materials due to their high atomic number.

According to the definition in Ref.~\cite{1995NIMPA.366..161C}, Compton polarimeters can be classified into two categories, depending on the materials used in the scatterer and in the absorber. A single-phase polarimeter has both the scatterer and the absorber made of the same low $Z$ material, whereas a dual-phase polarimeter comprises the scatterer made of a low $Z$ material coupled to an absorber made of a high $Z$ material. This design maximizes scattering and absorption probability in both materials, respectively.

One can choose as scatterer a passive light material, like Li, LiH or Be. Detectors based on these materials are usually called Rayleigh or Thomson polarimeters. They operate in the medium hard X-ray energy band between 10 to 35 keV typically. It is not possible to read out a signal from such a passive scatterer. On the other hand, if an active scattering material is used, the capability to read out small amplitude signals of Compton deposit in the scatterer is required. In this case, an efficient light collection and an adequate quantum efficiency are crucial to reach a low energy threshold of detection. For example a 20 keV photon scattered with a polar angle $\theta = 90^\circ$ in a plastic scintillator like BC-404 deposits about 750 eV in the scatterer (see Equation~\ref{eq:deltaE}). This energy is converted into a few photons (e.g. $\sim$10 photons in BC-404~\cite{Fabiani.2013vom}) that need to be efficiently collected and detected. Such a capability, also known as {\it tagging efficiency}~\cite{Chattopadhyay2012,Fabiani.2013vom}, is the efficiency of detecting a signal in the scatterer when the scattered photon is detected in coincidence in the absorber. Light attenuation and reflection losses in the scatterer material reduce the amplitude of the light signal before readout. In addition, the light-yield becomes non-linear at low energy, as described by Birks' Law~\cite{2009NIMPA.600..609M}.

Depending on the aspect ratio of the scintillating element, a fraction of the scintillation photons reaches the photo-sensor by means of total internal reflection. To improve light collection a wrapping with reflecting or diffusing materials~\cite{Janecek2012} is employed. In addition to wrapping, the optical contact between the scintillating element and the photo-sensor surface is crucial. A coupling material is used in order to minimize optical loss. This is achieved by matching the refraction index between the sensor window and the scintillating material and by selecting a material that provides a good transmittance over the emission spectrum of the scintillator. Another feature to take into account is the cross-talk between channels e.g. in a multi anode photomultiplier tube. This effect arises mainly due to the broadening of the beam of photoelectrons flowing from the photocathode to the first dynode, but it can be induced also by incident light spreading in the glass entrance window\footnote{``PHOTOMULTIPLIER TUBES Basics and Application'', 4th edition by Hamamatsu}.

We report in Table~\ref{tab:scatpoltab} some relevant examples of scattering polarimeters or instruments with polarimetric capability that underwent some polarimetric calibration before launch. An example of single-phase polarimeter, PoGOLite, is shown in Figure~\ref{fig:single_phase}. A dual-phase scattering polarimeter is shown in Figure~\ref{fig:Compton_schematic} where the central scattering stage (yellow) and surrounding absorption stage (grey) are indicated with different colors. In case an active scatterer is included in the detector, the central yellow rod can produce a signal after the interaction of a photon. If the scatterer is passive, no signal will be emitted.

\subsection{3.2 Read out sensors for scattering polarimeters}\label{sec:ReadoutSensor}
As described in the previous section, different materials and detectors can be employed in the design of a scattering polarimeter. When designing the
readout for a light signal from a scintillating element, the amplitude of the signal is a crucial parameter to take into account. Typically one is interested in reading the small amplitude signals resulting from the energy deposited in the scattering stage (low threshold). This can be achieved by employing multiplication detectors like Photo-Multiplier Tubes (PMTs). When the design includes a matrix of scatterers, it can be convenient to consider pixelated Multi-Anode Photo-Multiplier Tubes (MAPMTs). PMTs and MAPMTs can reach a gain of the order of 10$^6$ and are characterized by a very low dark current ($\sim$ nA). They are not particularly affected by radiation damage, that can induce the opalescence of the glass entrance window. The silica glass windows are less affected by the transmittance reduction. There is no virtually variation below 300 nm (and $<10$\% between 200 and 300 nm) after an irradiation with a dose of $2.0 \times 10^5$ Gy with X-rays ($^{60}$Co) and a dose of $1.4 \times 10^{14}$ Gy with neutrons\footnotemark[4].

Reinforced (rugged) designs are typically available to survive launch vibrations without reduction of performance. It may be difficult to fit PMTs in size-constrained applications (e.g. CubeSats) because they require a high voltage power supply ($\sim$1~kV) and the smallest devices have a minimum volume of some cubic cm. Moreover, the response of the PMT is strongly affected by external magnetic fields, and high-permeability shielding (e.g. mu-metal) is usually required.

Patented in 1996, silicon photomultipliers (SiPM) have attracted considerable attention as feasible replacements for PMTs. Their strengths are a smaller dimension (essentially the thickness of a Si wafer), a medium voltage ($< 100$ V) power supply and insensitivity to external magnetic fields. However, SiPMs are typically characterized by a larger dark current~\cite{2019NIMPA.94362376K}, which can be reduced by operating the sensor at a low temperature. The signal amplitude of the absorption phase, if a threshold higher than 40 -- 50 keV is acceptable, can be read out by such devices. The radiation damage increases the dark current of SiPMs thus worsening their noise performance  and spectroscopic capability \cite{Andreotti2014,Lacombe2019,Qiang2013,Ulyanov2020}. For this reason, radiation damage is a limitation that keeps PMTs and MAPMTs still competitive for in orbit use if an energy threshold lower than about 40 -- 50 keV is required~\cite{Mitchell2019}.

Other silicon sensors used to read out organic scintillators of the absorption stage in a polarimeter in orbit are the Avalanche Photo-Diodes (APDs) \cite{Kataoka2010,Kawai2012,Yatsu2014}. They do not present the criticalities caused by the radiation damage that occurs for SiPMs \cite{Chen2007,Yang2019}. Silicon Drift Detectors (SDD) have recently been proposed as readout sensors for inorganic scintillators \cite{Evangelista2020,Fuschino2020}.

Silicon detectors can be exploited as a scattering volume in a Compton polarimeter. This is the case of Double Sided Silicon Strip Detectors (DSSDs) used in Compton Telescopes that allow the Compton interaction point to be identified. Usually these types of telescopes are not designed as polarimeters but their polarisation sensitivity naturally arises due to the dependence on Compton scattering. Cross-strip Ge detectors will be used for the Compton Spectrometer and Imager (COSI) NASA Small Explorer mission~\cite{Tomsick2021}. In this case they act both as scatterer and absorber.

Sensors based on CdTe/CdZnTe are employed both in the absorbing stage to read out scattered photon, but also in single-phase detectors providing both scattering and absorbing signals~\cite{Caroli2018}. Such designs include 2D and 3D CdZnTe/CdTe spectroscopic imagers with coincidence readout logic to handle scattering events and to perform simultaneously polarisation, spectroscopy, imaging, and timing measurements. However, the mass attenuation coefficient of Compton scattering in CdTe and CdZnTe detectors equals the photoelectric absorption at $\sim$260 keV, thus the scattering process starts to be effective at energy about one order of magnitude higher than with plastic scintillators.
Gas detectors are also suitable to perform Compton polarimetry, both as detectors of the absorbing stage for photon energy lower than few tens of keV and as detector of the scattering stage \cite{Komura2017,Tanimori2004,Tanimori2015,Tanimori2017}.

In the next Sections we will discuss some relevant examples of instruments that include previously discussed sensor technologies.

\subsubsection{3.2.1 Single-phase scattering polarimeters}
\label{sec:SinglePhase}

POLAR \cite{2017ICRC...35..668X,2018APh...103...74X} and its upgrade POLAR-2 \cite{2022icrc.confE.580D,Hulsman2020} are typical examples of single-phase polarimeters. POLAR was a compact wide-field polarimeter developed by an international collaboration~of Switzerland, China and Poland, launched on 15 September 2016 on-board the Chinese space laboratory Tiangong-2. POLAR measured the linear polarisation of hard X-rays from GRBs and transient sources between 50 keV and 500 keV using 25 identical modules comprising 64 plastic scintillator bars read out by MAPMTs. The energy range was optimized for the detection of the prompt emission of the gamma ray bursts. The design is now being upgraded towards POLAR-2 with the aim to increase the effective area by an order of magnitude. POLAR-2 will replace the MAPMTs with SiPMs, increase of sensitive volume and will include many technological upgrades. POLAR-2 is expected to be operative on board the Chinese space station in 2024.

\begin{figure}[tb]
\begin{center}
    \includegraphics[width=.75\linewidth]{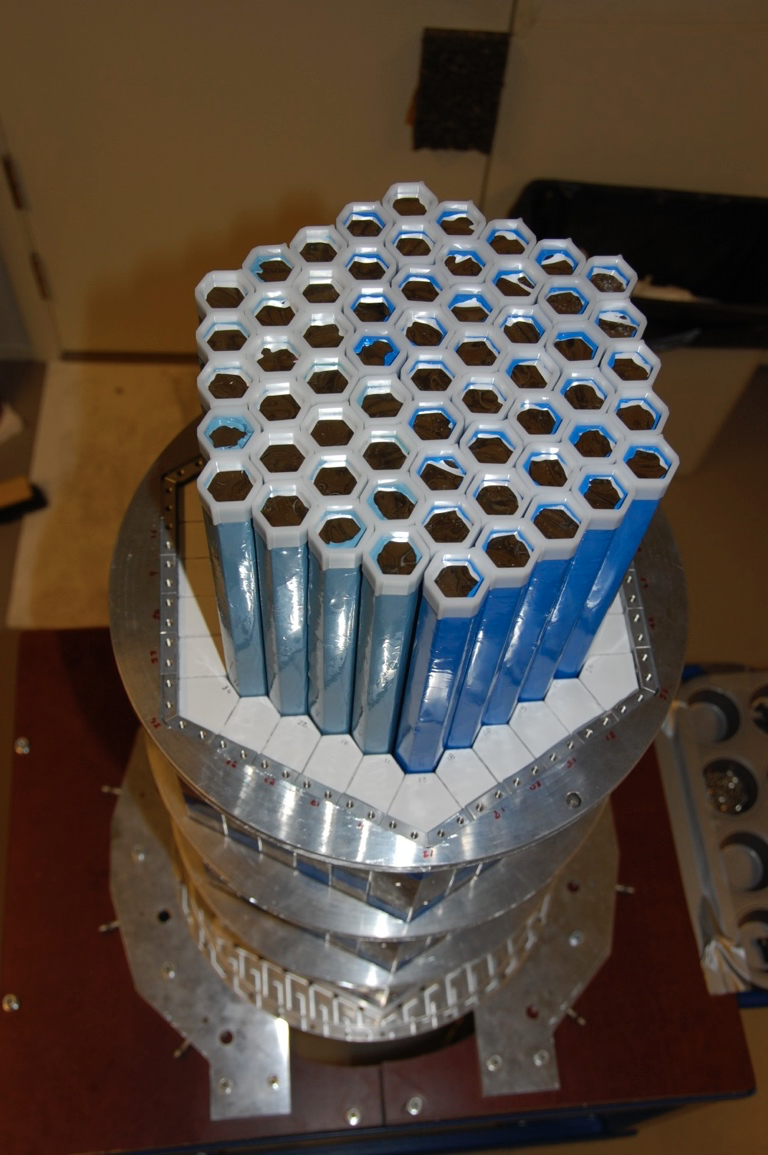}
\end{center}
\caption{The PoGOLite detector is an  example of a single-phase polarimeter. An array of hollow active plastic scintillator collimators ($\sim$3~cm opening) is seen protruding from the BGO anticoincidence shield. A solid plastic scintillator scatterer (not visible) is attached to the base of each collimator.}
\label{fig:single_phase}
\end{figure}

Other examples of single-phase polarimeters are PoGOLite~\cite{Chauvin2016b} and PoGO+ \cite{Chauvin2016,Chauvin.2017,2018NatAs...2..652C,Chauvin2018}, balloon-borne experiments that flew respectively in 2013 and 2016. They were conceived for pointed observations, with a collimator placed in front of an array of low $Z$ plastic scintillator bars. High $Z$ scintillators were present, but employed as anti-coincidence system for background rejection. A picture of the PoGOLite detector is shown in Figure~\ref{fig:single_phase}.

An example of CdZnTe detector is the AstroSat imager CZTI, a coded aperture telescope designed for hard X-ray observations, calibrated also on ground for polarisation measurements~\cite{Vadawale2015}. It consists of a pixelated detector plane with a geometric area of 976 cm$^2$, a pixel thickness of 5 mm and a size of 2.5~mm$\times$2.5~mm. The polarisation measurement is performed by detecting coincident events among neighbouring pixels. In the 100 -- 380 keV energy range the recoil electron has an energy much lower than the scattered photon. Therefore, the pixel with the lowest energy deposition is considered as the scattering pixel, while absorption is assumed to take place in the other fired pixel. Pixels have a square geometry and the instrument does not rotate. This detector reported the measurement of the Crab pulsar and nebula polarisation in the 100 -- 380~keV energy band~\cite{Vadawale2018a} and the measurement of polarisation of the prompt emission of GRBs~\cite{Chattopadhyay2019}. The sensitivity of the AstroSat CZTI as polarimeter is estimated with Geant4 simulations and is given in terms of Modulation Factor and MDP in Ref.~\cite{2014ExA....37..555C}.

Compton telescopes are a class of hard X-ray / $\gamma$-ray detectors that exploits Compton scattering for imaging and spectroscopy and have polarimetry as a by-product of the detection technique. COSI~\cite{Tomsick2021}, approved at the end of 2021 by NASA in the SMEX program, employs double sided Ge detectors both as scatterers and absorbers to perform polarimetry in the  0.2 -- 5 MeV energy range.

Compton polarimetery has also been pursued in instruments not designed nor calibrated for polarimetry. For example, the simulation of the INTEGRAL/SPI response to a linearly polarised emission at any position angle is reported in Ref.~\cite{2013ApJ...769..137C}. The Compton mode of the pixellated detectors in the INTEGRAL/IBIS instrument has been used for galactic sources \cite{2008ApJ...688L..29F,2015ApJ...807...17R} and GRBs \cite{2009ApJ...695L.208G}.

\subsubsection{3.2.2 Dual-phase scattering polarimeters}\label{sec:DualPhase}

Dual-phase scattering polarimeters can be subdivided into two sub-groups depending on whether the scatterer signal is read out (active scatterer) or not (passive scatterer).
Polarimeters with passive scatterers operate in the elastic (Rayleigh) scattering regime. Identifying the temporal coincidence of signals from the scatterer and the absorber is not possible. These instruments are sensitive to lower energies and the lower energy threshold depends on the transition between the photoelectric and Rayleigh scattering probabilities (see Figure~\ref{fig:ComptonVsPhoto}). Materials suitable for scattering elements are typically light elements like Be, Li, LiH and different geometries are possible~\cite{Vadawale2010}. Due to the fact that only the signal in the absorber is read out, background is typically relevant.

POLIX \cite{Paul2010,2016cosp...41E1533P} is an example of such a kind of polarimeter operating in the 5 -- 30 keV energy range. The absorption stage consists of proportional counters. X-Calibur \cite{Beilicke2014,Beilicke2015,Endsley2015,Guo2013,Kislat2017} and the upgrade XL-Calibur \cite{Abarr.2021} employ a Be rod surrounded by a circumadjacent array of CdZnTe detectors as absorbers. SPR-N is another example of a polarimeter with a passive scatterer and was launched in 2001 on board the past mission CORONAS-F. SPR-N measured the X-ray polarisation in the energy ranges 20 -- 40 keV, 40 -- 60 keV and 60 -- 100 keV \cite{Zhitnik2006,Zhitnik2014}. The scattering elements were beryllium plates surrounded by six scintillation detectors. Background discrimination due to charged particles was performed by means of a phoswich detector (CsI(Na)/plastic scintillator). This detector was devoted to solar flares X-ray polarimetry.

In the energy range where Compton scattering is effective, the capability to read out the energy deposit in an active scatterer allows to significantly reduce the background of charged particles. There are many examples of such detectors as for example PolariS~\cite{Hayashida2016} (based on the PHENEX design \cite{Gunji2007,Gunji2008,Kishimoto2009}) that consists of two kinds of scintillator pillars, plastic and GSO, read out by MAPMTs.

New polarimeters are planned on board CubeSat missions like CUSP (CUbeSat Solar Polarimeter~\cite{Fabiani_et_2022}), approved in 2021 for a phase A study by the Italian Space Agency. Plastic scintillator bars read out by MAPMTs act as scatters. The absorption stage consists of GAGG(Ce) elements coupled to APDs.

Multi-purpose instruments eAstrogam~\cite{DeAngelis2018} and AMEGO~\cite{McEnery2019} are Compton telescopes based on the dual-phase design. The trackers of both experiments are based on layers of double sided Si strip detectors that allow to identify the Compton interaction point and read out the track of the recoil electron. The absorption of the scattered photon is performed by a segmented calorimeter. In this context it is also worth notice the SGD Compton camera on board Hitomi \cite{Aharonian2018,Tajima2010} consisting of Si (scatterer) and CdTe (absorber) detectors. The performance of a prototype of the Hitomi SGD in measuring the X-ray polarisation using polarised photons at the Super Photon Ring – 8 GeV (Spring8) synchrotron radiation facility is shown in Ref.~\cite{2010NIMPA.622..619T}. The photon energy was 168 $\pm$ 1.4 keV, with a degree of polarisation of 92.5 $\pm$ 0.3\% . The authors find a modulation factor of $\sim 82$\%.

Compton scattering can be exploited not only in solid state detectors, but also in gas detectors such as the Electron Tracking Compton Camera (ETCC) \cite{Komura2017,Tanimori2004,Tanimori2015,Tanimori2017}. The ETCC is a gaseous three dimensional Compton camera with a gas cell some tens of cubic centimetres filled with an Ar-based gas mixture. A pixel scintillator array made of Gd$_2$SiO$_5$:Ce (GSO) acts as an absorber of the Compton scattered photons, while the track of the recoil electron is read out by means of a TPC based on a micro-pattern gas detector (cross reference to the chapter by Black and Zajczyk in the same volume).
The ETCC is expected to measure the initial direction of the recoil electron more finely than solid-state trackers allowing a better angular resolution and thus a significant reduction of background, thereby improving the polarimetric sensitivity. Moreover, the coincident measurement of the scattered photon direction and the recoil electron allow the incoming direction to be constrained event by event and thus allow to correct the spurious component in the modulation due to inclined penetration for off-axis observations \cite{Lei1997,Muleri2014}. A wide field of view of up to 2$\pi$ sr allows the observations of persistent sources as well as transient objects including GRBs.
Ref.~\cite{Tanimori2015} estimates for an ETCC comprising four 50 cm$^3$ cells an effective area of 280 cm$^2$ at 200 keV and a Minimum Detectable Polariation of about 10\% for a 13 mCrab for an observation of 10 Ms.

Finally, the performance of stacked imaging detectors for hard X-rays used as polarimeters is discussed in Ref.~\cite{2012ApJ...751...88M}. In the article, the stack is composed of two imaging detectors, a low energy imager e.g. made of Si and a high energy one e.g. composed of CdTe or CdZnTe. The authors find that this geometry has an extremely low sensitivity to polarisation, with a quality factor $Q$ (defined in Section 2) of the order of $10^{-2}$ due to the small scattering probability in the detector for low energy photons.

\begin{table}
\small
\caption{Examples of scattering polarimeters. Acronyms in the column Science Objectives: AGN is Active Galactic Nucleus, BH is Black Hole, GRB is Gamma Ray Burst, NS is Neutron Star, PWN is Pulsar-Wind Nebula, SNR is Supernova Remnant, XRB is X-ray Burst} \label{tab:scatpoltab}
\centering
\scalebox{0.95}[0.95]{\begin{tabular}{p{1cm}p{1.5cm}p{2cm}p{1.5cm}p{2.2cm}cp{2cm}c}
\toprule
\textbf{ }      &\textbf{Name}  &\textbf{Time Schedule} &\textbf{Optics}& \textbf{Field of view} & \textbf{Energy Range}& \textbf{Science Object.}& \textbf{References}\\
\midrule
\textbf{Thomson}        &\textbf{ }     & \textbf{ }    & \textbf{ }& \textbf{ }& \textbf{ }& \textbf{ }& \textbf{ }\\
\midrule
& \multirow{2}{*}{POLIX}        & \multirow{2}{*}{launch 2019}& \multirow{2}{*}{no}      & \multirow{2}{*}{$3^\circ \times 3^\circ$}    & \multirow{2}{*}{5--30~keV}& accretion powered pulsars, BH & \multirow{2}{*}{\cite{Paul2010,2016cosp...41E1533P}}\\ \midrule
& SPR-N & launched 2001 & no     & -   &20--100 keV& solar flares       & \cite{Zhitnik2006,Zhitnik2014}\\
\midrule
\textbf{Compton~single~phase}        &\textbf{ }     & \textbf{ }    & \textbf{ }& \textbf{ }& \textbf{ }& \textbf{ }& \textbf{ }\\
\midrule
& POLAR &launched 2016  & no            & $\sim$1/3 of full sky & 50--500 keV   & GRB & \cite{2017ICRC...35..668X,2018APh...103...74X}\\ \midrule
& POLAR-2       &launch 2024    & no            & - & 20--800 keV       & GRB & \cite{2022icrc.confE.580D,Hulsman2020}\\ \midrule
& PoGOLite &launched 2013       & no     &$\sim 2^\circ \times 2^\circ$ &20--240 keV &Crab emission     & \cite{Chauvin2016, Chauvin.2017, 2018NatAs...2..652C}\\ \midrule
& \multirow{3}{*}{PoGO+}        & \multirow{3}{*}{launched 2016} & \multirow{3}{*}{no}  & \multirow{3}{*}{$\sim 2^\circ \times 2^\circ$}  &\multirow{3}{*}{20--150 keV} &Crab pulsar and nebula, Cygnus X-1     & \multirow{3}{*}{\cite{Chauvin2016, Chauvin.2017, 2018NatAs...2..652C}}\\ \midrule
& AstroSat CZTI                 & \multirow{3}{*}{launched 2015} & \multirow{3}{*}{no}  & \multirow{3}{*}{-}  &\multirow{3}{*}{100--380 keV} &Crab~pulsar~and nebula,  bright X-ray sources & \multirow{3}{*}{\cite{Vadawale2015,Vadawale2018a}}\\ \midrule
& COSI (Compton telescope)       & \multirow{3}{*}{launch 2025} & \multirow{3}{*}{no}    & \multirow{3}{*}{wide}  &\multirow{3}{*}{0.2--5 MeV} &Galactic center and bulge, SNR, GRB, AGN & \multirow{3}{*}{\cite{Tomsick2021}}\\ \midrule
\textbf{Compton~dual~phase}&\textbf{ }    & \textbf{ }    & \textbf{ }& \textbf{ }& \textbf{ }& \textbf{ }& \textbf{ }\\
\midrule
& \multirow{2}{*}{X-Calibur}    &launched 2014, 2016    & \multirow{2}{*}{yes} & \multirow{2}{*}{8 arcmin at 20 keV} & \multirow{2}{*}{20--60 keV} & BHs,~NSs,~magnetars, AGN jets &\multirow{2}{*}{\cite{Beilicke2014,Beilicke2015,Endsley2015,Guo2013,Kislat2017}}\\ \midrule
& \multirow{2}{*}{XL-Calibur}   & Launched 2022 & \multirow{2}{*}{yes} & \multirow{2}{*}{10 arcmin} & \multirow{2}{*}{15--80 keV} & BHs,~NSs,~magnetars, AGN jets &\multirow{2}{*}{\cite{Abarr.2021}}\\ \midrule
& \multirow{3}{*}{PolariS} &\multirow{3}{*}{assessment (2022)}  & yes/no~(also~a wide field polarimeter) & \multirow{3}{*}{10 $\times$ 10 arcmin$^2$} & \multirow{3}{*}{10--80 keV}      & SNRs,~BHs,~accretion in X-ray pulsars, GRBs  & \multirow{3}{*}{\cite{Hayashida2016}}\\ \midrule
& \multirow{3}{*}{GRAPE}        & launched 2014 and 2016; LEAP new design~assessment & \multirow{3}{*}{no}       & \multirow{3}{*}{wide} & \multirow{3}{*}{50--500 keV}  & transient sources, GRBs, solar flares & \multirow{3}{*}{\cite{Bloser2010,Bloser2009,Connor2010,Kishimoto2007,McConnell2013,McConnell2014}}\\ \midrule
& \multirow{3}{*}{LEAP} & Phase A (2022) & \multirow{3}{*}{no}      & \multirow{3}{*}{wide} & \multirow{3}{*}{50--500 keV}  & transient sources, GRBs, solar flares & \multirow{3}{*}{\cite{McConnell2016b,Onate2021}}\\ \midrule
& PHENEX &launched 2006         & no     & 4.8$^\circ$ & 40--200 keV    & Crab Nebula & \cite{Gunji2007,Gunji2008,Kishimoto2009}\\
\midrule
& \multirow{2}{*}{GAP}  & \multirow{2}{*}{launched 2010} & \multirow{2}{*}{no}  & \multirow{2}{*}{$\pi$ sr} & \multirow{2}{*}{50--300 keV} & GRBs,~Crab~pulsar and nebula       & \multirow{2}{*}{\cite{Yonetoku2010,2011PASJ...63..625Y,Yonetoku.2011,Yonetoku2012}}\\ \midrule
& $\multirow{2}{*}{SPHiNX}$ & Phase-A/B1 (2019)    & \multirow{2}{*}{no}    & \multirow{2}{*}{$\pm 60^\circ$} & \multirow{2}{*}{50--500 keV} & \multirow{2}{*}{GRBs}       & \multirow{2}{*}{\cite{Xie2018}}\\ \midrule
& PENGUIN-M     & launched~2009, lost& no & - & 20--150 keV & solar flares & \cite{Dergachev2009}\\ \midrule
& PING-P        & launch~after~2025 & no & -    & 20--150 keV &solar flares & \cite{Kotov2016}\\ \midrule
& Hitomi SGD (Compton camera)& launched 2016 & yes & narrow      & 50 keV -- 200 keV& Pulsars, BH XRBs,AGNs, supernova remnanats & \cite{Aharonian2018,Tajima2010}\\  \midrule
& eASTROGAM (Compton telescope)& assessment (2022) & no & wide        & 300 keV -- 3 GeV& persistent and transient sources & \cite{DeAngelis2018}\\  \midrule
& AMEGO (Compton telescope)& assessment  (2022) & no & wide    & 200 keV -- 1 GeV & Pulsars, XRBs, AGNs,
PWNs, GRBs & \cite{McEnery2019}\\  \midrule
& ECCT (Compton telescope)& assessment (2022) & no & 2$\pi$   & $>$ 100 keV & Pulsars, XRBs, AGNs,
PWNs, GRBs & \cite{Komura2017,Tanimori2004,Tanimori2015,Tanimori2017}\\
\bottomrule
\end{tabular}}
\end{table}

\subsection{3.3 Electronics}

The front-end electronics (FEE) is the analogue stage directly connected to the detector, with the function to amplify and shape the signals produced by the detector elements. Given the variety of detector types described above for the polarimeter, it is impossible to sketch the design of a front-end electronics that may fit for all because the design needs to be adjusted on the specific characteristics (amplitude and shape) of the signals produced by the detector elements. The use of miniaturised Application Specific Integrated Circuits (ASICs) is widespread. The choice of circuit depends on the required functionality, the number of independent channels to read and the power constraints.

Since the Compton effect involves two interactions, the use of an active scattering stage in coincidence with the absorption stage for a time window $\le 1 \; \mu$s allows a large fraction of the background to be rejected. Thus one of the main functions of the Back-end Electronics (BEE) is the realisation of the coincidence window between scattering and absorption stages. The interested reader is addressed to Ref.~\cite{Leo_1987} for details about the coincidence systems. If the detector includes PMTs, another function of the BEE is the generation of the required high voltage, of the order of hundreds to few thousands Volts. Ref.~\cite{Leo_1987} explains how the high voltage stability affects the gain stability of a PMT and shows examples of voltage divider networks. In case a solid state sensor is involved, e.g. an APD or a SiPM, the bias voltage is much lower, of the order of 10 -- 100 V. Due to the strong variation of the gain with temperature, SiPMs require a compensation circuit. Other functions of the BEE are e.g. analogue to digital conversion of the signals from the FEE, configuration of the ASICs by programming the registers or generation of the low voltages.

\section{4 Systematic effects and calibration}

\begin{figure}[tb]
\begin{center}
    \includegraphics[width=0.9\linewidth]{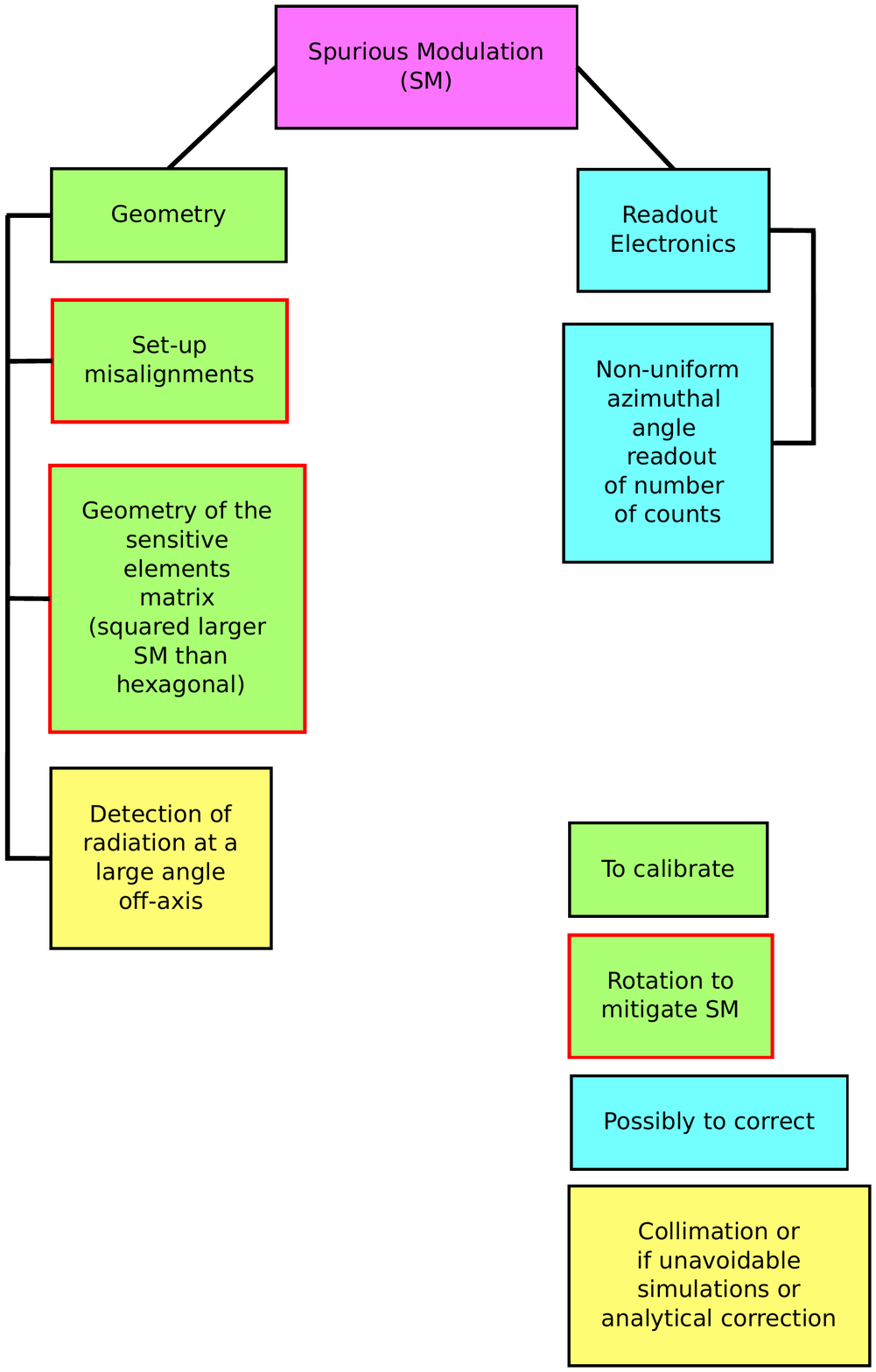}
\end{center}
\caption{Diagram summarising the main causes of the spurious modulation in a Compton-based polarimeter and the mitigation methods.}
\label{fig:SM}
\end{figure}

Laboratory calibration of a Compton scattering polarimeter with both polarised and unpolarised beams is required to ensure that the uncertainty on polarisation parameters is limited by the number of registered Compton scattering events, and not by instrument systematics.
If the azimuthal detector response is not uniform, systematic effects will generate a spurious polarisation signal.
For example, a source with $p_0 = 10 \%$ will generate a signal with 5\% modulation in a polarimeter with $\mu_{100}=0.5$. This means that systematic effects must be identified and calibrated to 1\%-level or better. The polarimetric response of flight instruments should therefore be studied in detail in the laboratory before observations of celestial sources are undertaken. We summarise in Figure~\ref{fig:SM} the main causes of spurious modulation and the most important mitigation methods described in this section.

\begin{figure}[tb]
\begin{center}
    \includegraphics[width=0.75\linewidth]{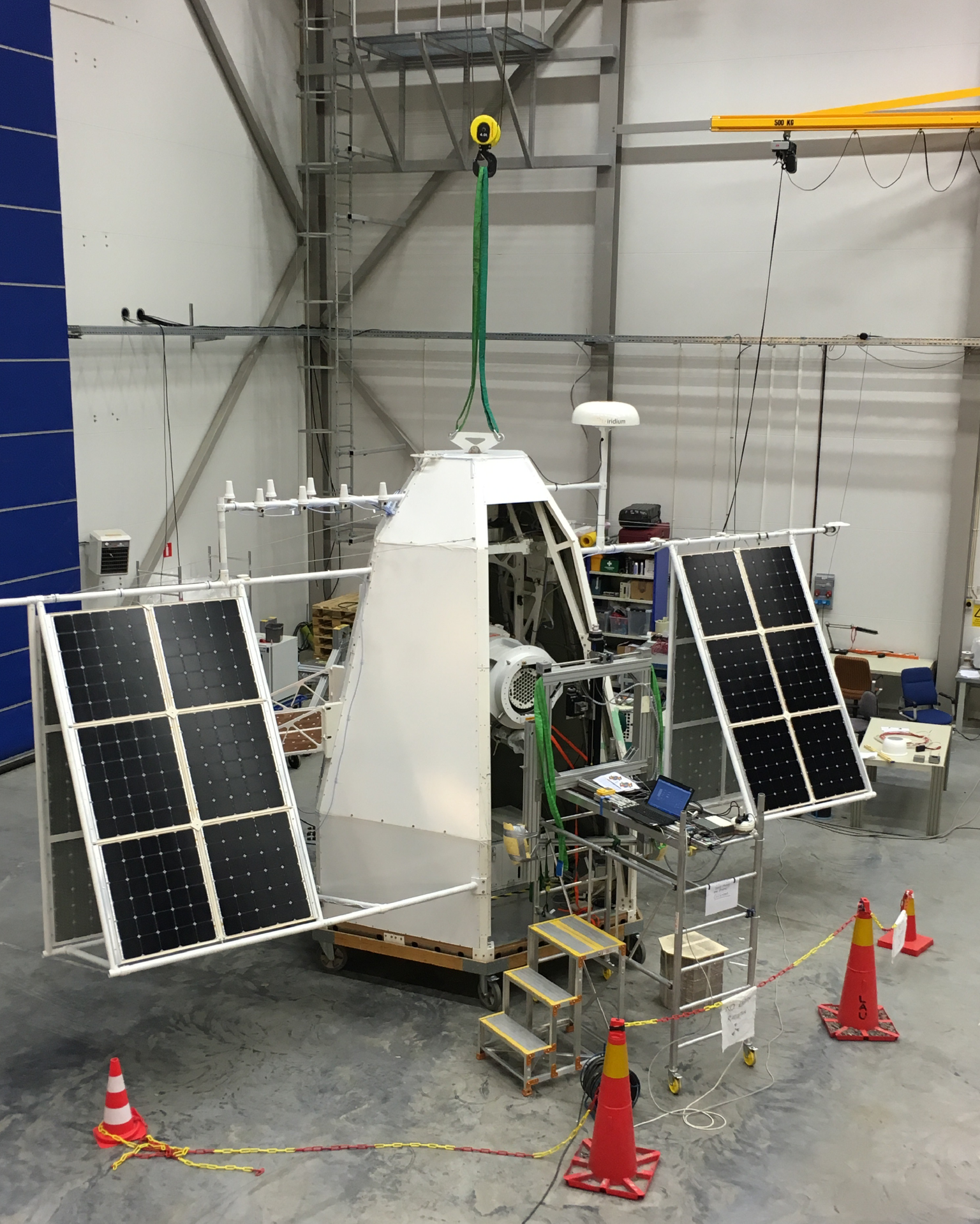}
\end{center}
\caption{The PoGO+ balloon-borne polarimeter undergoing final calibration tests using a scattered $^{241}$Am beam. A picture of the detector is shown in Figure \ref{fig:single_phase}. The source is mounted on a x-y scanning table, to allow the entire sensitive area of the polarimeter ($\sim$ 300 ~cm$^2$) to be systematically illuminated by the small diameter ($\sim1$~cm) beam. The calibration was performed at the Esrange Space Centre in the days leading up to the launch, in order to confirm that transport operations did not affect the polarimeter response.}
\label{fig:calib}
\end{figure}

There is a variety of methods for generating beams of calibration X-rays. A convenient approach is to use radionuclide sources. The type of radionuclei dictates the photon energy. Some sources provide predominantly monoenergetic emission, e.g. $^{241}$Am has dominant emission at 59.5~keV, while others emit at a number of energies, e.g. $^{152}$Eu.
Photons from the source can be directed onto a low atomic number material (e.g. plastic scintillator), whence photons Compton scattered through 90$^\circ$ are 100\% polarised. For a $^{241}$Am source, the scattered beam ($\sim 53$~keV) can be selected with a collimator and illuminate the polarimeter under test \cite{Chauvin.20160pr,Chauvin.2017qw,Vadawale2015}. The calibration of a recent balloon-borne polarimeter is exemplified in Figure~\ref{fig:calib}.
During prototyping it may be useful if the emission of a polarised source photon generates a trigger signal for the polarimeter data acquisition system. This is possible if the plastic scintillator scatterer is read out by a photosensor, although, depending on the source emission energy, the detection efficiency may be low due to the small energy deposited in the Compton scattering interaction. At higher energies, a $^{22}$Na source emits 511~keV photons back-to-back, and the polarisation of the photons is correlated. If the polarisation properties of one beam are determined, the other beam can be used to characterise an instrument~\cite{Orsi.2011}.

X-ray tubes are suitable sources to produce X-rays with a high flux in a laboratory environment. These devices produce characteristic lines, depending of the anode material, superimposed to a continuum energy spectrum. The radiation polarisation depends on the energy and the geometry of the Bremsstrahlung process. The polarisation vector of a Bremsstrahlung photon is parallel (low photon energy) or orthogonal (high photon energy) to the plane identified by the photon wave vector and the electron momentum. The measured polarisation degree depends on the integral of the observed radiation along the line of sight~\cite{Gluckstern1953}.

The generation of unpolarised X-rays (more precisely with a very low polarisation degree typically $<0.5 \% $) can be achieved by means of X-ray tubes with an edge-on geometry. The electron beam impinging on the anode is orthogonal to the output window. Due to the circular symmetry with respect to the line of sight, the polarisation cancel out and the observed beam is unpolarised. In general X-ray tubes have a geometry such that the electron beam is orthogonal to the line of sight. In this case the polarisation degree varies across the energy spectrum. Bragg diffraction at $45^\circ$, by matching the photon energy with the crystal lattice, allows $100 \%$ polarised photons~\cite{Muleri2022} to produced. Another possible technique that allows to polarise X-rays involves scattering on a passive scatterer element at $90^\circ$ from the line of sight.

At a synchrotron beam facility the intensity and energy of the beam can be varied, allowing a more systematic calibration process. The monoenergetic synchrotron beam can be polarised using a double-scattering silicon monochromator \cite{Beilicke2014,Bloser2009,Kamae.2008,Kole.2017}, resulting in a very high polarisation fraction (typically, $\sim$99\%). A practical challenge is that the beam diameter is small, $\sim1$~mm, compared to the typical dimensions of a polarimeter sensitive area. For this reason, the polarimeter needs to be mounted on a positioning stage to allow the beam to be scanned over the polarimeter.

A Compton polarimeter is often constructed in a modular fashion from an assembly of photon detectors. Before constructing the polarimeter, the photon detection efficiency of individual detector elements can be determined and used to inform the assembly arrangement and optimise the flat-field response. Residual systematics are still inevitable due to, e.g., uncertainties in the detector element response and positioning tolerances during assembly. Moreover, during flight operations, the detector response can change for the following reasons: if detector elements malfunction; due to variations in temperature or magnetic field environment; and (for longer missions) as a consequence of radiation damage and ageing effects. The response of detector elements should ideally be monitored as observations proceed. This is possible using radioactive calibration sources integrated with the polarimeter, 511~keV gamma-rays generated by cosmic-ray positrons impinging on the payload, as well as stimuli which mimic the photon energy deposition, e.g. LED light injected into scintillators.
The POLAR GRB polarimetry mission~\cite{Li.2018otl} included four 100~Bq $^{22}$Na sources, which were used in-flight to calibrate the instrument energy response but did not address the polarimetric response. To date, in-flight calibration of a Compton polarimeter using an integrated polarised source of photons has not been reported, although it is done at lower energies for the IXPE photoelectric polarimeter~\cite{Ferrazzoli.2020}. Using celestial sources for calibration of a null polarimetric response is problematic, since emission from all sources is expected to be polarised at some level (with the probable exception of symmetric clusters of galaxies).

For instruments designed for pointed observation of sources, a straight-forward way to mitigate response systematics is to continuously rotate the polarimeter around the viewing axis. The rotation period should be shorter than the source (or background) variability and many rotations should be accumulated during observations. Rotation provides a smooth distribution of reconstructed scattering angles rather than the discretized curve that would otherwise arise when assuming center-to-center trajectory between hit detector elements.

Rotation is not practical when observing brief transient sources, e.g. second to minute duration prompt phase of gamma ray bursts, and response systematics must instead be addressed using computer simulations. An exception was GAP~\cite{2011PASJ...63..625Y}, whose spacecraft rotated with an angular speed of 1 -- 2 rotations per minute to tension a solar-sail spread using centrifugal effects.

Many polarimeter designs comprise a pixelated detector volume in which time-coincident energy deposits are used to determine the azimuthal scattering angle. If the detector pixels are not azimuthally symmetric, the modulation response will not be harmonic.
In this respect, a close-packed array of hexagonal elements is preferable but difficult to realise for practical reasons, i.e. multi-pixel photosensors and semiconductor sensors usually have square pixels.
For pointed observations, rotation will address the systematics generated by an array of square pixels, otherwise response corrections are neeeded.

A spurious modulation response is also generated if incident photons impinge on the polarimeter at some angle from the viewing axis (`off-axis'). Well-collimated instruments can minimise this problem. For some types of observation this effect is unavoidable.
For example, GRB events occur randomly on the sky and observations require a large field-of-view polarimeter. Since there is no time to slew the viewing axis, the source is likely to be off-axis during observations. A variety of approaches has been followed to remove the resulting response systematics.
For example, POLAR measured the polarisation properties for the prompt emission of 5 GRBs~\cite{Zhang.2019}. Following the scheme used by the GAP mission~\cite{Yonetoku.2011}, polarisation parameters were extracted from the measured modulation curve through a $\chi^2$ comparison with 6060 simulated modulation curves which accounted for the instrument response~\cite{Kole.2017}. This approach required that the POLAR simulation model was well-calibrated~\cite{Li.2018otl}, and included a detailed mass modelling of the host Tiangong-2 space station on which POLAR is mounted. The energy spectrum and sky location of each GRB was determined by other missions, and was used as input to the simulation.

The CZTI instrument on-board the AstroSat mission is primarily designed for X-ray spectroscopy, but can be used as a GRB polarimeter owing to the X-ray transparency of the CZTI support structure~\cite{Chattopadhyay2019}. In this case, modulation curves were corrected by normalising the azimuthal distribution of the GRB by that for simulated 100\% unpolarised radiation, of the same spectrum and incident at the same off-axis angle as the source.

\begin{figure}[tb]
\begin{center}
    \includegraphics[width=0.95\linewidth]{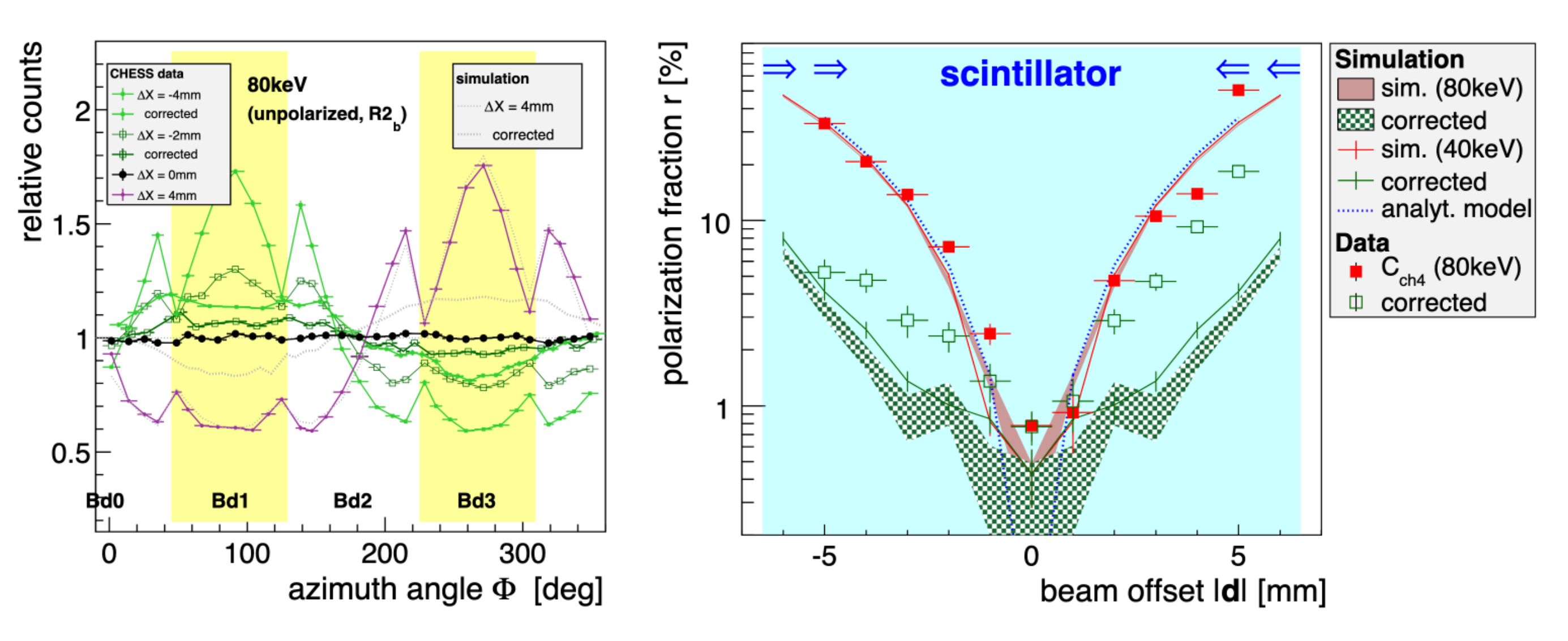}
\end{center}
\caption{The left panel shows distributions of the azimuthal scattering angle measured by the X-Calibur polarimeter as a function of beam off-set (with respect to the optical axis of the polarimeter) for a non-polarised X-ray beam at the CHESS synchrotron facility. Beam offsets of $\sim$2~mm introduce systematic asymmetries in the scattering angle distribution. Note that the asymmetries are not sinusoidal, as expected from a polarised beam, and can therefore be identified through inspection of the modulation curve. In the right panel, the results of a Stokes analysis of the data is presented in order to quantify the level of artificial polarisation created when the beam offset is not accounted for. The reconstructed polarisation fraction is shown as a function of the beam offset. The magnitude of the artificial polarisation signal is greatly reduced for offset magnitudes $<3$~mm, when a correction procedure based on the reconstructed first moments of the scattering events is followed. This observation drives requirements on the alignment stability of the X-ray mirror/polarimeter during flight.
The figure is reprinted from Journal of Astronomical Instrumentation, Vol 3, Matthias Beilicke et al., ``Design and Performance of the X-ray Polarimeter X-Calibur'', 1440008 (2014) \cite{Beilicke2014} with permission from World Scientific Publishing. In the same publication, a similar study is present for a polarised incident beam.
}
\label{fig:optics}
\end{figure}

As a complement to Monte Carlo simulations, a detailed analytical approach has also been developed to study response systematics~\cite{Muleri2014}.
For polarimeters located in the focal plane of X-ray mirror/optics, the impact on polarimetric response of misalignment between the X-ray mirror/optics and the polarimeter needs to be determined. The modulation signal will be affected by unpolarised sources which lie within the telescope field-of-view, but offset from the pointing direction~\cite{Elsner.1990}. As depicted in Figure~\ref{fig:optics} and Ref.~\cite{Beilicke2014}, a synchrotron beam was used to study the effect of an offset unpolarised beam on the X-Calibur polarimeter, which is mounted at the focal point of an X-ray mirror with a 8~m focal length. The arrangement is similar to the follow-on mission, XL-Calibur, shown in Figure~\ref{fig:anti}. X-Calibur observed the accreting neutron star GX301-2~\cite{Abarr.2020p1n} in 2018.
Due to the long focal length of hard X-rays, it is a considerable experimental challenge to conduct end-to-end polarimetric calibration of complete telescope assemblies.
Since X-rays are focussed using grazing incidence reflection, the impact on the polarisation properties of the incident beam has been studied, and found to be negligible for the sensitivity of current instruments~\cite{Almeida.1993,Katsuta.2009}.

Knowledge of $\mu$ directly affects the precision of $p_r$. Since pre-flight calibration is conducted at discrete energies, the determination of $\mu$ for emission from celestial sources requires computer simulations. Since instrument response is complex and often non-intuitive, it is important to benchmark simulations using calibration data. There are many instrumental effects which can act to modify $\mu$, e.g. non-uniform energy thresholds for detector units, low energy quenching effects in scintillators~\cite{2009NIMPA.600..609M}, poorly modelled detector spectral and spatial response, cross-talk between detector channels~\cite{Xiao.2016}, and the presence of passive materials. Balloon-borne instruments, which operate at altitudes of $\sim$ 40~km, must account the effect of changing atmospheric absorption much as the source elevation changes during an observation.
For instruments designed for the observation of GRBs, $\mu$ will depend on the location of the GRB with respect to the polarimeter and the GRB energy spectrum. Since a GRB polarimeter may not be optimised for such measurements, independent observations of the GRB by standard instrumentation may be required. GRBs may therefore be observed with $\mu$ poorly determined. This is exemplified in Ref.~\cite{Mikhalev.2018}, where it is shown that the relative increase in MDP is 1\% (80\%) for a relative uncertainty on $\mu$ of 5\% (25\%).

\section{5 Background estimation and mitigation}

Payloads operated outside of the Earth’s atmosphere are subjected to primary cosmic rays (p, e$^{\pm}$, $\alpha$), secondary cosmic rays (p, e$^{\pm}$, $\gamma$), albedo particles created in the atmosphere (n, $\gamma$), and the cosmic X-ray background. These components and their effect on satellite-borne instrumentation are described in the chapter by Riccardo Campana (cross reference). Here, we only give a brief description of the background components for instruments on balloons and we concentrate on the mitigation measures to reduce the background in Compton scattering polarimeters.

In the hard X-ray band, observations are possible using balloon-borne missions operating at an altitude of $\sim 40$ km, where there are a few g/cm$^2$ of residual atmosphere overburden above the payload. The primary fluxes of particles are present, but attenuated by the atmosphere. The dominant source of background therefore stems from secondary particles produced during interactions of the primary radiation with residual gas molecules in the atmosphere. For this reason, the nature of the background depends on altitude and geomagnetic latitude. The energy spectra and flux of background species depends on the location of the launch site, e.g. background rates for payloads launched from the NASA facility at Palestine, New Mexico ($\lambda \sim 50 \degree$), are lower than those experienced by payloads launched from high latitude facilities, e.g. Esrange in northern Sweden ($\lambda \sim 65\degree$) or McMurdo base in Antarctica ($\lambda \sim 80\degree$). These latter two sites, however, permit long duration flights which provides better signal statistics.

\begin{figure}[tb]
\begin{center}
    \includegraphics[width=0.95\linewidth]{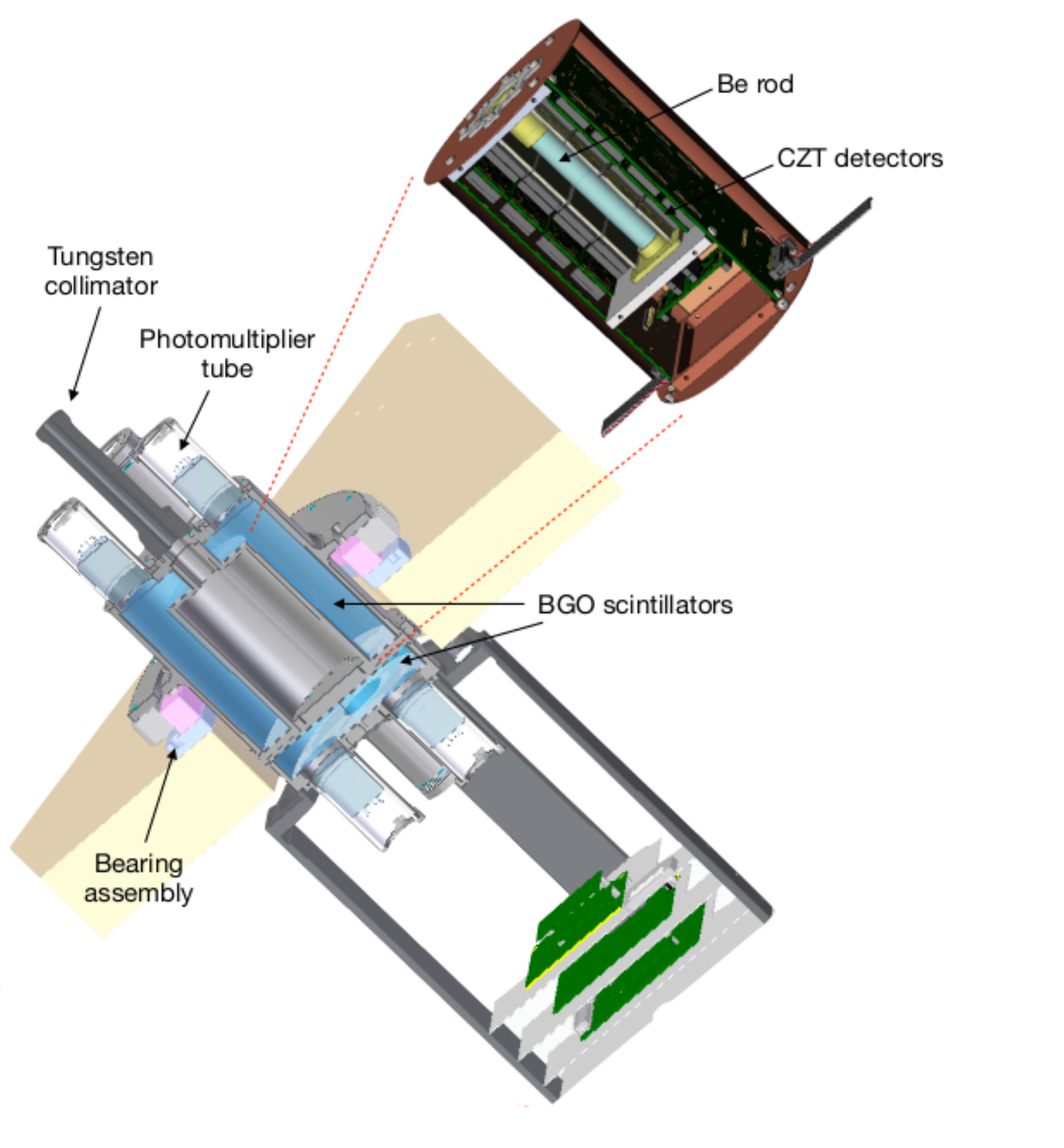}
\end{center}
\caption{An overview of XL-Calibur polarimeter and anticoincidence system~\cite{Abarr.2021}. X-rays incident on the X-ray mirror (not shown) are focussed onto the 8~cm long Be rod inside the polarimeter. An array of CdZnTe detectors, which surrounds the rod allows the azimuthal scattering angle to be measured. The polarimeter is mounted inside an anticoincidence shield comprising an inverted well, which covers the top (3 cm BGO thickness) and sides (4~cm BGO thickness) of the polarimeter, and a puck, which covers the bottom of the polarimeter (3~cm BGO thickness). The high stopping power of BGO (7.1~g/cm$^2$ density) mitigates particle and X-/$\gamma$-ray backgrounds present in the stratosphere. More images of the XL-Calibur polarimeter can be found in Ref.~\cite{Abarr.2021}.}
\label{fig:anti}
\end{figure}

There are a variety of approaches to reduce the instrument background rate. As for standard pointed X-ray instruments, aperture backgrounds can be reduced using collimators. The production of fluorescence X-rays in the collimator material can be problematic and is mitigated using a multi-layer approach, e.g. a layer of tin foil will absorb the 88 keV line X-ray produced in lead. A second standard solution is to house the polarimeter inside an anticoincidence shield, as demonstrated in Figure~\ref{fig:anti}. High-Z scintillators, such as CsI ($\mathrm{Z_{eff}}$ = 54) and BGO ($\mathrm{Z_{eff}}$ = 75) are often used, and if thick enough will provide effective shielding against photon and charged particle background. Even for thick shields, a background is likely from high energy photons which interact in the shield and forward scatter (e.g. Figure~\ref{fig:peanut}) into the sensitive volume of the polarimeter. This will introduce additional harmonic components to the modulation curve, with the details depending on the spatial properties of the background, and ambiguity in resolving the scattering angle due to the finite energy resolution of detector elements (i.e. confusing the Compton scatter and photoabsorption site). If the shield is segmented, background anisotropies can be independently studied in flight, which provides a useful diagnostic for understanding systematic effects in measured modulation curves~\cite{Chauvin.2017}. In instruments with an active scatterer and absorber, a narrow ($\le 1 \; \mu$s) temporal coincidence condition can be placed on energy deposits when identifying a Compton scatter event, thereby greatly reducing the background generated by random coincidences. In a similar manner, the background in instruments utilising a passive scattering element can be reduced by requiring a single hit in the surrounding pixelated absorber stage.

The neutron background can be reduced by encapsulating the instrument in hydrogen-rich passive polyethylene shielding, which reduces the mean energy of the neutron flux through repeated scatterings. The residual background will be highly asymmetric since the neutron flux is predominately created below the payload. This will impart additional harmonic components on the modulation curve in a similar manner to photon leakage through an anticoincidence shield. The resulting increase in instrument mass often renders this approach as impractical, although it was successfully used in the PoGO+ mission~\cite{Chauvin2016}. Although ultimately dependent on instrument design, albedo neutrons (generated
through secondary interactions in the atmosphere) and photons (primary or secondary in origin) often dominate the background rate.

Background should be measured during observations. Since backgrounds may vary with time, the best approach is to conduct on- and off-source observations simultaneously. Although feasible with some instrument designs, e.g. coded-mask, it is usually more practical to intersperse on-source observations with off-source (blank field) observations close to the observed X-ray source, with the pointing offset dictated by the instrument field-of-view (see for example Ref.~\cite{Frontera1997}). In Ref.~\cite{Kislat.2015}, a prescription for optimising $\alpha = t_{off} / t_{on} = f_{off} / (1 - f_{off}) = (1 + R_s / R_b)^{-0.5}$ is presented, where a source (background) field is observed for a duration $t_{on}$ ($t_{off}$), such that $f_{on}$ and $f_{off}$ are the fractions of total observation time spent on- and off-source respectively. When summing Stokes parameters derived for individual scattering events, $\alpha$ can be incorporated in the weighting factor. The resulting MDP in Equation~\ref{eq:MDP_1} is then,

\begin{equation}\label{eq:MDP_2}
  \mathrm{MDP} = \frac{4.29}{\mu R_s} \sqrt{ \frac{R_b + f_{off} R_s }{(1 - f_{off}) f_{off}} t_{obs}}
\end{equation}

\noindent Since Stokes parameters are additive, the polarisation parameters for a source can be determined by subtracting the off-source Stokes parameter from the on-source ones, where the off-source parameters are scaled according to $\alpha$.

Two balloon missions are used to illustrate the consequence of this strategy for observation of the Crab during a flight from the Esrange Space Centre in Sweden, with an atmospheric overburden of $\sim 5$ g/cm$^2$. For the PoGO+ mission which flew in 2016, $R_s / R_b = 0.14$ requiring on- and off-source of equal duration. For the XL-Calibur mission, which is scheduled for launch in 2022, benefits from a design incorporating an X-ray mirror and a compact polarimeter, to obtain $R_s / R_b = 6.6$, such that the off-source pointing is $\sim 35$\% of the total observation time. For observations of transient sources, such as GRBs, background can be determined with a data collection scheme which allows polarisation parameters to be determined for a period of time before the instrument is triggered by the transient event, e.g. by continuously
storing triggers in a ring buffer arrangement.

Background estimates rely on simulating the interaction of photons and particles from the models of the background components in a mass model of the instrument using the Monte Carlo method. Among the most widely used software codes for this purpose are GEANT\footnote{https://www.geant.org} and MCNP\footnote{https://mcnp.lanl.gov}. The GEANT4 Monte Carlo toolkit~\cite{Agostinelli.2003,Allison.2016} is a widely-used C++ software framework which allows the interaction of radiation in detector materials to be modelled. This tool has a broad range of application during mission development, covering the study of early-stage prototypes to performance simulations for complete instruments mounted on a spacecraft. The scope and nature of the physical interactions used in the simulation are specified in so-called physics lists chosen at run-time. The physics lists are subject to revision as benchmarking experiments are performed. The energy spectrum and geometry of the generation surface for signal and background sources are specified by the user.

\section{6 Operational Issues}

Possible platforms to carry an instrument above the Earth's atmosphere are, in ascending order of available observation time, sounding rockets ($\sim$ minutes), high-altitude balloons ($\sim$ days) and satellites ($\sim$ years). A particular case of satellites are space stations, e.g. the International Space Station (ISS) or the Chinese Tiangong, which host a human crew, fly in Low Earth Orbit and may have onboard research programs with scientific instrumentation. The choice of a platform is a trade-off involving the instrument mass, volume and power, the time available for the scientific observation, the technology readiness of the instrument and finally the available funds. The interested reader is referred to Ref.~\cite{1999smad.book.....W} for a thorough discussion on this topic.

Assuming for a Compton polarimeter an energy threshold around 15 -- 20 keV, the Earth's atmosphere has a 1/e transparency (i.e. 37\%) at 40 -- 50 km altitude~\cite{Zombeck_2006}. This corresponds to the typical altitude of high-altitude balloons, that can thus be effectively used as platforms for this type of polarimeter.

We have seen in the previous sections how the altitude and inclination of an instrument orbit affect the background of the detectors. In addition, detector elements like scintillators and solid state read-out sensors may undergo radiation damage due to particles in orbit. For example, some types of inorganic scintillators are ``activated'' by nuclear reactions with neutrons and charged particles. The decay electrons and gamma rays produced by the reaction, with half-lives between less than a second and days~\cite{1992NIMPA.313..513G}, give signals in the detector thus increasing the background. Various techniques based on the coincidence between detector elements can be employed to mitigate this background component, as suggested in Ref.~\cite{1992NIMPA.313..513G}. Solid state readout sensors are extremely sensitive to the radiation damage in orbit, both Total Ionising Dose and Displacement Damage (see for example Ref.~\cite{Lutz_1999,2018ITNS...65.1561M}). For this reason, the instrument orbit and the type of detectors should be carefully optimised when designing a Compton polarimeter.

Finally, as explained in Ref.~\cite{1999smad.book.....W}, the selection of the platform for an instrument is also affected by the location of the ground segment where the data will be downlinked.

\section{7 Future perspectives}

We have shown how to exploit the Compton scattering to measure the linear polarisation of hard X-rays. We have discussed the main types of detectors involved in the measurement and we have seen that scintillators, both low-Z and high-Z, as well as thin CdZnTe detectors, are an important tool for this type of science.

From the decade of the 2000s, research and development activities, mostly motivated by medical physics, have improved the performance of solid state sensors (e.g. SiPMs, APDs and SDDs) as read-out elements for scintillators by reducing the noise and driving the miniaturisation of the device. The performance of this technology is certainly adequate for the large number of photons emitted by the inorganic scintillators employed as absorption stage, but not yet competitive with ``classical'' photomultipliers for the small numbers of photons in the scattering stage. Radiation damage is still a concern for some types of devices, e.g. SDDs and SiPMs. Further development of solid state sensors aims to reduce the electronic (dark) noise and the sensitivity to radiation damage.

Developed in 1999, the CubeSat standard is gaining importance as an inexpensive and easy to adapt platform to build satellites with mass in the range 2 -- 24 kg. CubeSats are not only employed as technological demonstrators, but also as building blocks of satellites constellations for scientific research. CubeSats are already competitive with small satellites due to their low cost and short development time and are expected to play an important role in the future for these reasons. We have discussed the examples of PolarLight, which measured the polarisation of the Crab Nebula and Sco X-1 in the soft X-ray band, and CUSP, a constellation aiming to study the polarisation of solar flares in hard X-rays.

Stratospheric balloons will continue to play an important role in the development of Compton polarimetry and allow initial scientific studies to be undertaken. Instrument designs can be assessed in the near-space environment to inform the design of future satellite missions. New scientific opportunities may be afforded by the development of larger volume zero-pressure balloons, which can fly at altitudes above 40 km, as well as super-pressure balloons, which maintain a constant day/night altitude during long-duration flights.

%

\bibliography{Section_Compton_polarimetry_v2_preprint}

\begin{thebibliography}{118}
\providecommand{\natexlab}[1]{#1}
\providecommand{\url}[1]{{#1}}
\providecommand{\urlprefix}{URL }
\expandafter\ifx\csname urlstyle\endcsname\relax
  \providecommand{\doi}[1]{DOI~\discretionary{}{}{}#1}\else
  \providecommand{\doi}{DOI~\discretionary{}{}{}\begingroup
  \urlstyle{rm}\Url}\fi
\providecommand{\eprint}[2][]{\url{#2}}

\bibitem[{Abarr et~al.(2020)Abarr, Baring, Beheshtipour, Beilicke, Geronimo,
  Dowkontt, Errando, Guarino, Iyer, Kislat, Kiss, Kitaguchi, Krawczynski,
  Lanzi, Li, Lisalda, Okajima, Pearce, Press, Rauch, Stuchlik, Takahashi, Tang,
  Uchida, West, Jenke, Krimm, Lien, Malacaria, Miller, and
  Wilson-Hodge}]{Abarr.2020p1n}
Abarr Q, Baring M, Beheshtipour B, Beilicke M, Geronimo Gd, Dowkontt P, Errando
  M, Guarino V, Iyer N, Kislat F, Kiss M, Kitaguchi T, Krawczynski H, Lanzi J,
  Li S, Lisalda L, Okajima T, Pearce M, Press L, Rauch B, Stuchlik D, Takahashi
  H, Tang J, Uchida N, West A, Jenke P, Krimm H, Lien A, Malacaria C, Miller
  JM, Wilson-Hodge C (2020) {Observations of a GX 301–2 Apastron Flare with
  the X-Calibur Hard X-Ray Polarimeter Supported by NICER , the Swift XRT and
  BAT, and Fermi GBM}. The Astrophysical Journal 891(1):70,
  \doi{10.3847/1538-4357/ab672c}, \eprint{2001.03581}

\bibitem[{Abarr et~al.(2021)Abarr, Awaki, Baring, Bose, Geronimo, Dowkontt,
  Errando, Guarino, Hattori, Hayashida, Imazato, Ishida, Iyer, Kislat, Kiss,
  Kitaguchi, Krawczynski, Lisalda, Matake, Maeda, Matsumoto, Mineta, Miyazawa,
  Mizuno, Okajima, Pearce, Rauch, Ryde, Shreves, Spooner, Stana, Takahashi,
  Takeo, Tamagawa, Tamura, Tsunemi, Uchida, Uchida, West, Wulf, and
  Yamamoto}]{Abarr.2021}
Abarr Q, Awaki H, Baring M, Bose R, Geronimo GD, Dowkontt P, Errando M, Guarino
  V, Hattori K, Hayashida K, Imazato F, Ishida M, Iyer N, Kislat F, Kiss M,
  Kitaguchi T, Krawczynski H, Lisalda L, Matake H, Maeda Y, Matsumoto H, Mineta
  T, Miyazawa T, Mizuno T, Okajima T, Pearce M, Rauch B, Ryde F, Shreves C,
  Spooner S, Stana TA, Takahashi H, Takeo M, Tamagawa T, Tamura K, Tsunemi H,
  Uchida N, Uchida Y, West A, Wulf E, Yamamoto R (2021) {XL-Calibur – a
  second-generation balloon-borne hard X-ray polarimetry mission}.
  Astroparticle Physics 126:102529, \doi{10.1016/j.astropartphys.2020.102529},
  \eprint{2010.10608}

\bibitem[{Agostinelli et~al.(2003)Agostinelli, Allison, Amako, Apostolakis,
  Araujo, Arce, Asai, Axen, Banerjee, Barrand, Behner, Bellagamba, Boudreau,
  Broglia, Brunengo, Burkhardt, Chauvie, Chuma, Chytracek, Cooperman, Cosmo,
  Degtyarenko, Dell'Acqua, Depaola, Dietrich, Enami, Feliciello, Ferguson,
  Fesefeldt, Folger, Foppiano, Forti, Garelli, Giani, Giannitrapani, Gibin,
  Cadenas, González, Abril, Greeniaus, Greiner, Grichine, Grossheim, Guatelli,
  Gumplinger, Hamatsu, Hashimoto, Hasui, Heikkinen, Howard, Ivanchenko,
  Johnson, Jones, Kallenbach, Kanaya, Kawabata, Kawabata, Kawaguti, Kelner,
  Kent, Kimura, Kodama, Kokoulin, Kossov, Kurashige, Lamanna, Lampén, Lara,
  Lefebure, Lei, Liendl, Lockman, Longo, Magni, Maire, Medernach, Minamimoto,
  Freitas, Morita, Murakami, Nagamatu, Nartallo, Nieminen, Nishimura, Ohtsubo,
  Okamura, O'Neale, Oohata, Paech, Perl, Pfeiffer, Pia, Ranjard, Rybin,
  Sadilov, Salvo, Santin, Sasaki, Savvas, Sawada, Scherer, Sei, Sirotenko,
  Smith, Starkov, Stoecker, Sulkimo, Takahata, Tanaka, Tcherniaev, Tehrani,
  Tropeano, Truscott, Uno, Urban, Urban, Verderi, Walkden, Wander, Weber,
  Wellisch, Wenaus, Williams, Wright, Yamada, Yoshida, Zschiesche, and
  Collaboration}]{Agostinelli.2003}
Agostinelli S, Allison J, Amako K, Apostolakis J, Araujo H, Arce P, Asai M,
  Axen D, Banerjee S, Barrand G, Behner F, Bellagamba L, Boudreau J, Broglia L,
  Brunengo A, Burkhardt H, Chauvie S, Chuma J, Chytracek R, Cooperman G, Cosmo
  G, Degtyarenko P, Dell'Acqua A, Depaola G, Dietrich D, Enami R, Feliciello A,
  Ferguson C, Fesefeldt H, Folger G, Foppiano F, Forti A, Garelli S, Giani S,
  Giannitrapani R, Gibin D, Cadenas JG, González I, Abril GG, Greeniaus G,
  Greiner W, Grichine V, Grossheim A, Guatelli S, Gumplinger P, Hamatsu R,
  Hashimoto K, Hasui H, Heikkinen A, Howard A, Ivanchenko V, Johnson A, Jones
  F, Kallenbach J, Kanaya N, Kawabata M, Kawabata Y, Kawaguti M, Kelner S, Kent
  P, Kimura A, Kodama T, Kokoulin R, Kossov M, Kurashige H, Lamanna E, Lampén
  T, Lara V, Lefebure V, Lei F, Liendl M, Lockman W, Longo F, Magni S, Maire M,
  Medernach E, Minamimoto K, Freitas PMd, Morita Y, Murakami K, Nagamatu M,
  Nartallo R, Nieminen P, Nishimura T, Ohtsubo K, Okamura M, O'Neale S, Oohata
  Y, Paech K, Perl J, Pfeiffer A, Pia M, Ranjard F, Rybin A, Sadilov S, Salvo
  ED, Santin G, Sasaki T, Savvas N, Sawada Y, Scherer S, Sei S, Sirotenko V,
  Smith D, Starkov N, Stoecker H, Sulkimo J, Takahata M, Tanaka S, Tcherniaev
  E, Tehrani ES, Tropeano M, Truscott P, Uno H, Urban L, Urban P, Verderi M,
  Walkden A, Wander W, Weber H, Wellisch J, Wenaus T, Williams D, Wright D,
  Yamada T, Yoshida H, Zschiesche D, Collaboration G (2003) {Geant4—a
  simulation toolkit}. Nuclear Instruments and Methods in Physics Research
  Section A: Accelerators, Spectrometers, Detectors and Associated Equipment
  506(3):250--303, \doi{10.1016/s0168-9002(03)01368-8}

\bibitem[{Allison et~al.(2016)Allison, Amako, Apostolakis, Arce, Asai, Aso,
  Bagli, Bagulya, Banerjee, Barrand, Beck, Bogdanov, Brandt, Brown, Burkhardt,
  Canal, Cano-Ott, Chauvie, Cho, Cirrone, Cooperman, Cortés-Giraldo, Cosmo,
  Cuttone, Depaola, Desorgher, Dong, Dotti, Elvira, Folger, Francis, Galoyan,
  Garnier, Gayer, Genser, Grichine, Guatelli, Guèye, Gumplinger, Howard,
  Hřivnáčová, Hwang, Incerti, Ivanchenko, Ivanchenko, Jones, Jun,
  Kaitaniemi, Karakatsanis, Karamitrosi, Kelsey, Kimura, Koi, Kurashige,
  Lechner, Lee, Longo, Maire, Mancusi, Mantero, Mendoza, Morgan, Murakami,
  Nikitina, Pandola, Paprocki, Perl, Petrović, Pia, Pokorski, Quesada, Raine,
  Reis, Ribon, Fira, Romano, Russo, Santin, Sasaki, Sawkey, Shin, Strakovsky,
  Taborda, Tanaka, Tomé, Toshito, Tran, Truscott, Urban, Uzhinsky, Verbeke,
  Verderi, Wendt, Wenzel, Wright, Wright, Yamashita, Yarba, and
  Yoshida}]{Allison.2016}
Allison J, Amako K, Apostolakis J, Arce P, Asai M, Aso T, Bagli E, Bagulya A,
  Banerjee S, Barrand G, Beck B, Bogdanov A, Brandt D, Brown J, Burkhardt H,
  Canal P, Cano-Ott D, Chauvie S, Cho K, Cirrone G, Cooperman G,
  Cortés-Giraldo M, Cosmo G, Cuttone G, Depaola G, Desorgher L, Dong X, Dotti
  A, Elvira V, Folger G, Francis Z, Galoyan A, Garnier L, Gayer M, Genser K,
  Grichine V, Guatelli S, Guèye P, Gumplinger P, Howard A, Hřivnáčová I,
  Hwang S, Incerti S, Ivanchenko A, Ivanchenko V, Jones F, Jun S, Kaitaniemi P,
  Karakatsanis N, Karamitrosi M, Kelsey M, Kimura A, Koi T, Kurashige H,
  Lechner A, Lee S, Longo F, Maire M, Mancusi D, Mantero A, Mendoza E, Morgan
  B, Murakami K, Nikitina T, Pandola L, Paprocki P, Perl J, Petrović I, Pia M,
  Pokorski W, Quesada J, Raine M, Reis M, Ribon A, Fira AR, Romano F, Russo G,
  Santin G, Sasaki T, Sawkey D, Shin J, Strakovsky I, Taborda A, Tanaka S,
  Tomé B, Toshito T, Tran H, Truscott P, Urban L, Uzhinsky V, Verbeke J,
  Verderi M, Wendt B, Wenzel H, Wright D, Wright D, Yamashita T, Yarba J,
  Yoshida H (2016) {Recent developments in Geant4}. Nuclear Instruments and
  Methods in Physics Research Section A: Accelerators, Spectrometers, Detectors
  and Associated Equipment 835:186--225, \doi{10.1016/j.nima.2016.06.125}

\bibitem[{Almeida and Pillet(1993)}]{Almeida.1993}
Almeida JS, Pillet VM (1993) {Polarizing properties of grazing-incidence x-ray
  mirrors: comment}. Applied Optics 32(22):4231, \doi{10.1364/ao.32.004231}

\bibitem[{{Andreotti} et~al.(2014){Andreotti}, {Baldini}, {Calabrese},
  {Cibinetto}, {Cotta Ramusino}, {De Donato}, {Faccini}, {Fiorini}, {Luppi},
  {Malaguti}, {Montanari}, {Pietropaolo}, {Santoro}, {Tellarini}, {Tomassetti},
  and {Tosi}}]{Andreotti2014}
{Andreotti} M, {Baldini} W, {Calabrese} R, {Cibinetto} G, {Cotta Ramusino} A,
  {De Donato} C, {Faccini} R, {Fiorini} M, {Luppi} E, {Malaguti} R, {Montanari}
  A, {Pietropaolo} A, {Santoro} V, {Tellarini} G, {Tomassetti} L, {Tosi} N
  (2014) {Study of the radiation damage of Silicon Photo-Multipliers at the
  GELINA facility}. Journal of Instrumentation 9(4):P04004,
  \doi{10.1088/1748-0221/9/04/P04004}

\bibitem[{{Beilicke} et~al.(2014){Beilicke}, {Kislat}, {Zajczyk}, {Guo},
  {Endsley}, {Stork}, {Cowsik}, {Dowkontt}, {Barthelmy}, {Hams}, {Okajima},
  {Sasaki}, {Zeiger}, {de Geronimo}, {Baring}, and
  {Krawczynski}}]{Beilicke2014}
{Beilicke} M, {Kislat} F, {Zajczyk} A, {Guo} Q, {Endsley} R, {Stork} M,
  {Cowsik} R, {Dowkontt} P, {Barthelmy} S, {Hams} T, {Okajima} T, {Sasaki} M,
  {Zeiger} B, {de Geronimo} G, {Baring} MG, {Krawczynski} H (2014) {Design and
  Performance of the X-ray Polarimeter X-Calibur}. Journal of Astronomical
  Instrumentation 3:1440008, \doi{10.1142/S225117171440008X},
  \eprint{1412.6457}

\bibitem[{Beilicke et~al.(2015)Beilicke, Kislat, Zajczyk, Guo, Endsley, Cowsik,
  Dowkontt, Krawczynski, Barthelmy, Hams, Okajima, Sasaki, De~Geronimo, Haba,
  and Saji}]{Beilicke2015}
Beilicke M, Kislat F, Zajczyk A, Guo Q, Endsley R, Cowsik R, Dowkontt P,
  Krawczynski H, Barthelmy S, Hams T, Okajima T, Sasaki M, De~Geronimo G, Haba
  Y, Saji S (2015) First flight of the x-ray polarimeter x-calibur. In: 2015
  IEEE Aerospace Conference, pp 1--10, \doi{10.1109/AERO.2015.7118915}

\bibitem[{{Bloser} and {et al.}(2010)}]{Bloser2010}
{Bloser} PF, {et al} (2010) {The Gamma-RAy Polarimeter Experiment (GRAPE)
  balloon payload}. In: X-ray Polarimetry: A New Window in Astrophysics by
  Ronaldo Bellazzini, p 314, \doi{10.1017/CBO9780511750809.047}

\bibitem[{{Bloser} et~al.(2009){Bloser}, {Legere}, {McConnell}, {Macri},
  {Bancroft}, {Connor}, and {Ryan}}]{Bloser2009}
{Bloser} PF, {Legere} JS, {McConnell} ML, {Macri} JR, {Bancroft} CM, {Connor}
  TP, {Ryan} JM (2009) {Calibration of the Gamma-RAy Polarimeter Experiment
  (GRAPE) at a polarized hard X-ray beam}. Nuclear Instruments and Methods in
  Physics Research A 600:424, \doi{10.1016/j.nima.2008.11.118},
  \eprint{0812.0782}

\bibitem[{{Caroli} et~al.(2018){Caroli}, {Moita}, {da Silva}, {Del Sordo}, {de
  Cesare}, {Maia}, and {P{\`a}scoa}}]{Caroli2018}
{Caroli} E, {Moita} M, {da Silva} R, {Del Sordo} S, {de Cesare} G, {Maia} J,
  {P{\`a}scoa} M (2018) {Hard X-ray and Soft Gamma Ray Polarimetry with
  CdTe/CZT Spectro-Imager}. Galaxies 6(3):69, \doi{10.3390/galaxies6030069}

\bibitem[{{Chattopadhyay} et~al.(2012){Chattopadhyay}, {Vadawale}, and
  {Pendharkar}}]{Chattopadhyay2012}
{Chattopadhyay} T, {Vadawale} SV, {Pendharkar} J (2012) {Compton polarimeter as
  a focal plane detector for hard X-ray telescope: sensitivity estimation with
  Geant4 simulations}. Experimental Astronomy p~39,
  \doi{10.1007/s10686-012-9312-3}

\bibitem[{{Chattopadhyay} et~al.(2014){Chattopadhyay}, {Vadawale}, {Rao},
  {Sreekumar}, and {Bhattacharya}}]{2014ExA....37..555C}
{Chattopadhyay} T, {Vadawale} SV, {Rao} AR, {Sreekumar} S, {Bhattacharya} D
  (2014) {Prospects of hard X-ray polarimetry with Astrosat-CZTI}. Experimental
  Astronomy 37(3):555--577, \doi{10.1007/s10686-014-9386-1}

\bibitem[{{Chattopadhyay} et~al.(2019){Chattopadhyay}, {Vadawale}, {Aarthy},
  {Mithun}, {Chand}, {Ratheesh}, {Basak}, {Rao}, {Bhalerao}, {Mate}, {Arvind},
  {Sharma}, and {Bhattacharya}}]{Chattopadhyay2019}
{Chattopadhyay} T, {Vadawale} SV, {Aarthy} E, {Mithun} NPS, {Chand} V,
  {Ratheesh} A, {Basak} R, {Rao} AR, {Bhalerao} V, {Mate} S, {Arvind} B,
  {Sharma} V, {Bhattacharya} D (2019) {Prompt Emission Polarimetry of Gamma-Ray
  Bursts with the AstroSat CZT Imager}. \apj 884(2):123,
  \doi{10.3847/1538-4357/ab40b7}, \eprint{1707.06595}

\bibitem[{{Chauvin} et~al.(2016{\natexlab{a}}){Chauvin}, {Jackson}, {Kamae},
  {Kawano}, {Kiss}, {Kole}, {Mikhalev}, {Moretti}, {Olofsson}, {Rydstr{\"o}m},
  {Takahashi}, {Lind}, {Str{\"o}mberg}, {Welin}, {Iyudin}, {Shifrin}, and
  {Pearce}}]{Chauvin2016b}
{Chauvin} HG Mand~{Flor{\'e}n}, {Jackson} M, {Kamae} T, {Kawano} T, {Kiss} M,
  {Kole} M, {Mikhalev} V, {Moretti} E, {Olofsson} G, {Rydstr{\"o}m} S,
  {Takahashi} H, {Lind} J, {Str{\"o}mberg} JE, {Welin} O, {Iyudin} A, {Shifrin}
  D, {Pearce} M (2016{\natexlab{a}}) {The design and flight performance of the
  PoGOLite Pathfinder balloon-borne hard X-ray polarimeter}. Experimental
  Astronomy 41:17--41, \doi{10.1007/s10686-015-9474-x}, \eprint{1508.03345}

\bibitem[{{Chauvin} et~al.(2013){Chauvin}, {Roques}, {Clark}, and
  {Jourdain}}]{2013ApJ...769..137C}
{Chauvin} M, {Roques} JP, {Clark} DJ, {Jourdain} E (2013) {Polarimetry in the
  Hard X-Ray Domain with INTEGRAL SPI}. \apj 769(2):137,
  \doi{10.1088/0004-637X/769/2/137}, \eprint{1305.0802}

\bibitem[{Chauvin et~al.(2015)Chauvin, Jackson, Kawano, Kiss, Kole, Mikhalev,
  Moretti, Takahashi, and Pearce}]{Chauvin.20160pr}
Chauvin M, Jackson M, Kawano T, Kiss M, Kole M, Mikhalev V, Moretti E,
  Takahashi H, Pearce M (2015) {Preflight performance studies of the PoGOLite
  hard X-ray polarimeter}. Astroparticle Physics 72:1--10,
  \doi{10.1016/j.astropartphys.2015.05.003},
  \urlprefix\url{http://cds.cern.ch/record/2020661}, \eprint{1505.08093}

\bibitem[{{Chauvin} et~al.(2016{\natexlab{b}}){Chauvin}, {Jackson}, {Kawano},
  {Kiss}, {Kole}, {Mikhalev}, {Moretti}, {Takahashi}, and
  {Pearce}}]{Chauvin2016}
{Chauvin} M, {Jackson} M, {Kawano} T, {Kiss} M, {Kole} M, {Mikhalev} V,
  {Moretti} E, {Takahashi} H, {Pearce} M (2016{\natexlab{b}}) {Optimising a
  balloon-borne polarimeter in the hard X-ray domain: From the PoGOLite
  Pathfinder to PoGO+}. Astroparticle Physics 82:99--107,
  \doi{10.1016/j.astropartphys.2016.06.005}, \eprint{1606.04504}

\bibitem[{Chauvin et~al.(2017{\natexlab{a}})Chauvin, Florén, Friis, Jackson,
  Kamae, Kataoka, Kawano, Kiss, Mikhalev, Mizuno, Ohashi, Stana, Tajima,
  Takahashi, Uchida, and Pearce}]{Chauvin.2017}
Chauvin M, Florén HG, Friis M, Jackson M, Kamae T, Kataoka J, Kawano T, Kiss
  M, Mikhalev V, Mizuno T, Ohashi N, Stana T, Tajima H, Takahashi H, Uchida N,
  Pearce M (2017{\natexlab{a}}) {Shedding new light on the Crab with polarized
  X-rays}. Scientific Reports 7(1):7816, \doi{10.1038/s41598-017-07390-7}

\bibitem[{Chauvin et~al.(2017{\natexlab{b}})Chauvin, Friis, Jackson, Kawano,
  Kiss, Mikhalev, Ohashi, Stana, Takahashi, and Pearce}]{Chauvin.2017qw}
Chauvin M, Friis M, Jackson M, Kawano T, Kiss M, Mikhalev V, Ohashi N, Stana T,
  Takahashi H, Pearce M (2017{\natexlab{b}}) {Calibration and performance
  studies of the balloon-borne hard X-ray polarimeter PoGO+}. Nuclear
  Instruments and Methods in Physics Research Section A: Accelerators,
  Spectrometers, Detectors and Associated Equipment 859:125--133,
  \doi{10.1016/j.nima.2017.03.027}, \eprint{1703.07627}

\bibitem[{{Chauvin} et~al.(2018{\natexlab{a}}){Chauvin}, {Flor{\'e}n}, {Friis},
  {Jackson}, {Kamae}, {Kataoka}, {Kawano}, {Kiss}, {Mikhalev}, {Mizuno},
  {Ohashi}, {Stana}, {Tajima}, {Takahashi}, {Uchida}, and
  {Pearce}}]{2018NatAs...2..652C}
{Chauvin} M, {Flor{\'e}n} HG, {Friis} M, {Jackson} M, {Kamae} T, {Kataoka} J,
  {Kawano} T, {Kiss} M, {Mikhalev} V, {Mizuno} T, {Ohashi} N, {Stana} T,
  {Tajima} H, {Takahashi} H, {Uchida} N, {Pearce} M (2018{\natexlab{a}})
  {Accretion geometry of the black-hole binary Cygnus X-1 from X-ray
  polarimetry}. Nature Astronomy 2:652--655, \doi{10.1038/s41550-018-0489-x},
  \eprint{1812.09907}

\bibitem[{{Chauvin} et~al.(2018{\natexlab{b}}){Chauvin}, {Flor{\'e}n}, {Friis},
  {Jackson}, {Kamae}, {Kataoka}, {Kawano}, {Kiss}, {Mikhalev}, {Mizuno},
  {Tajima}, {Takahashi}, {Uchida}, and {Pearce}}]{Chauvin2018}
{Chauvin} M, {Flor{\'e}n} HG, {Friis} M, {Jackson} M, {Kamae} T, {Kataoka} J,
  {Kawano} T, {Kiss} M, {Mikhalev} V, {Mizuno} T, {Tajima} H, {Takahashi} H,
  {Uchida} N, {Pearce} M (2018{\natexlab{b}}) {The PoGO+ view on Crab off-pulse
  hard X-ray polarisation}. \mnras \doi{10.1093/mnrasl/sly027},
  \eprint{1802.07775}

\bibitem[{{Chen} et~al.(2007){Chen}, {Mao}, {Zhang}, and {Zhu}}]{Chen2007}
{Chen} J, {Mao} R, {Zhang} L, {Zhu} Ry (2007) {Gamma-Ray Induced Radiation
  Damage in Large Size LSO and LYSO Crystal Samples}. IEEE Transactions on
  Nuclear Science 54(4):1319--1326, \doi{10.1109/TNS.2007.902370}

\bibitem[{{Connor} et~al.(2010){Connor}, {Bancroft}, {Bloser}, {Legere},
  {McConnell}, and {Ryan}}]{Connor2010}
{Connor} TP, {Bancroft} CM, {Bloser} PF, {Legere} JS, {McConnell} ML, {Ryan} JM
  (2010) {Plans for the first balloon flight of the gamma-ray polarimeter
  experiment (GRAPE)}. In: Space Telescopes and Instrumentation 2010:
  Ultraviolet to Gamma Ray, \procspie, vol 7732, p 77324E,
  \doi{10.1117/12.857467}

\bibitem[{{Costa} et~al.(1995){Costa}, {Cinti}, {Feroci}, {Matt}, and
  {Rapisarda}}]{1995NIMPA.366..161C}
{Costa} E, {Cinti} MN, {Feroci} M, {Matt} G, {Rapisarda} M (1995) {Design of a
  scattering polarimeter for hard X-ray astronomy}. Nuclear Instruments and
  Methods in Physics Research A 366:161--172,
  \doi{10.1016/0168-9002(95)00460-2}

\bibitem[{De~Angelis et~al.(2018)De~Angelis, Tatischeff, Grenier, McEnery,
  Mallamaci, Tavani, Oberlack, Hanlon, Walter, Argan, and
  et~al.}]{DeAngelis2018}
De~Angelis A, Tatischeff V, Grenier I, McEnery J, Mallamaci M, Tavani M,
  Oberlack U, Hanlon L, Walter R, Argan A, et~al (2018) Science with
  e-astrogam. Journal of High Energy Astrophysics 19:1–106,
  \doi{10.1016/j.jheap.2018.07.001},
  \urlprefix\url{http://dx.doi.org/10.1016/j.jheap.2018.07.001}

\bibitem[{{de Angelis} and {Polar-2 Collaboration}(2022)}]{2022icrc.confE.580D}
{de Angelis} N, {Polar-2 Collaboration} (2022) {Development and science
  perspectives of the POLAR-2 instrument: a large scale GRB polarimeter}. In:
  37th International Cosmic Ray Conference. 12-23 July 2021. Berlin, p 580,
  \eprint{2109.02978}

\bibitem[{{Dergachev} et~al.(2009){Dergachev}, {Matveev}, {Kruglov},
  {Lazutkov}, {Savchenko}, {Skorodumov}, {Pyatigorsky}, {Chichikaluk},
  {Shishov}, {Khmylko}, {Vasiliev}, {Dranevich}, {Krut'Kov}, {Stepanov},
  {Kotov}, {Yurov}, {Glyanenko}, {Arkhangelsky}, {Gorelyi}, and
  {Rubtsov}}]{Dergachev2009}
{Dergachev} VA, {Matveev} GA, {Kruglov} EM, {Lazutkov} VP, {Savchenko} MI,
  {Skorodumov} DV, {Pyatigorsky} GA, {Chichikaluk} YA, {Shishov} II, {Khmylko}
  VV, {Vasiliev} GI, {Dranevich} VA, {Krut'Kov} SY, {Stepanov} SV, {Kotov} YD,
  {Yurov} VN, {Glyanenko} AS, {Arkhangelsky} AI, {Gorelyi} YA, {Rubtsov} IV
  (2009) {Hard X-ray compton polarimetry with the PENGUIN-M instrument in the
  spaceborne experiment CORONAS-PHOTON}. Bulletin of the Russian Academy of
  Sciences, Physics 73:419--421, \doi{10.3103/S1062873809030423}

\bibitem[{Elsner et~al.(1990)Elsner, Weisskopf, Kaaret, Novick, and
  Silver}]{Elsner.1990}
Elsner RF, Weisskopf MC, Kaaret PE, Novick R, Silver EH (1990) {Off-axis
  effects on the performance of a scattering polarimeter at the focus of an
  x-ray telescope}. Optical Engineering 29(7):767--772, \doi{10.1117/12.55662}

\bibitem[{{Endsley} et~al.(2015){Endsley}, {Beilicke}, {Kislat}, {Krawczynski},
  and {X-Calibur/InFOCuS}}]{Endsley2015}
{Endsley} R, {Beilicke} M, {Kislat} F, {Krawczynski} H, {X-Calibur/InFOCuS}
  (2015) {Systematic and Performance Tests of the Hard X-ray Polarimeter
  X-Calibur}. In: American Astronomical Society Meeting Abstracts, American
  Astronomical Society Meeting Abstracts, vol 225, p 337.15

\bibitem[{{Evangelista} et~al.(2020){Evangelista}, {Fiore}, {Fuschino},
  {Campana}, {Ceraudo}, {Demenev}, {Guzman}, {Labanti}, {La Rosa}, {Fiorini},
  {Gandola}, {Grassi}, {Mele}, {Morgante}, {Nogara}, {Piazzolla}, {Pliego
  Caballero}, {Rashevskaya}, {Russo}, {Sciarrone}, {Sottile}, {Milankovich},
  {P{\'a}l}, {Ambrosino}, {Auricchio}, {Barbera}, {Bellutti}, {Bertuccio},
  {Borghi}, {Cao}, {Chen}, {Dilillo}, {Feroci}, {Ficorella}, {Lo Cicero},
  {Malcovati}, {Morbidini}, {Pauletta}, {Picciotto}, {Rachevski}, {Santangelo},
  {Tenzer}, {Vacchi}, {Wang}, {Xu}, {Zampa}, {Zampa}, {Zorzi}, {Burderi},
  {Lavagna}, {Bertacin}, {Lunghi}, {Monge}, {Negri}, {Pirrotta}, {Puccetti},
  {Sanna}, {Amarilli}, {Amelino-Camelia}, {Bechini}, {Citossi}, {Colagrossi},
  {Curzel}, {Della Casa}, {Cinelli}, {Del Santo}, {Di Salvo}, {Feruglio},
  {Ferrandi}, {Fiorito}, {Gacnik}, {Galg{\'o}czi}, {Gambino}, {Ghirlanda},
  {Gomboc}, {Karlica}, {Efremov}, {Kostic}, {Clerici}, {Lopez Fernandez},
  {Maselli}, {Nava}, {Ohno}, {Ottolina}, {Pasquale}, {Perri}, {Piccinin},
  {Prinetto}, {Riggio}, {Ripa}, {Papitto}, {Piranomonte}, {Scala}, {Selcan},
  {Silvestrini}, {Rotovnik}, {Virgilli}, {Troisi}, {Werner}, {Zanotti},
  {Anitra}, {Manca}, and {Clerici}}]{Evangelista2020}
{Evangelista} Y, {Fiore} F, {Fuschino} F, {Campana} R, {Ceraudo} F, {Demenev}
  E, {Guzman} A, {Labanti} C, {La Rosa} G, {Fiorini} M, {Gandola} M, {Grassi}
  M, {Mele} F, {Morgante} G, {Nogara} P, {Piazzolla} R, {Pliego Caballero} S,
  {Rashevskaya} I, {Russo} F, {Sciarrone} G, {Sottile} G, {Milankovich} D,
  {P{\'a}l} A, {Ambrosino} F, {Auricchio} N, {Barbera} M, {Bellutti} P,
  {Bertuccio} G, {Borghi} G, {Cao} J, {Chen} T, {Dilillo} G, {Feroci} M,
  {Ficorella} F, {Lo Cicero} U, {Malcovati} P, {Morbidini} A, {Pauletta} G,
  {Picciotto} A, {Rachevski} A, {Santangelo} A, {Tenzer} C, {Vacchi} A, {Wang}
  L, {Xu} Y, {Zampa} G, {Zampa} N, {Zorzi} N, {Burderi} L, {Lavagna} M,
  {Bertacin} R, {Lunghi} P, {Monge} A, {Negri} B, {Pirrotta} S, {Puccetti} S,
  {Sanna} A, {Amarilli} F, {Amelino-Camelia} G, {Bechini} M, {Citossi} M,
  {Colagrossi} A, {Curzel} S, {Della Casa} G, {Cinelli} M, {Del Santo} M, {Di
  Salvo} T, {Feruglio} C, {Ferrandi} F, {Fiorito} M, {Gacnik} D, {Galg{\'o}czi}
  G, {Gambino} AF, {Ghirlanda} G, {Gomboc} A, {Karlica} M, {Efremov} P,
  {Kostic} U, {Clerici} A, {Lopez Fernandez} B, {Maselli} A, {Nava} L, {Ohno}
  M, {Ottolina} D, {Pasquale} A, {Perri} M, {Piccinin} M, {Prinetto} J,
  {Riggio} A, {Ripa} J, {Papitto} A, {Piranomonte} S, {Scala} F, {Selcan} D,
  {Silvestrini} S, {Rotovnik} T, {Virgilli} E, {Troisi} I, {Werner} N,
  {Zanotti} G, {Anitra} A, {Manca} A, {Clerici} A (2020) {The scientific
  payload on-board the HERMES-TP and HERMES-SP CubeSat missions}. In: Society
  of Photo-Optical Instrumentation Engineers (SPIE) Conference Series, Society
  of Photo-Optical Instrumentation Engineers (SPIE) Conference Series, vol
  11444, p 114441T, \doi{10.1117/12.2561018}, \eprint{2101.03032}

\bibitem[{Fabiani et~al.(2013)Fabiani, Campana, Costa, Del~Monte, Muleri,
  Rubini, and Soffitta}]{Fabiani.2013vom}
Fabiani S, Campana R, Costa E, Del~Monte E, Muleri F, Rubini A, Soffitta P
  (2013) {Characterization of scatterers for an active focal plane Compton
  polarimeter}. Astroparticle Physics 44:91--101,
  \doi{10.1016/j.astropartphys.2012.12.008}, \eprint{1301.1161}

\bibitem[{{Fabiani} et~al.(2022){Fabiani}, {Baffo}, {Bonomo}, {Contini},
  {Costa}, {Cucinella}, {De Cesare}, {Del Monte}, {Del Re}, {Di Cosimo}, {Di
  Filippo}, {Di Marco}, {Fanelli}, {La Monaca}, {Locarini}, {Loffredo},
  {Lombardi}, {Minervini}, {Modenini}, {Muleri}, {Negri}, {Perelli}, {Rankin},
  {Rubini}, {Soffitta}, {Strollo}, {Tortora}, and {Zambardi}}]{Fabiani_et_2022}
{Fabiani} S, {Baffo} I, {Bonomo} S, {Contini} G, {Costa} E, {Cucinella} G, {De
  Cesare} G, {Del Monte} E, {Del Re} A, {Di Cosimo} S, {Di Filippo} S, {Di
  Marco} A, {Fanelli} P, {La Monaca} F, {Locarini} A, {Loffredo} P, {Lombardi}
  G, {Minervini} G, {Modenini} D, {Muleri} F, {Negri} A, {Perelli} M, {Rankin}
  J, {Rubini} A, {Soffitta} P, {Strollo} E, {Tortora} P, {Zambardi} A (2022)
  {CUSP: a two cubesats constellation for Space Weather and solar flares X-ray
  polarimetry}. arXiv e-prints arXiv:2208.06211, \eprint{2208.06211}

\bibitem[{{Feng} and {Bellazzini}(2020)}]{2020NatAs...4..547F}
{Feng} H, {Bellazzini} R (2020) {The X-ray polarimetry window reopens}. Nature
  Astronomy 4:547--547, \doi{10.1038/s41550-020-1103-6}

\bibitem[{Ferrazzoli et~al.(2020)Ferrazzoli, Muleri, Lefevre, Morbidini, Amici,
  Brienza, Costa, Monte, Marco, Persio, Donnarumma, Fabiani, Monaca, Loffredo,
  Maiolo, Maita, Piazzolla, Ramsey, Rankin, Ratheesh, Rubini, Sarra, Soffitta,
  Tobia, and Xie}]{Ferrazzoli.2020}
Ferrazzoli R, Muleri F, Lefevre C, Morbidini A, Amici F, Brienza D, Costa E,
  Monte ED, Marco AD, Persio GD, Donnarumma I, Fabiani S, Monaca FL, Loffredo
  P, Maiolo L, Maita F, Piazzolla R, Ramsey B, Rankin J, Ratheesh A, Rubini A,
  Sarra P, Soffitta P, Tobia A, Xie F (2020) {In-flight calibration system of
  imaging x-ray polarimetry explorer}. Journal of Astronomical Telescopes,
  Instruments, and Systems 6(04), \doi{10.1117/1.jatis.6.4.048002},
  \eprint{2010.14185}

\bibitem[{{Forot} et~al.(2008){Forot}, {Laurent}, {Grenier}, {Gouiff{\`e}s},
  and {Lebrun}}]{2008ApJ...688L..29F}
{Forot} M, {Laurent} P, {Grenier} IA, {Gouiff{\`e}s} C, {Lebrun} F (2008)
  {Polarization of the Crab Pulsar and Nebula as Observed by the INTEGRAL/IBIS
  Telescope}. \apjl 688(1):L29, \doi{10.1086/593974}, \eprint{0809.1292}

\bibitem[{{Frontera} et~al.(1997){Frontera}, {Costa}, {dal Fiume}, {Feroci},
  {Nicastro}, {Orlandini}, {Palazzi}, and {Zavattini}}]{Frontera1997}
{Frontera} F, {Costa} E, {dal Fiume} D, {Feroci} M, {Nicastro} L, {Orlandini}
  M, {Palazzi} E, {Zavattini} G (1997) {The high energy instrument PDS on-board
  the BeppoSAX X--ray astronomy satellite}. \aaps 122:357,
  \doi{10.1051/aas:1997140}

\bibitem[{{Fuschino} et~al.(2020){Fuschino}, {Campana}, {Labanti},
  {Evangelista}, {Fiore}, {Gandola}, {Grassi}, {Mele}, {Ambrosino}, {Ceraudo},
  {Demenev}, {Fiorini}, {Morgante}, {Piazzolla}, {Bertuccio}, {Malcovati},
  {Bellutti}, {Borghi}, {Dilillo}, {Feroci}, {Ficorella}, {La Rosa}, {Nogara},
  {Pauletta}, {Picciotto}, {Rashevskaya}, {Rashevsky}, {Giuseppe}, {Vacchi},
  {Virgilli}, {Zampa}, {Zampa}, {Zorzi}, {Chen}, {Xu}, {Gao}, {Cao}, and
  {Wang}}]{Fuschino2020}
{Fuschino} F, {Campana} R, {Labanti} C, {Evangelista} Y, {Fiore} F, {Gandola}
  M, {Grassi} M, {Mele} F, {Ambrosino} F, {Ceraudo} F, {Demenev} E, {Fiorini}
  M, {Morgante} G, {Piazzolla} R, {Bertuccio} G, {Malcovati} P, {Bellutti} P,
  {Borghi} G, {Dilillo} G, {Feroci} M, {Ficorella} F, {La Rosa} G, {Nogara} P,
  {Pauletta} G, {Picciotto} A, {Rashevskaya} I, {Rashevsky} A, {Giuseppe} S,
  {Vacchi} A, {Virgilli} E, {Zampa} G, {Zampa} N, {Zorzi} N, {Chen} T, {Xu} Y,
  {Gao} N, {Cao} J, {Wang} L (2020) {An innovative architecture for wide band
  transient monitor on board the HERMES nano-satellite constellation}. In:
  Society of Photo-Optical Instrumentation Engineers (SPIE) Conference Series,
  Society of Photo-Optical Instrumentation Engineers (SPIE) Conference Series,
  vol 11444, p 114441S, \doi{10.1117/12.2561037}, \eprint{2101.03035}

\bibitem[{{Gehrels}(1992)}]{1992NIMPA.313..513G}
{Gehrels} N (1992) {Instrumental background in gamma-ray spectrometers flown in
  low Earth orbit}. Nuclear Instruments and Methods in Physics Research A
  313(3):513--528, \doi{10.1016/0168-9002(92)90832-O}

\bibitem[{{Gluckstern} and {Hull}(1953)}]{Gluckstern1953}
{Gluckstern} RL, {Hull} MH (1953) {Polarization Dependence of the Integrated
  Bremsstrahlung Cross Section}. Physical Review 90:1030--1035,
  \doi{10.1103/PhysRev.90.1030}

\bibitem[{{G{\"o}tz} et~al.(2009){G{\"o}tz}, {Laurent}, {Lebrun}, {Daigne}, and
  {Bo{\v{s}}njak}}]{2009ApJ...695L.208G}
{G{\"o}tz} D, {Laurent} P, {Lebrun} F, {Daigne} F, {Bo{\v{s}}njak} {\v{Z}}
  (2009) {Variable Polarization Measured in the Prompt Emission of GRB 041219A
  Using IBIS on Board INTEGRAL}. \apjl 695(2):L208--L212,
  \doi{10.1088/0004-637X/695/2/L208}, \eprint{0903.1712}

\bibitem[{{Gunji} et~al.(2007){Gunji}, {Sakurai}, {Tokanai}, {Kishimoto},
  {Kanno}, {Ishikawa}, {Hayashida}, {Anabuki}, {Tsunemi}, {Mihara}, {Kohama},
  {Suzuki}, {Saito}, and {Yamagami}}]{Gunji2007}
{Gunji} S, {Sakurai} H, {Tokanai} F, {Kishimoto} Y, {Kanno} M, {Ishikawa} Y,
  {Hayashida} K, {Anabuki} N, {Tsunemi} H, {Mihara} T, {Kohama} M, {Suzuki} M,
  {Saito} Y, {Yamagami} T (2007) {Observation of Crab Nebula with hard x-ray
  polarimeter: PHENEX}. In: UV, X-Ray, and Gamma-Ray Space Instrumentation for
  Astronomy XV, \procspie, vol 6686, p 668618, \doi{10.1117/12.738449}

\bibitem[{{Gunji} et~al.(2008){Gunji}, {Kishimoto}, {Sakurai}, {Tokanai},
  {Kanno}, {Ishikawa}, {Hayashida}, {Anabuke}, {Tsunemi}, {Mihara}, {Kohama},
  {Suzuki}, and {Saito}}]{Gunji2008}
{Gunji} S, {Kishimoto} Y, {Sakurai} H, {Tokanai} F, {Kanno} M, {Ishikawa} Y,
  {Hayashida} K, {Anabuke} N, {Tsunemi} H, {Mihara} T, {Kohama} M, {Suzuki} M,
  {Saito} Y (2008) {The PHENEX experiment result}. In: Polarimetry days in
  Rome: Crab status, theory and prospects

\bibitem[{{Guo} et~al.(2013){Guo}, {Beilicke}, {Garson}, {Kislat}, {Fleming},
  and {Krawczynski}}]{Guo2013}
{Guo} Q, {Beilicke} M, {Garson} A, {Kislat} F, {Fleming} D, {Krawczynski} H
  (2013) {Optimization of the design of the hard X-ray polarimeter X-Calibur}.
  Astroparticle Physics 41:63--72, \doi{10.1016/j.astropartphys.2012.11.006},
  \eprint{1212.4509}

\bibitem[{{Hayashida} et~al.(2016){Hayashida}, {Kim}, {Sadamoto}, {Yoshinaga},
  {Gunji}, {Mihara}, {Kishimoto}, {Kubo}, {Mizuno}, {Takahashi}, {Dotani},
  {Yonetoku}, {Nakamori}, {Yoneyama}, {Ikeyama}, and
  {Kamitsukasa}}]{Hayashida2016}
{Hayashida} K, {Kim} J, {Sadamoto} M, {Yoshinaga} K, {Gunji} S, {Mihara} T,
  {Kishimoto} Y, {Kubo} H, {Mizuno} T, {Takahashi} H, {Dotani} T, {Yonetoku} D,
  {Nakamori} T, {Yoneyama} T, {Ikeyama} Y, {Kamitsukasa} F (2016) {Hard x-ray
  imaging polarimeter for PolariS}. In: Space Telescopes and Instrumentation
  2016: Ultraviolet to Gamma Ray, \procspie, vol 9905, p 99051A,
  \doi{10.1117/12.2232472}

\bibitem[{{Hitomi Collaboration} et~al.(2018){Hitomi Collaboration},
  {Aharonian}, {Akamatsu}, {Akimoto}, {Allen}, {Angelini}, {Audard}, {Awaki},
  {Axelsson}, {Bamba}, {Bautz}, {Blandford}, {Brenneman}, {Brown}, {Bulbul},
  {Cackett}, {Chernyakova}, {Chiao}, {Coppi}, {Costantini}, {de Plaa}, {de
  Vries}, {den Herder}, {Done}, {Dotani}, {Ebisawa}, {Eckart}, {Enoto}, {Ezoe},
  {Fabian}, {Ferrigno}, {Foster}, {Fujimoto}, {Fukazawa}, {Furuzawa},
  {Galeazzi}, {Gallo}, {Gandhi}, {Giustini}, {Goldwurm}, {Gu}, {Guainazzi},
  {Haba}, {Hagino}, {Hamaguchi}, {Harrus}, {Hatsukade}, {Hayashi}, {Hayashi},
  {Hayashida}, {Hiraga}, {Hornschemeier}, {Hoshino}, {Hughes}, {Ichinohe},
  {Iizuka}, {Inoue}, {Inoue}, {Ishida}, {Ishikawa}, {Ishisaki}, {Iwai},
  {Kaastra}, {Kallman}, {Kamae}, {Kataoka}, {Katsuda}, {Kawai}, {Kelley},
  {Kilbourne}, {Kitaguchi}, {Kitamoto}, {Kitayama}, {Kohmura}, {Kokubun},
  {Koyama}, {Koyama}, {Kretschmar}, {Krimm}, {Kubota}, {Kunieda}, {Laurent},
  {Lee}, {Leutenegger}, {Limousin}, {Loewenstein}, {Long}, {Lumb}, {Madejski},
  {Maeda}, {Maier}, {Makishima}, {Markevitch}, {Matsumoto}, {Matsushita},
  {McCammon}, {McNamara}, {Mehdipour}, {Miller}, {Miller}, {Mineshige},
  {Mitsuda}, {Mitsuishi}, {Miyazawa}, {Mizuno}, {Mori}, {Mori}, {Mukai},
  {Murakami}, {Mushotzky}, {Nakagawa}, {Nakajima}, {Nakamori}, {Nakashima},
  {Nakazawa}, {Nobukawa}, {Nobukawa}, {Noda}, {Odaka}, {Ohashi}, {Ohno},
  {Okajima}, {Ota}, {Ozaki}, {Paerels}, {Paltani}, {Petre}, {Pinto}, {Porter},
  {Pottschmidt}, {Reynolds}, {Safi-Harb}, {Saito}, {Sakai}, {Sasaki}, {Sato},
  {Sato}, {Sato}, {Sawada}, {Schartel}, {Serlemtsos}, {Seta}, {Shidatsu},
  {Simionescu}, {Smith}, {Soong}, {Stawarz}, {Sugawara}, {Sugita},
  {Szymkowiak}, {Tajima}, {Takahashi}, {Takahashi}, {Takeda}, {Takei},
  {Tamagawa}, {Tamura}, {Tanaka}, {Tanaka}, {Tanaka}, {Tashiro}, {Tawara},
  {Terada}, {Terashima}, {Tombesi}, {Tomida}, {Tsuboi}, {Tsujimoto}, {Tsunemi},
  {Tsuru}, {Uchida}, {Uchiyama}, {Uchiyama}, {Ueda}, {Ueda}, {Uno}, {Urry},
  {Ursino}, {Watanabe}, {Werner}, {Wilkins}, {Williams}, {Yamada}, {Yamaguchi},
  {Yamaoka}, {Yamasaki}, {Yamauchi}, {Yamauchi}, {Yaqoob}, {Yatsu}, {Yonetoku},
  {Zhuravleva}, {Zoghbi}, and {Uchida}}]{Aharonian2018}
{Hitomi Collaboration}, {Aharonian} F, {Akamatsu} H, {Akimoto} F, {Allen} SW,
  {Angelini} L, {Audard} M, {Awaki} H, {Axelsson} M, {Bamba} A, {Bautz} MW,
  {Blandford} R, {Brenneman} LW, {Brown} GV, {Bulbul} E, {Cackett} EM,
  {Chernyakova} M, {Chiao} MP, {Coppi} PS, {Costantini} E, {de Plaa} J, {de
  Vries} CP, {den Herder} JW, {Done} C, {Dotani} T, {Ebisawa} K, {Eckart} ME,
  {Enoto} T, {Ezoe} Y, {Fabian} AC, {Ferrigno} C, {Foster} AR, {Fujimoto} R,
  {Fukazawa} Y, {Furuzawa} A, {Galeazzi} M, {Gallo} LC, {Gandhi} P, {Giustini}
  M, {Goldwurm} A, {Gu} L, {Guainazzi} M, {Haba} Y, {Hagino} K, {Hamaguchi} K,
  {Harrus} IM, {Hatsukade} I, {Hayashi} K, {Hayashi} T, {Hayashida} K, {Hiraga}
  JS, {Hornschemeier} A, {Hoshino} A, {Hughes} JP, {Ichinohe} Y, {Iizuka} R,
  {Inoue} H, {Inoue} Y, {Ishida} M, {Ishikawa} K, {Ishisaki} Y, {Iwai} M,
  {Kaastra} J, {Kallman} T, {Kamae} T, {Kataoka} J, {Katsuda} S, {Kawai} N,
  {Kelley} RL, {Kilbourne} CA, {Kitaguchi} T, {Kitamoto} S, {Kitayama} T,
  {Kohmura} T, {Kokubun} M, {Koyama} K, {Koyama} S, {Kretschmar} P, {Krimm} HA,
  {Kubota} A, {Kunieda} H, {Laurent} P, {Lee} SH, {Leutenegger} MA, {Limousin}
  O, {Loewenstein} M, {Long} KS, {Lumb} D, {Madejski} G, {Maeda} Y, {Maier} D,
  {Makishima} K, {Markevitch} M, {Matsumoto} H, {Matsushita} K, {McCammon} D,
  {McNamara} BR, {Mehdipour} M, {Miller} ED, {Miller} JM, {Mineshige} S,
  {Mitsuda} K, {Mitsuishi} I, {Miyazawa} T, {Mizuno} T, {Mori} H, {Mori} K,
  {Mukai} K, {Murakami} H, {Mushotzky} RF, {Nakagawa} T, {Nakajima} H,
  {Nakamori} T, {Nakashima} S, {Nakazawa} K, {Nobukawa} KK, {Nobukawa} M,
  {Noda} H, {Odaka} H, {Ohashi} T, {Ohno} M, {Okajima} T, {Ota} N, {Ozaki} M,
  {Paerels} F, {Paltani} S, {Petre} R, {Pinto} C, {Porter} FS, {Pottschmidt} K,
  {Reynolds} CS, {Safi-Harb} S, {Saito} S, {Sakai} K, {Sasaki} T, {Sato} G,
  {Sato} K, {Sato} R, {Sawada} M, {Schartel} N, {Serlemtsos} PJ, {Seta} H,
  {Shidatsu} M, {Simionescu} A, {Smith} RK, {Soong} Y, {Stawarz} {\L},
  {Sugawara} Y, {Sugita} S, {Szymkowiak} A, {Tajima} H, {Takahashi} H,
  {Takahashi} T, {Takeda} S, {Takei} Y, {Tamagawa} T, {Tamura} T, {Tanaka} T,
  {Tanaka} Y, {Tanaka} YT, {Tashiro} MS, {Tawara} Y, {Terada} Y, {Terashima} Y,
  {Tombesi} F, {Tomida} H, {Tsuboi} Y, {Tsujimoto} M, {Tsunemi} H, {Tsuru} TG,
  {Uchida} H, {Uchiyama} H, {Uchiyama} Y, {Ueda} S, {Ueda} Y, {Uno} S, {Urry}
  CM, {Ursino} E, {Watanabe} S, {Werner} N, {Wilkins} DR, {Williams} BJ,
  {Yamada} S, {Yamaguchi} H, {Yamaoka} K, {Yamasaki} NY, {Yamauchi} M,
  {Yamauchi} S, {Yaqoob} T, {Yatsu} Y, {Yonetoku} D, {Zhuravleva} I, {Zoghbi}
  A, {Uchida} Y (2018) {Detection of polarized gamma-ray emission from the Crab
  nebula with the Hitomi Soft Gamma-ray Detector}. \pasj 70(6):113,
  \doi{10.1093/pasj/psy118}, \eprint{1810.00704}

\bibitem[{{Hulsman}(2020)}]{Hulsman2020}
{Hulsman} J (2020) {POLAR-2: a large scale gamma-ray polarimeter for GRBs}. In:
  Society of Photo-Optical Instrumentation Engineers (SPIE) Conference Series,
  Society of Photo-Optical Instrumentation Engineers (SPIE) Conference Series,
  vol 11444, p 114442V, \doi{10.1117/12.2559374}, \eprint{2101.03084}

\bibitem[{{Janecek}(2012)}]{Janecek2012}
{Janecek} M (2012) {Reflectivity Spectra for Commonly Used Reflectors}. IEEE
  Transactions on Nuclear Science 59(3):490--497,
  \doi{10.1109/TNS.2012.2183385}

\bibitem[{Kamae et~al.(2008)Kamae, Andersson, Arimoto, Axelsson, Bettolo,
  Björnsson, Bogaert, Carlson, Craig, Ekeberg, Engdegård, Fukazawa, Gunji,
  Hjalmarsdotter, Iwan, Kanai, Kataoka, Kawai, Kazejev, Kiss, Klamra, Larsson,
  Madejski, Mizuno, Ng, Pearce, Ryde, Suhonen, Tajima, Takahashi, Takahashi,
  Tanaka, Thurston, Ueno, Varner, Yamamoto, Yamashita, Ylinen, and
  Yoshida}]{Kamae.2008}
Kamae T, Andersson V, Arimoto M, Axelsson M, Bettolo CM, Björnsson CI, Bogaert
  G, Carlson P, Craig W, Ekeberg T, Engdegård O, Fukazawa Y, Gunji S,
  Hjalmarsdotter L, Iwan B, Kanai Y, Kataoka J, Kawai N, Kazejev J, Kiss M,
  Klamra W, Larsson S, Madejski G, Mizuno T, Ng J, Pearce M, Ryde F, Suhonen M,
  Tajima H, Takahashi H, Takahashi T, Tanaka T, Thurston T, Ueno M, Varner G,
  Yamamoto K, Yamashita Y, Ylinen T, Yoshida H (2008) {PoGOLite A high
  sensitivity balloon-borne soft gamma-ray polarimeter}. Astroparticle Physics
  30(2):72 -- 84, \doi{10.1016/j.astropartphys.2008.07.004},
  \urlprefix\url{http://adsabs.harvard.edu/cgi-bin/nph-data\_query?bibcode=2008APh....30...72K\&link\_type=ABSTRACT},
  \eprint{0709.1278v2}

\bibitem[{{Kataoka} et~al.(2010){Kataoka}, {Toizumi}, {Nakamori}, {Yatsu},
  {Tsubuku}, {Kuramoto}, {Enomoto}, {Usui}, {Kawai}, {Ashida}, {Omagari},
  {Fujihashi}, {Inagawa}, {Miura}, {Konda}, {Miyashita}, {Matsunaga},
  {Ishikawa}, {Matsunaga}, and {Kawabata}}]{Kataoka2010}
{Kataoka} J, {Toizumi} T, {Nakamori} T, {Yatsu} Y, {Tsubuku} Y, {Kuramoto} Y,
  {Enomoto} T, {Usui} R, {Kawai} N, {Ashida} H, {Omagari} K, {Fujihashi} K,
  {Inagawa} S, {Miura} Y, {Konda} Y, {Miyashita} N, {Matsunaga} S, {Ishikawa}
  Y, {Matsunaga} Y, {Kawabata} N (2010) {In-orbit performance of avalanche
  photodiode as radiation detector on board the picosatellite Cute-1.7+APD II}.
  Journal of Geophysical Research (Space Physics) 115(A5):A05204,
  \doi{10.1029/2009JA014699}

\bibitem[{Katsuta et~al.(2009)Katsuta, Mizuno, Ogasaka, Yoshida, Takahashi,
  Kano, Iwahara, Sasaki, Kamae, Kokubun, Takahashi, Hayashida, and
  Uesugi}]{Katsuta.2009}
Katsuta J, Mizuno T, Ogasaka Y, Yoshida H, Takahashi H, Kano Y, Iwahara T,
  Sasaki N, Kamae T, Kokubun M, Takahashi T, Hayashida K, Uesugi K (2009)
  {Evaluation of polarization characteristics of multilayer mirror for hard
  X-ray observation of astrophysical objects}. Nuclear Instruments and Methods
  in Physics Research Section A 603:393, \doi{10.1016/j.nima.2009.02.039},
  \urlprefix\url{http://adsabs.harvard.edu/cgi-bin/nph-data\_query?bibcode=2009NIMPA.603..393K\&link\_type=ABSTRACT}

\bibitem[{{Kawai} and {Yatsu}(2012)}]{Kawai2012}
{Kawai} N, {Yatsu} Y (2012) {TSUBAME: a small satellite for GRB polarimetry}.
  In: 39th COSPAR Scientific Assembly, vol~39, p 901

\bibitem[{{Kishimoto} et~al.(2007){Kishimoto}, {Gunji}, {Ishigaki}, {Kanno},
  {Murayama}, {Ito}, {Tokanai}, {Suzuki}, {Sakurai}, {Mihara}, {Kohama},
  {Suzuki}, {Hayato}, {Hayashida}, {Anabuki}, {Morimoto}, {Tsunemi}, {Saito},
  {Yamagami}, and {Kishimoto}}]{Kishimoto2007}
{Kishimoto} Y, {Gunji} S, {Ishigaki} Y, {Kanno} M, {Murayama} H, {Ito} C,
  {Tokanai} F, {Suzuki} K, {Sakurai} H, {Mihara} T, {Kohama} M, {Suzuki} M,
  {Hayato} A, {Hayashida} K, {Anabuki} N, {Morimoto} M, {Tsunemi} H, {Saito} Y,
  {Yamagami} T, {Kishimoto} S (2007) {Basic Performance of PHENEX: A
  Polarimeter for High ENErgy X rays}. IEEE Transactions on Nuclear Science
  54:561--566, \doi{10.1109/TNS.2007.897827}

\bibitem[{{Kishimoto} et~al.(2009){Kishimoto}, {Gunji}, {Ishikawa}, {Takada},
  {Toukairin}, {Tanaka}, {Tokanai}, {Sakurai}, {Mihara}, {Sato}, {Hayashida},
  {Anabuki}, {Ota}, {Tsunemi}, {Narita}, {Saito}, {Kohama}, {Suzuki}, and
  {Kishimoto}}]{Kishimoto2009}
{Kishimoto} Y, {Gunji} S, {Ishikawa} Y, {Takada} M, {Toukairin} N, {Tanaka} Y,
  {Tokanai} F, {Sakurai} H, {Mihara} T, {Sato} T, {Hayashida} K, {Anabuki} N,
  {Ota} M, {Tsunemi} H, {Narita} T, {Saito} Y, {Kohama} M, {Suzuki} M,
  {Kishimoto} S (2009) {Observation of Polarization in Hard X-Ray Region with
  PHENEX Polarimeter}. In: {Kawai} N, {Mihara} T, {Kohama} M, {Suzuki} M (eds)
  Astrophysics with All-Sky X-Ray Observations, p 380

\bibitem[{Kislat et~al.(2015)Kislat, Clark, Beilicke, and
  Krawczynski}]{Kislat.2015}
Kislat F, Clark B, Beilicke M, Krawczynski H (2015) {Analyzing the data from
  X-ray polarimeters with Stokes parameters}. Astroparticle Physics 68:45--51,
  \doi{10.1016/j.astropartphys.2015.02.007},
  \urlprefix\url{http://arxiv.org/pdf/1409.6214v1.pdf}, \eprint{1409.6214}

\bibitem[{{Kislat} et~al.(2017){Kislat}, {Beheshtipour}, {Dowkontt}, {Guarino},
  {Lanzi}, {Okajima}, {Braun}, {Cannon}, {de Geronimo}, {Heatwole}, {Hoorman},
  {Li}, {Mori}, {Shreves}, {Stuchlik}, and {Krawczynski}}]{Kislat2017}
{Kislat} F, {Beheshtipour} B, {Dowkontt} P, {Guarino} V, {Lanzi} RJ, {Okajima}
  T, {Braun} D, {Cannon} S, {de Geronimo} G, {Heatwole} S, {Hoorman} J, {Li} S,
  {Mori} H, {Shreves} CM, {Stuchlik} D, {Krawczynski} H (2017) {Design of the
  Telescope Truss and Gondola for the Balloon-Borne X-ray Polarimeter
  X-Calibur}. Journal of Astronomical Instrumentation 6:1740003,
  \doi{10.1142/S2251171717400037}, \eprint{1701.04536}

\bibitem[{{Kislat} et~al.(2019){Kislat}, {Abarr}, {Bose}, {Braun}, {de
  Geronimo}, {Dowkontt}, {Errando}, {Gadson}, {Guarino}, {Heatwole}, {Iyer},
  {Kiss}, {Kitaguchi}, {Krawczynski}, {Kushwah}, {Lanzi}, {Li}, {Lisalda},
  {Okajima}, {Pearce}, {Peterson}, {Rauch}, {Stuchlik}, {Takahashi}, {Tang},
  {Uchida}, and {West}}]{2019AAS...23421508K}
{Kislat} F, {Abarr} Q, {Bose} R, {Braun} D, {de Geronimo} G, {Dowkontt} P,
  {Errando} M, {Gadson} TA, {Guarino} V, {Heatwole} SE, {Iyer} N, {Kiss} M,
  {Kitaguchi} T, {Krawczynski} H, {Kushwah} R, {Lanzi} J, {Li} S, {Lisalda} L,
  {Okajima} T, {Pearce} M, {Peterson} Z, {Rauch} B, {Stuchlik} D, {Takahashi}
  H, {Tang} J, {Uchida} N, {West} A (2019) {Results From the Antarctic Balloon
  Flight of the Hard X-ray Polarimeter X-Calibur}. In: American Astronomical
  Society Meeting Abstracts \#234, American Astronomical Society Meeting
  Abstracts, vol 234, p 215.08

\bibitem[{Klein and Nishina(1929)}]{Klein.1929}
Klein O, Nishina Y (1929) {Über die Streuung von Strahlung durch freie
  Elektronen nach der neuen relativistischen Quantendynamik von Dirac}.
  Zeitschrift für Physik 52(11-12):853--868, \doi{10.1007/bf01366453}

\bibitem[{Kole et~al.(2017)Kole, Li, Produit, Tymieniecka, Zhang, Zwolinska,
  Bao, Bernasconi, Cadoux, Feng, Gauvin, Hajdas, Kong, Li, Li, Liu,
  Marcinkowski, Orsi, Pohl, Rybka, Sun, Song, Szabelski, Wang, Wang, Wen, Wu,
  Wu, Xiao, Xiong, Zhang, Zhang, Zhang, Zhang, Zhang, and Zhao}]{Kole.2017}
Kole M, Li Z, Produit N, Tymieniecka T, Zhang J, Zwolinska A, Bao T, Bernasconi
  T, Cadoux F, Feng M, Gauvin N, Hajdas W, Kong S, Li H, Li L, Liu X,
  Marcinkowski R, Orsi S, Pohl M, Rybka D, Sun J, Song L, Szabelski J, Wang R,
  Wang Y, Wen X, Wu B, Wu X, Xiao H, Xiong S, Zhang L, Zhang L, Zhang S, Zhang
  X, Zhang Y, Zhao Y (2017) {Instrument performance and simulation verification
  of the POLAR detector}. Nuclear Instruments and Methods in Physics Research
  Section A: Accelerators, Spectrometers, Detectors and Associated Equipment
  872:28--40, \doi{10.1016/j.nima.2017.07.070},
  \urlprefix\url{https://arxiv.org/pdf/1708.00664.pdf}, \eprint{1708.00664}

\bibitem[{{Komura} et~al.(2017){Komura}, {Takada}, {Mizumura}, {Miyamoto},
  {Takemura}, {Kishimoto}, {Kubo}, {Kurosawa}, {Matsuoka}, {Miuchi},
  {Mizumoto}, {Nakamasu}, {Nakamura}, {Oda}, {Parker}, {Sawano}, {Sonoda},
  {Tanimori}, {Tomono}, and {Yoshikawa}}]{Komura2017}
{Komura} S, {Takada} A, {Mizumura} Y, {Miyamoto} S, {Takemura} T, {Kishimoto}
  T, {Kubo} H, {Kurosawa} S, {Matsuoka} Y, {Miuchi} K, {Mizumoto} T, {Nakamasu}
  Y, {Nakamura} K, {Oda} M, {Parker} JD, {Sawano} T, {Sonoda} S, {Tanimori} T,
  {Tomono} D, {Yoshikawa} K (2017) {Imaging Polarimeter for a Sub-MeV Gamma-Ray
  All-sky Survey Using an Electron-tracking Compton Camera}. \apj 839(1):41,
  \doi{10.3847/1538-4357/aa68dc}, \eprint{1703.07600}

\bibitem[{{Kotov} et~al.(2016){Kotov}, {Yurov}, {Glyanenko}, {Lupar},
  {Kochemasov}, {Trofimov}, {Zakharov}, {Faradzhaev}, {Tyshkevich}, {Rubtsov},
  {Dergachev}, {Kruglov}, {Lazutkov}, {Savchenko}, and
  {Skorodumov}}]{Kotov2016}
{Kotov} YD, {Yurov} VN, {Glyanenko} AS, {Lupar} EE, {Kochemasov} AV, {Trofimov}
  YA, {Zakharov} MS, {Faradzhaev} RM, {Tyshkevich} VG, {Rubtsov} IV,
  {Dergachev} VA, {Kruglov} EM, {Lazutkov} VP, {Savchenko} MI, {Skorodumov} DV
  (2016) {Solar X-ray polarimetry and spectrometry instrument PING-M for the
  Interhelioprobe mission}. Advances in Space Research 58:635--643,
  \doi{10.1016/j.asr.2016.05.024}

\bibitem[{{Krawczynski} et~al.(2011){Krawczynski}, {Garson}, {Guo}, {Baring},
  {Ghosh}, {Beilicke}, and {Lee}}]{2011APh....34..550K}
{Krawczynski} H, {Garson} A, {Guo} Q, {Baring} MG, {Ghosh} P, {Beilicke} M,
  {Lee} K (2011) {Scientific prospects for hard X-ray polarimetry}.
  Astroparticle Physics 34(7):550--567,
  \doi{10.1016/j.astropartphys.2010.12.001}, \eprint{1012.0321}

\bibitem[{{Kushwah} et~al.(2019){Kushwah}, {Iyer}, {Kiss}, {Stana}, and
  {Pearce}}]{2019NIMPA.94362376K}
{Kushwah} R, {Iyer} NK, {Kiss} M, {Stana} TA, {Pearce} M (2019) {A Compton
  polarimeter using scintillators read out with MPPCs through Citiroc ASIC}.
  Nuclear Instruments and Methods in Physics Research A 943:162376,
  \doi{10.1016/j.nima.2019.162376}, \eprint{1907.09030}

\bibitem[{{Lacombe} et~al.(2019){Lacombe}, {Belkacem}, {Houret},
  {Kn{\"o}dlseder}, {Ramon}, {Bardoux}, and {Gimenez}}]{Lacombe2019}
{Lacombe} K, {Belkacem} I, {Houret} B, {Kn{\"o}dlseder} J, {Ramon} P, {Bardoux}
  A, {Gimenez} T (2019) {Impact of Proton Irradiation on SiPM Dark Current for
  High-Energy Space Instruments}. In: 5th International Workshop on New
  Photon-Detectors (PD18), p 012006, \doi{10.7566/JPSCP.27.012006}

\bibitem[{{Lei} et~al.(1997){Lei}, {Dean}, and {Hills}}]{Lei1997}
{Lei} F, {Dean} AJ, {Hills} GL (1997) {Compton Polarimetry in Gamma-Ray
  Astronomy}. \ssr 82:309--388, \doi{10.1023/A:1005027107614}

\bibitem[{Leo(1987)}]{Leo_1987}
Leo WR (1987) Techniques for Nuclear and Particle Physics Experiments.
  Springer-Verlag

\bibitem[{Li et~al.(2018)Li, Kole, Sun, Song, Produit, Wu, Bao, Bernasconi,
  Cadoux, Dong, Feng, Gauvin, Hajdas, Li, Li, Liu, Marcinkowski, Pohl, Rybka,
  Shi, Szabelski, Tymieniecka, Wang, Wang, Wen, Wu, Xiong, Zwolinska, Zhang,
  Zhang, Zhang, Zhang, and Zhao}]{Li.2018otl}
Li Z, Kole M, Sun J, Song L, Produit N, Wu B, Bao T, Bernasconi T, Cadoux F,
  Dong Y, Feng M, Gauvin N, Hajdas W, Li H, Li L, Liu X, Marcinkowski R, Pohl
  M, Rybka DK, Shi H, Szabelski J, Tymieniecka T, Wang R, Wang Y, Wen X, Wu X,
  Xiong S, Zwolinska A, Zhang L, Zhang L, Zhang S, Zhang Y, Zhao Y (2018)
  {In-orbit instrument performance study and calibration for POLAR polarization
  measurements}. Nuclear Instruments and Methods in Physics Research Section A:
  Accelerators, Spectrometers, Detectors and Associated Equipment 900:8--24,
  \doi{10.1016/j.nima.2018.05.041}, \eprint{1805.07605}

\bibitem[{{Long} et~al.(2021){Long}, {Feng}, {Li}, {Zhu}, {Wu}, {Huang},
  {Minuti}, {Jiang}, {Wang}, {Xu}, {Costa}, {Yang}, {Citraro}, {Nasimi}, {Yu},
  {Jin}, {Zeng}, {An}, {Baldini}, {Bellazzini}, {Brez}, {Latronico},
  {Sgr{\`o}}, {Spandre}, {Pinchera}, {Muleri}, and
  {Soffitta}}]{2021ApJ...912L..28L}
{Long} X, {Feng} H, {Li} H, {Zhu} J, {Wu} Q, {Huang} J, {Minuti} M, {Jiang} W,
  {Wang} W, {Xu} R, {Costa} E, {Yang} D, {Citraro} S, {Nasimi} H, {Yu} J, {Jin}
  G, {Zeng} M, {An} P, {Baldini} L, {Bellazzini} R, {Brez} A, {Latronico} L,
  {Sgr{\`o}} C, {Spandre} G, {Pinchera} M, {Muleri} F, {Soffitta} P (2021)
  {X-Ray Polarimetry of the Crab Nebula with PolarLight: Polarization Recovery
  after the Glitch and a Secular Position Angle Variation}. \apjl 912(2):L28,
  \doi{10.3847/2041-8213/abfb00}, \eprint{2104.11391}

\bibitem[{{Long} et~al.(2022){Long}, {Feng}, {Li}, {Zhu}, {Wu}, {Huang},
  {Minuti}, {Jiang}, {Yang}, {Citraro}, {Nasimi}, {Yu}, {Jin}, {Zeng}, {An},
  {Jiang}, {Costa}, {Baldini}, {Bellazzini}, {Brez}, {Latronico}, {Sgr{\`o}},
  {Spandre}, {Pinchera}, {Muleri}, and {Soffitta}}]{2022ApJ...924L..13L}
{Long} X, {Feng} H, {Li} H, {Zhu} J, {Wu} Q, {Huang} J, {Minuti} M, {Jiang} W,
  {Yang} D, {Citraro} S, {Nasimi} H, {Yu} J, {Jin} G, {Zeng} M, {An} P, {Jiang}
  J, {Costa} E, {Baldini} L, {Bellazzini} R, {Brez} A, {Latronico} L,
  {Sgr{\`o}} C, {Spandre} G, {Pinchera} M, {Muleri} F, {Soffitta} P (2022) {A
  Significant Detection of X-ray Polarization in Sco X-1 with PolarLight and
  Constraints on the Corona Geometry}. \apjl 924(1):L13,
  \doi{10.3847/2041-8213/ac4673}, \eprint{2112.02837}

\bibitem[{{Lutz}(1999, 207)}]{Lutz_1999}
{Lutz} G (1999, 207) {Semiconductor Radiation Detectors}. Springer

\bibitem[{{McConnell} and {LEAP Collaboration}(2016)}]{McConnell2016b}
{McConnell} ML, {LEAP Collaboration} (2016) {LEAP - A LargE Area Burst
  Polarimeter for the ISS}. In: Eighth Huntsville Gamma-Ray Burst Symposium,
  LPI Contributions, vol 1962, p 4051

\bibitem[{{McConnell} et~al.(2013){McConnell}, {Bloser}, {Connor}, {Ertley},
  {Legere}, {Ryan}, and {Wasti}}]{McConnell2013}
{McConnell} ML, {Bloser} PF, {Connor} TP, {Ertley} C, {Legere} J, {Ryan} JM,
  {Wasti} SK (2013) {Plans for the next GRAPE balloon flight}. In: UV, X-Ray,
  and Gamma-Ray Space Instrumentation for Astronomy XVIII, \procspie, vol 8859,
  p 885909, \doi{10.1117/12.2024145}

\bibitem[{{McConnell} et~al.(2014){McConnell}, {Bloser}, {Ertley}, {Legere},
  {Ryan}, and {Wasti}}]{McConnell2014}
{McConnell} ML, {Bloser} PF, {Ertley} C, {Legere} J, {Ryan} JM, {Wasti} SK
  (2014) {Current status of the GRAPE balloon program}. In: Space Telescopes
  and Instrumentation 2014: Ultraviolet to Gamma Ray, \procspie, vol 9144, p
  91443P, \doi{10.1117/12.2056886}

\bibitem[{{McEnery} et~al.(2019){McEnery}, {van der Horst}, {Dominguez},
  {Moiseev}, {Marcowith}, {Harding}, {Lien}, {Giuliani}, {Inglis}, {Ansoldi},
  {Stamerra}, {Manousakis}, {Strong}, {Bambi}, {Patricelli}, {Baring},
  {Barrio}, {Bastieri}, {Fields}, {Beacom}, {Beckmann}, {Bednarek}, {Rani},
  {Boggs}, {Bolotnikov}, {Cenko}, {Buckley}, {Grefenstette}, {Hui}, {Pittori},
  {Prescod-Weinstein}, {Shrader}, {Gouiffes}, {Kierans}, {Wilson-Hodge},
  {D'Ammando}, {Castro}, {Kocveski}, {Gasparrini}, {Thompson}, {Williams}, {De
  Angelis}, {Bernard}, {Digel}, {Morcuende}, {Charles}, {Bissaldi}, {Hays},
  {Ferrara}, {Bozzo}, {Grove}, {Wulf}, {Bottacini}, {Caroli}, {Kislat},
  {Oikonomou}, {Giordano}, {Longo}, {Fryer}, {Fukazawa}, {Georganopoulos}, {De
  Nolfo}, {Vianello}, {Kanbach}, {Younes}, {Blumer}, {Hartmann}, {Hernanz},
  {Takahashi}, {Li}, {Agudo}, {Moskalenko}, {Stumke}, {Grenier}, {Smith},
  {Rodi}, {Perkins}, {Gelfand}, {Holder}, {Knodlseder}, {Kopp}, {Lenain},
  {{\'A}lvarez}, {Metcalfe}, {Krizmanic}, {Stephen}, {Hewitt}, {Mitchell},
  {Harding}, {Tomsick}, {Racusin}, {Finke}, {Kargaltsev}, {Klimenko},
  {Krawczynski}, {Smith}, {Kubo}, {Di Venere}, {Marcotulli}, {Lommler},
  {Parker}, {Baldini}, {Foffano}, {Zampieri}, {Tibaldo}, {Petropoulou},
  {Ajello}, {Meyer}, {L{\'o}pez}, {McConnell}, {Boettcher}, {Cardillo},
  {Martinez}, {Kerr}, {Mazziotta}, {McEnery}, {Di Mauro}, {Wood}, {Meyer},
  {Briggs}, {De Becker}, {Lovellette}, {Doro}, {Sanchez-Conde}, {Moss},
  {Mizuno}, {Rib{\'o}}, {Nakazawa}, {Neilson}, {Auricchio}, {Omodei},
  {Oberlack}, {Ohno}, {Orlando}, {Otte}, {Coppi}, {Bloser}, {Zhang}, {Laurent},
  {Pohl}, {Prandini}, {Shawhan}, {Caputo}, {Campana}, {Rando}, {Woolf},
  {Johnson}, {Mignani}, {Walter}, {Ojha}, {da Silva}, {Dietrich}, {Funk},
  {Zane}, {Anton}, {Buson}, {Cutini}, {Saz Parkinson}, {Schirato}, {Griffin},
  {Kaufmann}, {Stawarz}, {Ciprini}, {Del Sordo}, {Jones}, {Guiriec}, {Tajima},
  {Cheung}, {The}, {Venters}, {Porter}, {Linden}, {Barres}, {Paliya},
  {Bozhilov}, {Vestrand}, {Tatischeff}, {Chen}, {Wang}, {Tanaka}, {Uhm},
  {Zhang}, {Zimmer}, {Zoglauer}, and {Wadiasingh}}]{McEnery2019}
{McEnery} J, {van der Horst} A, {Dominguez} A, {Moiseev} A, {Marcowith} A,
  {Harding} A, {Lien} A, {Giuliani} A, {Inglis} A, {Ansoldi} S, {Stamerra} A,
  {Manousakis} A, {Strong} A, {Bambi} C, {Patricelli} B, {Baring} M, {Barrio}
  JA, {Bastieri} D, {Fields} B, {Beacom} J, {Beckmann} V, {Bednarek} W, {Rani}
  B, {Boggs} S, {Bolotnikov} A, {Cenko} SB, {Buckley} J, {Grefenstette} B,
  {Hui} M, {Pittori} C, {Prescod-Weinstein} C, {Shrader} C, {Gouiffes} C,
  {Kierans} C, {Wilson-Hodge} C, {D'Ammando} F, {Castro} D, {Kocveski} D,
  {Gasparrini} D, {Thompson} D, {Williams} D, {De Angelis} A, {Bernard} D,
  {Digel} S, {Morcuende} D, {Charles} E, {Bissaldi} E, {Hays} E, {Ferrara} E,
  {Bozzo} E, {Grove} E, {Wulf} E, {Bottacini} E, {Caroli} E, {Kislat} F,
  {Oikonomou} F, {Giordano} F, {Longo} F, {Fryer} C, {Fukazawa} Y,
  {Georganopoulos} M, {De Nolfo} G, {Vianello} G, {Kanbach} G, {Younes} G,
  {Blumer} H, {Hartmann} D, {Hernanz} M, {Takahashi} H, {Li} H, {Agudo} I,
  {Moskalenko} I, {Stumke} I, {Grenier} I, {Smith} J, {Rodi} J, {Perkins} J,
  {Gelfand} J, {Holder} J, {Knodlseder} J, {Kopp} J, {Lenain} JP, {{\'A}lvarez}
  JM, {Metcalfe} J, {Krizmanic} J, {Stephen} JB, {Hewitt} J, {Mitchell} J,
  {Harding} P, {Tomsick} J, {Racusin} J, {Finke} J, {Kargaltsev} O, {Klimenko}
  AV, {Krawczynski} H, {Smith} K, {Kubo} H, {Di Venere} L, {Marcotulli} L,
  {Lommler} J, {Parker} L, {Baldini} L, {Foffano} L, {Zampieri} L, {Tibaldo} L,
  {Petropoulou} M, {Ajello} M, {Meyer} M, {L{\'o}pez} M, {McConnell} M,
  {Boettcher} M, {Cardillo} M, {Martinez} M, {Kerr} M, {Mazziotta} MN,
  {McEnery} J, {Di Mauro} M, {Wood} M, {Meyer} E, {Briggs} M, {De Becker} M,
  {Lovellette} M, {Doro} M, {Sanchez-Conde} MA, {Moss} M, {Mizuno} T,
  {Rib{\'o}} M, {Nakazawa} K, {Neilson} NK, {Auricchio} N, {Omodei} N,
  {Oberlack} U, {Ohno} M, {Orlando} E, {Otte} N, {Coppi} P, {Bloser} P, {Zhang}
  H, {Laurent} P, {Pohl} M, {Prandini} E, {Shawhan} P, {Caputo} R, {Campana} R,
  {Rando} R, {Woolf} R, {Johnson} R, {Mignani} R, {Walter} R, {Ojha} R, {da
  Silva} RC, {Dietrich} S, {Funk} S, {Zane} S, {Anton} S, {Buson} S, {Cutini}
  S, {Saz Parkinson} P, {Schirato} R, {Griffin} S, {Kaufmann} S, {Stawarz} L,
  {Ciprini} S, {Del Sordo} S, {Jones} S, {Guiriec} S, {Tajima} H, {Cheung} T,
  {The} LS, {Venters} T, {Porter} T, {Linden} T, {Barres} U, {Paliya} VS,
  {Bozhilov} V, {Vestrand} T, {Tatischeff} V, {Chen} W, {Wang} X, {Tanaka} Y,
  {Uhm} L, {Zhang} B, {Zimmer} S, {Zoglauer} A, {Wadiasingh} Z (2019) {All-sky
  Medium Energy Gamma-ray Observatory: Exploring the Extreme Multimessenger
  Universe}. In: Bulletin of the American Astronomical Society, vol~51, p 245,
  \eprint{1907.07558}

\bibitem[{Mikhalev(2018)}]{Mikhalev.2018}
Mikhalev V (2018) {Pitfalls of statistics-limited X-ray polarization analysis}.
  Astronomy \& Astrophysics 615:A54, \doi{10.1051/0004-6361/201731971},
  \eprint{1803.10120}

\bibitem[{{Mitchell} et~al.(2019){Mitchell}, {Phlips}, {Woolf}, {Finne}, and
  {Johnson}}]{Mitchell2019}
{Mitchell} LJ, {Phlips} BF, {Woolf} RS, {Finne} TT, {Johnson} WN (2019)
  {Strontium iodide radiation instrumentation II (SIRI-2)}. In: UV, X-Ray, and
  Gamma-Ray Space Instrumentation for Astronomy XXI, Society of Photo-Optical
  Instrumentation Engineers (SPIE) Conference Series, vol 11118, p 111180I,
  \doi{10.1117/12.2528073}

\bibitem[{{Mizuno} et~al.(2009){Mizuno}, {Kanai}, {Kataoka}, {Kiss}, {Kurita},
  {Pearce}, {Tajima}, {Takahashi}, {Tanaka}, {Ueno}, {Umeki}, {Yoshida},
  {Arimoto}, {Axelsson}, {Marini Bettolo}, {Bogaert}, {Chen}, {Craig},
  {Fukazawa}, {Gunji}, {Kamae}, {Katsuta}, {Kawai}, {Kishimoto}, {Klamra},
  {Larsson}, {Madejski}, {Ng}, {Ryde}, {Rydstr{\"o}m}, {Takahashi}, {Thurston},
  and {Varner}}]{2009NIMPA.600..609M}
{Mizuno} T, {Kanai} Y, {Kataoka} J, {Kiss} M, {Kurita} K, {Pearce} M, {Tajima}
  H, {Takahashi} H, {Tanaka} T, {Ueno} M, {Umeki} Y, {Yoshida} H, {Arimoto} M,
  {Axelsson} M, {Marini Bettolo} C, {Bogaert} G, {Chen} P, {Craig} W,
  {Fukazawa} Y, {Gunji} S, {Kamae} T, {Katsuta} J, {Kawai} N, {Kishimoto} S,
  {Klamra} W, {Larsson} S, {Madejski} G, {Ng} JST, {Ryde} F, {Rydstr{\"o}m} S,
  {Takahashi} T, {Thurston} TS, {Varner} G (2009) {A Monte Carlo method for
  calculating the energy response of plastic scintillators to polarized photons
  below 100 keV}. Nuclear Instruments and Methods in Physics Research A
  600(3):609--617, \doi{10.1016/j.nima.2008.11.148}

\bibitem[{{Moll}(2018)}]{2018ITNS...65.1561M}
{Moll} M (2018) {Displacement Damage in Silicon Detectors for High Energy
  Physics}. IEEE Transactions on Nuclear Science 65(8):1561--1582,
  \doi{10.1109/TNS.2018.2819506}

\bibitem[{{Muleri}(2014)}]{Muleri2014}
{Muleri} F (2014) {On the Operation of X-Ray Polarimeters with a Large Field of
  View}. \apj 782:28, \doi{10.1088/0004-637X/782/1/28}, \eprint{1312.2497}

\bibitem[{{Muleri} and {Campana}(2012)}]{2012ApJ...751...88M}
{Muleri} F, {Campana} R (2012) {Sensitivity of Stacked Imaging Detectors to
  Hard X-Ray Polarization}. \apj 751(2):88, \doi{10.1088/0004-637X/751/2/88},
  \eprint{1204.0681}

\bibitem[{{Muleri} et~al.(2022){Muleri}, {Piazzolla}, {Di Marco}, {Fabiani},
  {La Monaca}, {Lefevre}, {Morbidini}, {Rankin}, {Soffitta}, {Tobia}, {Xie},
  {Amici}, {Attin{\`a}}, {Bachetti}, {Brienza}, {Centrone}, {Costa}, {Del
  Monte}, {Di Cosimo}, {Di Persio}, {Evangelista}, {Ferrazzoli}, {Loffredo},
  {Perri}, {Pilia}, {Puccetti}, {Ratheesh}, {Rubini}, {Santoli}, {Scalise}, and
  {Trois}}]{Muleri2022}
{Muleri} F, {Piazzolla} R, {Di Marco} A, {Fabiani} S, {La Monaca} F, {Lefevre}
  C, {Morbidini} A, {Rankin} J, {Soffitta} P, {Tobia} A, {Xie} F, {Amici} F,
  {Attin{\`a}} P, {Bachetti} M, {Brienza} D, {Centrone} M, {Costa} E, {Del
  Monte} E, {Di Cosimo} S, {Di Persio} G, {Evangelista} Y, {Ferrazzoli} R,
  {Loffredo} P, {Perri} M, {Pilia} M, {Puccetti} S, {Ratheesh} A, {Rubini} A,
  {Santoli} F, {Scalise} E, {Trois} A (2022) {The IXPE instrument calibration
  equipment}. Astroparticle Physics 136:102658,
  \doi{10.1016/j.astropartphys.2021.102658}, \eprint{2111.02066}

\bibitem[{{O{\~n}ate Melecio} et~al.(2021){O{\~n}ate Melecio}, {Ertley},
  {McConnell}, {Legere}, {Bloser}, {Briggs}, {Gaskin}, {Goldstein}, {Grove},
  {Hui}, {Jenke}, {Kippen}, {Kislat}, {Kocevski}, {Krizmanic}, {Meegan},
  {Preece}, {Ryan}, {Sturner}, {Veres}, {Vestrand}, and
  {Wilson-Hodge}}]{Onate2021}
{O{\~n}ate Melecio} K, {Ertley} C, {McConnell} M, {Legere} J, {Bloser} P,
  {Briggs} M, {Gaskin} J, {Goldstein} A, {Grove} E, {Hui} M, {Jenke} P,
  {Kippen} MR, {Kislat} F, {Kocevski} D, {Krizmanic} JF, {Meegan} C, {Preece}
  R, {Ryan} J, {Sturner} SJ, {Veres} P, {Vestrand} WT, {Wilson-Hodge} C (2021)
  {Evaluation of a prototype detector for the LargE Area burst Polarimeter
  (LEAP)}. In: Society of Photo-Optical Instrumentation Engineers (SPIE)
  Conference Series, Society of Photo-Optical Instrumentation Engineers (SPIE)
  Conference Series, vol 11821, p 118210Q, \doi{10.1117/12.2594568}

\bibitem[{Orsi et~al.(2011)Orsi, Haas, Hajdas, Honkimaki, Lamanna,
  Marcinkowski, Pohl, Produit, Rapin, and Rybka}]{Orsi.2011}
Orsi, Haas, Hajdas, Honkimaki, Lamanna, Marcinkowski, Pohl, Produit, Rapin,
  Rybka (2011) {Response of the Compton polarimeter POLAR to polarized hard
  X-rays}. Nuclear Instruments and Methods in Physics Research Section A:
  Accelerators, Spectrometers, Detectors and Associated Equipment 648(1):16 --
  16, \doi{10.1016/j.nima.2011.04.012},
  \urlprefix\url{http://pubget.com/site/paper/pgtmp\_36a9b045a0ad5a85e465a23e7ab05487?institution=}

\bibitem[{{Paul} et~al.(2010){Paul}, {Rishin}, {Maitra}, {Gopalakrishna},
  {Duraichelvan}, {Ateequlla}, {Cowsik}, {Devasia}, and {Marykutty}}]{Paul2010}
{Paul} B, {Rishin} PV, {Maitra} C, {Gopalakrishna} MR, {Duraichelvan} R,
  {Ateequlla} CM, {Cowsik} R, {Devasia} J, {Marykutty} J (2010) {Thomson X-ray
  Polarimeter for a Small Satellite Mission}. In: The First Year of MAXI:
  Monitoring Variable X-ray Sources, p~68

\bibitem[{{Paul} et~al.(2016){Paul}, {Gopala Krishna}, and {Puthiya
  Veetil}}]{2016cosp...41E1533P}
{Paul} B, {Gopala Krishna} MR, {Puthiya Veetil} R (2016) {POLIX: A Thomson
  X-ray polarimeter for a small satellite mission}. In: 41st COSPAR Scientific
  Assembly, vol~41, pp E1.15--8--16

\bibitem[{{Qiang} et~al.(2013){Qiang}, {Zorn}, {Barbosa}, and
  {Smith}}]{Qiang2013}
{Qiang} Y, {Zorn} C, {Barbosa} F, {Smith} E (2013) {Radiation hardness tests of
  SiPMs for the JLab Hall D Barrel calorimeter}. Nuclear Instruments and
  Methods in Physics Research A 698:234--241, \doi{10.1016/j.nima.2012.10.015},
  \eprint{1207.3743}

\bibitem[{{Rodriguez} et~al.(2015){Rodriguez}, {Grinberg}, {Laurent}, {Cadolle
  Bel}, {Pottschmidt}, {Pooley}, {Bodaghee}, {Wilms}, and
  {Gouiff{\`e}s}}]{2015ApJ...807...17R}
{Rodriguez} J, {Grinberg} V, {Laurent} P, {Cadolle Bel} M, {Pottschmidt} K,
  {Pooley} G, {Bodaghee} A, {Wilms} J, {Gouiff{\`e}s} C (2015) {Spectral State
  Dependence of the 0.4-2 MeV Polarized Emission in Cygnus X-1 Seen with
  INTEGRAL/IBIS, and Links with the AMI Radio Data}. \apj 807(1):17,
  \doi{10.1088/0004-637X/807/1/17}, \eprint{1505.02786}

\bibitem[{{Soffitta} et~al.(2019){Soffitta}, {Bucciantini}, {Churazov},
  {Costa}, {Dovciak}, {Feng}, {Heyl}, {Ingram}, {Jahoda}, {Kaaret}, {Kallman},
  {Karas}, {Khabibullin}, {Krawczynski}, {Malzac}, {Marin}, {Marshall}, {Matt},
  {Muleri}, {Mundell}, {Pearce}, {Petrucci}, {Poutanen}, {Romani},
  {Santangelo}, {Tagliaferri}, {Taverna}, {Turolla}, {Vink}, and
  {Zane}}]{2019arXiv191010092S}
{Soffitta} P, {Bucciantini} N, {Churazov} E, {Costa} E, {Dovciak} M, {Feng} H,
  {Heyl} J, {Ingram} A, {Jahoda} K, {Kaaret} P, {Kallman} T, {Karas} V,
  {Khabibullin} I, {Krawczynski} H, {Malzac} J, {Marin} F, {Marshall} H, {Matt}
  G, {Muleri} F, {Mundell} C, {Pearce} M, {Petrucci} PO, {Poutanen} J, {Romani}
  R, {Santangelo} A, {Tagliaferri} G, {Taverna} R, {Turolla} R, {Vink} J,
  {Zane} S (2019) {ESA Voyage 2050 white paper: A Polarized View of the Hot and
  Violent Universe}. arXiv e-prints arXiv:1910.10092, \eprint{1910.10092}

\bibitem[{{Stokes}(1851)}]{1851TCaPS...9..399S}
{Stokes} GG (1851) {On the Composition and Resolution of Streams of Polarized
  Light from different Sources}. Transactions of the Cambridge Philosophical
  Society 9:399

\bibitem[{{Tajima} et~al.(2010){Tajima}, {Blandford}, {Enoto}, {Fukazawa},
  {Gilmore}, {Kamae}, {Kataoka}, {Kawaharada}, {Kokubun}, {Laurent}, {Lebrun},
  {Limousin}, {Madejski}, {Makishima}, {Mizuno}, {Nakazawa}, {Ohno}, {Ohta},
  {Sato}, {Sato}, {Takahashi}, {Takahashi}, {Tanaka}, {Tashiro}, {Terada},
  {Uchiyama}, {Watanabe}, {Yamaoka}, and {Yonetoku}}]{Tajima2010}
{Tajima} H, {Blandford} R, {Enoto} T, {Fukazawa} Y, {Gilmore} K, {Kamae} T,
  {Kataoka} J, {Kawaharada} M, {Kokubun} M, {Laurent} P, {Lebrun} F, {Limousin}
  O, {Madejski} G, {Makishima} K, {Mizuno} T, {Nakazawa} K, {Ohno} M, {Ohta} M,
  {Sato} G, {Sato} R, {Takahashi} H, {Takahashi} T, {Tanaka} T, {Tashiro} M,
  {Terada} Y, {Uchiyama} Y, {Watanabe} S, {Yamaoka} K, {Yonetoku} D (2010)
  {Soft gamma-ray detector for the ASTRO-H Mission}. In: {Arnaud} M, {Murray}
  SS, {Takahashi} T (eds) Space Telescopes and Instrumentation 2010:
  Ultraviolet to Gamma Ray, Society of Photo-Optical Instrumentation Engineers
  (SPIE) Conference Series, vol 7732, p 773216, \doi{10.1117/12.857531},
  \eprint{1010.4997}

\bibitem[{{Takeda} et~al.(2010){Takeda}, {Odaka}, {Katsuta}, {Ishikawa},
  {Sugimoto}, {Koseki}, {Watanabe}, {Sato}, {Kokubun}, {Takahashi}, {Nakazawa},
  {Fukazawa}, {Tajima}, and {Toyokawa}}]{2010NIMPA.622..619T}
{Takeda} S, {Odaka} H, {Katsuta} J, {Ishikawa} Sn, {Sugimoto} Si, {Koseki} Y,
  {Watanabe} S, {Sato} G, {Kokubun} M, {Takahashi} T, {Nakazawa} K, {Fukazawa}
  Y, {Tajima} H, {Toyokawa} H (2010) {Polarimetric performance of Si/CdTe
  semiconductor Compton camera}. Nuclear Instruments and Methods in Physics
  Research A 622(3):619--627, \doi{10.1016/j.nima.2010.07.077}

\bibitem[{Tanimori et~al.(2004)Tanimori, Kubo, Miuchi, Nagayoshi, Orito,
  Takada, Takeda, and Ueno}]{Tanimori2004}
Tanimori T, Kubo H, Miuchi K, Nagayoshi T, Orito R, Takada A, Takeda A, Ueno M
  (2004) Mev $\gamma$-ray imaging detector with micro-tpc. New Astronomy
  Reviews 48(1):263--268, \doi{https://doi.org/10.1016/j.newar.2003.11.038},
  \urlprefix\url{https://www.sciencedirect.com/science/article/pii/S1387647303003154},
  astronomy with Radioactivities IV and Filling the Sensitivity Gap in MeV
  Astronomy

\bibitem[{{Tanimori} et~al.(2015){Tanimori}, {Kubo}, {Takada}, {Iwaki},
  {Komura}, {Kurosawa}, {Matsuoka}, {Miuchi}, {Miyamoto}, {Mizumoto},
  {Mizumura}, {Nakamura}, {Nakamura}, {Oda}, {Parker}, {Sawano}, {Sonoda},
  {Takemura}, {Tomono}, and {Ueno}}]{Tanimori2015}
{Tanimori} T, {Kubo} H, {Takada} A, {Iwaki} S, {Komura} S, {Kurosawa} S,
  {Matsuoka} Y, {Miuchi} K, {Miyamoto} S, {Mizumoto} T, {Mizumura} Y,
  {Nakamura} K, {Nakamura} S, {Oda} M, {Parker} JD, {Sawano} T, {Sonoda} S,
  {Takemura} T, {Tomono} D, {Ueno} K (2015) {An Electron-Tracking Compton
  Telescope for a Survey of the Deep Universe by MeV Gamma-Rays}. \apj
  810(1):28, \doi{10.1088/0004-637X/810/1/28}, \eprint{1507.03850}

\bibitem[{{Tanimori} et~al.(2017){Tanimori}, {Mizumura}, {Takada}, {Miyamoto},
  {Takemura}, {Kishimoto}, {Komura}, {Kubo}, {Kurosawa}, {Matsuoka}, {Miuchi},
  {Mizumoto}, {Nakamasu}, {Nakamura}, {Parker}, {Sawano}, {Sonoda}, {Tomono},
  and {Yoshikawa}}]{Tanimori2017}
{Tanimori} T, {Mizumura} Y, {Takada} A, {Miyamoto} S, {Takemura} T, {Kishimoto}
  T, {Komura} S, {Kubo} H, {Kurosawa} S, {Matsuoka} Y, {Miuchi} K, {Mizumoto}
  T, {Nakamasu} Y, {Nakamura} K, {Parker} JD, {Sawano} T, {Sonoda} S, {Tomono}
  D, {Yoshikawa} K (2017) {Establishment of Imaging Spectroscopy of Nuclear
  Gamma-Rays based on Geometrical Optics}. Scientific Reports 7:41511,
  \doi{10.1038/srep41511}, \eprint{1702.01483}

\bibitem[{{Tomsick} and {COSI Collaboration}(2022)}]{Tomsick2021}
{Tomsick} J, {COSI Collaboration} (2022) {The Compton Spectrometer and Imager
  Project for MeV Astronomy}. In: 37th International Cosmic Ray Conference.
  12-23 July 2021. Berlin, p 652, \eprint{2109.10403}

\bibitem[{{Ulyanov} et~al.(2020){Ulyanov}, {Murphy}, {Mangan}, {Gupta},
  {Hajdas}, {de Faoite}, {Shortt}, {Hanlon}, and {McBreen}}]{Ulyanov2020}
{Ulyanov} A, {Murphy} D, {Mangan} J, {Gupta} V, {Hajdas} W, {de Faoite} D,
  {Shortt} B, {Hanlon} L, {McBreen} S (2020) {Radiation damage study of SensL
  J-series silicon photomultipliers using 101.4 MeV protons}. Nuclear
  Instruments and Methods in Physics Research A 976:164203,
  \doi{10.1016/j.nima.2020.164203}, \eprint{2007.10919}

\bibitem[{{Vadawale} et~al.(2010){Vadawale}, {Paul}, {Pendharkar}, and
  {Naik}}]{Vadawale2010}
{Vadawale} SV, {Paul} B, {Pendharkar} J, {Naik} S (2010) {Comparative study of
  different scattering geometries for the proposed Indian X-ray polarization
  measurement experiment using Geant4}. Nuclear Instruments and Methods in
  Physics Research A 618(1-3):182--189, \doi{10.1016/j.nima.2010.02.116},
  \eprint{1003.0519}

\bibitem[{{Vadawale} et~al.(2015){Vadawale}, {Chattopadhyay}, {Rao},
  {Bhattacharya}, {Bhalerao}, {Vagshette}, {Pawar}, and
  {Sreekumar}}]{Vadawale2015}
{Vadawale} SV, {Chattopadhyay} T, {Rao} AR, {Bhattacharya} D, {Bhalerao} VB,
  {Vagshette} N, {Pawar} P, {Sreekumar} S (2015) {Hard X-ray polarimetry with
  Astrosat-CZTI}. \aap 578:A73, \doi{10.1051/0004-6361/201525686}

\bibitem[{{Vadawale} et~al.(2018){Vadawale}, {Chattopadhyay}, {Mithun}, {Rao},
  {Bhattacharya}, {Vibhute}, {Bhalerao}, {Dewangan}, {Misra}, {Paul}, {Basu},
  {Joshi}, {Sreekumar}, {Samuel}, {Priya}, {Vinod}, and
  {Seetha}}]{Vadawale2018a}
{Vadawale} SV, {Chattopadhyay} T, {Mithun} NPS, {Rao} AR, {Bhattacharya} D,
  {Vibhute} A, {Bhalerao} VB, {Dewangan} GC, {Misra} R, {Paul} B, {Basu} A,
  {Joshi} BC, {Sreekumar} S, {Samuel} E, {Priya} P, {Vinod} P, {Seetha} S
  (2018) {Phase-resolved X-ray polarimetry of the Crab pulsar with the AstroSat
  CZT Imager}. Nature Astronomy 2:50--55, \doi{10.1038/s41550-017-0293-z}

\bibitem[{{Weisskopf} et~al.(1976){Weisskopf}, {Cohen}, {Kestenbaum}, {Long},
  {Novick}, and {Wolff}}]{Weisskopf1976}
{Weisskopf} MC, {Cohen} GG, {Kestenbaum} HL, {Long} KS, {Novick} R, {Wolff} RS
  (1976) {Measurement of the X-ray polarization of the Crab Nebula}. \apjl
  208:L125

\bibitem[{{Weisskopf} et~al.(1978){Weisskopf}, {Kestenbaum}, {Long}, {Novick},
  and {Silver}}]{1978ApJ...221L..13W}
{Weisskopf} MC, {Kestenbaum} HL, {Long} KS, {Novick} R, {Silver} EH (1978) {An
  upper limit to the linear X-ray polarization of Scorpius X-1.} \apjl
  221:L13--L16, \doi{10.1086/182655}

\bibitem[{Weisskopf et~al.(2010)Weisskopf, Elsner, and
  O'Dell}]{Weisskopf.2010855k}
Weisskopf MC, Elsner RF, O'Dell SL (2010) {On understanding the figures of
  merit for detection and measurement of x-ray polarization}. SPIE Astronomical
  Telescopes and Instrumentation: Observational Frontiers of Astronomy for the
  New Decade, SPIE Astronomical Telescopes and Instrumentation: Observational
  Frontiers of Astronomy for the New Decade, vol 7732, pp 77320E -- 77320E--5,
  \doi{10.1117/12.857357},
  \urlprefix\url{http://proceedings.spiedigitallibrary.org/proceeding.aspx?doi=10.1117/12.857357}

\bibitem[{{Weisskopf} et~al.(2022){Weisskopf}, {Soffitta}, {Baldini}, {Ramsey},
  {O'Dell}, {Romani}, {Matt}, {Deininger}, {Baumgartner}, {Bellazzini},
  {Costa}, {Kolodziejczak}, {Latronico}, {Marshall}, {Muleri}, {Bongiorno},
  {Tennant}, {Bucciantini}, {Dovciak}, {Marin}, {Marscher}, {Poutanen},
  {Slane}, {Turolla}, {Kalinowski}, {Di Marco}, {Fabiani}, {Minuti}, {La
  Monaca}, {Pinchera}, {Rankin}, {Sgro'}, {Trois}, {Xie}, {Alexander}, {Allen},
  {Amici}, {Andersen}, {Antonelli}, {Antoniak}, {Attin{\`a}}, {Barbanera},
  {Bachetti}, {Baggett}, {Bladt}, {Brez}, {Bonino}, {Boree}, {Borotto},
  {Breeding}, {Brienza}, {Bygott}, {Caporale}, {Cardelli}, {Carpentiero},
  {Castellano}, {Castronuovo}, {Cavalli}, {Cavazzuti}, {Ceccanti}, {Centrone},
  {Citraro}, {D'Amico}, {D'Alba}, {Di Gesu}, {Del Monte}, {Dietz}, {Di Lalla},
  {Persio}, {Dolan}, {Donnarumma}, {Evangelista}, {Ferrant}, {Ferrazzoli},
  {Ferrie}, {Footdale}, {Forsyth}, {Foster}, {Garelick}, {Gunji}, {Gurnee},
  {Head}, {Hibbard}, {Johnson}, {Kelly}, {Kilaru}, {Lefevre}, {Roy},
  {Loffredo}, {Lorenzi}, {Lucchesi}, {Maddox}, {Magazzu}, {Maldera},
  {Manfreda}, {Mangraviti}, {Marengo}, {Marrocchesi}, {Massaro}, {Mauger},
  {McCracken}, {McEachen}, {Mize}, {Mereu}, {Mitchell}, {Mitsuishi},
  {Morbidini}, {Mosti}, {Nasimi}, {Negri}, {Negro}, {Nguyen}, {Nitschke},
  {Nuti}, {Onizuka}, {Oppedisano}, {Orsini}, {Osborne}, {Pacheco}, {Paggi},
  {Painter}, {Pavelitz}, {Pentz}, {Piazzolla}, {Perri}, {Pesce-Rollins},
  {Peterson}, {Pilia}, {Profeti}, {Puccetti}, {Ranganathan}, {Ratheesh},
  {Reedy}, {Root}, {Rubini}, {Ruswick}, {Sanchez}, {Sarra}, {Santoli},
  {Scalise}, {Sciortino}, {Schroeder}, {Seek}, {Sosdian}, {Spandre}, {Speegle},
  {Tamagawa}, {Tardiola}, {Tobia}, {Thomas}, {Valerie}, {Vimercati}, {Walden},
  {Weddendorf}, {Wedmore}, {Welch}, {Zanetti}, and
  {Zanetti}}]{2022JATIS...8b6002W}
{Weisskopf} MC, {Soffitta} P, {Baldini} L, {Ramsey} BD, {O'Dell} SL, {Romani}
  RW, {Matt} G, {Deininger} WD, {Baumgartner} WH, {Bellazzini} R, {Costa} E,
  {Kolodziejczak} JJ, {Latronico} L, {Marshall} HL, {Muleri} F, {Bongiorno} SD,
  {Tennant} A, {Bucciantini} N, {Dovciak} M, {Marin} F, {Marscher} A,
  {Poutanen} J, {Slane} P, {Turolla} R, {Kalinowski} W, {Di Marco} A, {Fabiani}
  S, {Minuti} M, {La Monaca} F, {Pinchera} M, {Rankin} J, {Sgro'} C, {Trois} A,
  {Xie} F, {Alexander} C, {Allen} DZ, {Amici} F, {Andersen} J, {Antonelli} A,
  {Antoniak} S, {Attin{\`a}} P, {Barbanera} M, {Bachetti} M, {Baggett} RM,
  {Bladt} J, {Brez} A, {Bonino} R, {Boree} C, {Borotto} F, {Breeding} S,
  {Brienza} D, {Bygott} HK, {Caporale} C, {Cardelli} C, {Carpentiero} R,
  {Castellano} S, {Castronuovo} M, {Cavalli} L, {Cavazzuti} E, {Ceccanti} M,
  {Centrone} M, {Citraro} S, {D'Amico} F, {D'Alba} E, {Di Gesu} L, {Del Monte}
  E, {Dietz} KL, {Di Lalla} N, {Persio} GD, {Dolan} D, {Donnarumma} I,
  {Evangelista} Y, {Ferrant} K, {Ferrazzoli} R, {Ferrie} M, {Footdale} J,
  {Forsyth} B, {Foster} M, {Garelick} B, {Gunji} S, {Gurnee} E, {Head} M,
  {Hibbard} G, {Johnson} S, {Kelly} E, {Kilaru} K, {Lefevre} C, {Roy} SL,
  {Loffredo} P, {Lorenzi} P, {Lucchesi} L, {Maddox} T, {Magazzu} G, {Maldera}
  S, {Manfreda} A, {Mangraviti} E, {Marengo} M, {Marrocchesi} A, {Massaro} F,
  {Mauger} D, {McCracken} J, {McEachen} M, {Mize} R, {Mereu} P, {Mitchell} S,
  {Mitsuishi} I, {Morbidini} A, {Mosti} F, {Nasimi} H, {Negri} B, {Negro} M,
  {Nguyen} T, {Nitschke} I, {Nuti} A, {Onizuka} M, {Oppedisano} C, {Orsini} L,
  {Osborne} D, {Pacheco} R, {Paggi} A, {Painter} W, {Pavelitz} SD, {Pentz} C,
  {Piazzolla} R, {Perri} M, {Pesce-Rollins} M, {Peterson} C, {Pilia} M,
  {Profeti} A, {Puccetti} S, {Ranganathan} J, {Ratheesh} A, {Reedy} L, {Root}
  N, {Rubini} A, {Ruswick} S, {Sanchez} J, {Sarra} P, {Santoli} F, {Scalise} E,
  {Sciortino} A, {Schroeder} C, {Seek} T, {Sosdian} K, {Spandre} G, {Speegle}
  CO, {Tamagawa} T, {Tardiola} M, {Tobia} A, {Thomas} NE, {Valerie} R,
  {Vimercati} M, {Walden} AL, {Weddendorf} B, {Wedmore} J, {Welch} D, {Zanetti}
  D, {Zanetti} F (2022) {The Imaging X-Ray Polarimetry Explorer (IXPE):
  Pre-Launch}. Journal of Astronomical Telescopes, Instruments, and Systems
  8(2):026002, \doi{10.1117/1.JATIS.8.2.026002}, \eprint{2112.01269}

\bibitem[{{Wertz} and {Larson}(1999)}]{1999smad.book.....W}
{Wertz} JR, {Larson} WJ (1999) {Space mission analysis and design}

\bibitem[{Xiao et~al.(2016)Xiao, Hajdas, Wu, Produit, Bao, Batsch, Cadoux,
  Chai, Dong, Kong, Kong, Rybka, Leluc, Li, Liu, Liu, Marcinkowski, Paniccia,
  Pohl, Rapin, Shi, Song, Sun, Szabelski, Wang, Wen, Xu, Zhang, Zhang, Zhang,
  Zhang, Zhang, and Zwolinska}]{Xiao.2016}
Xiao H, Hajdas W, Wu B, Produit N, Bao T, Batsch T, Cadoux F, Chai J, Dong Y,
  Kong M, Kong S, Rybka DK, Leluc C, Li L, Liu J, Liu X, Marcinkowski R,
  Paniccia M, Pohl M, Rapin D, Shi H, Song L, Sun J, Szabelski J, Wang R, Wen
  X, Xu H, Zhang L, Zhang L, Zhang S, Zhang X, Zhang Y, Zwolinska A (2016) {A
  crosstalk and non-uniformity correction method for the space-borne Compton
  polarimeter POLAR}. Elsevier BV pp 1 -- 7,
  \doi{10.1016/j.astropartphys.2016.06.007}

\bibitem[{{Xiao} et~al.(2017){Xiao}, {Hajdas}, {Marcinkowski}, and {Polar
  Collaboration}}]{2017ICRC...35..668X}
{Xiao} H, {Hajdas} W, {Marcinkowski} R, {Polar Collaboration} (2017)
  {Optimization of the final settings for the Space-borne Hard X-ray Compton
  Polarimeter POLAR}. In: 35th International Cosmic Ray Conference (ICRC2017),
  International Cosmic Ray Conference, vol 301, p 668, \eprint{1707.02291}

\bibitem[{{Xiao} et~al.(2018){Xiao}, {Hajdas}, {Wu}, {Produit}, {Bao},
  {Bernasconi}, {Cadoux}, {Dong}, {Egli}, {Gauvin}, {Kole}, {Kramert}, {Kong},
  {Li}, {Li}, {Liu}, {Liu}, {Marcinkowski}, {Rybka}, {Pohl}, {Shi}, {Song},
  {Sun}, {Xiong}, {Szabelski}, {Socha}, {Wang}, {Wen}, {Wu}, {Zhang}, {Zhang},
  {Zhang}, {Zhang}, {Zhang}, and {Zwolinska}}]{2018APh...103...74X}
{Xiao} H, {Hajdas} W, {Wu} B, {Produit} N, {Bao} T, {Bernasconi} T, {Cadoux} F,
  {Dong} Y, {Egli} K, {Gauvin} N, {Kole} M, {Kramert} R, {Kong} S, {Li} L, {Li}
  Z, {Liu} J, {Liu} X, {Marcinkowski} R, {Rybka} DK, {Pohl} M, {Shi} H, {Song}
  L, {Sun} J, {Xiong} S, {Szabelski} J, {Socha} P, {Wang} R, {Wen} X, {Wu} X,
  {Zhang} L, {Zhang} P, {Zhang} S, {Zhang} X, {Zhang} Y, {Zwolinska} A (2018)
  {In-flight energy calibration of the space-borne Compton polarimeter POLAR}.
  Astroparticle Physics 103:74--86, \doi{10.1016/j.astropartphys.2018.07.009},
  \eprint{1710.08918}

\bibitem[{{Xie} and {Pearce}(2018)}]{Xie2018}
{Xie} F, {Pearce} M (2018) {A Study of Background Conditions for
  Sphinx{\textemdash}The Satellite-Borne Gamma-Ray Burst Polarimeter}. Galaxies
  6(2):50, \doi{10.3390/galaxies6020050}, \eprint{1809.06629}

\bibitem[{{Yang} et~al.(2019){Yang}, {Liao}, {Li}, and {Peng}}]{Yang2019}
{Yang} M, {Liao} S, {Li} Y, {Peng} C (2019) {Evaluating the Radiation Effects
  on the Characteristics of the Silicon Avalanche Photodiode with Protons}. In:
  Materials Science and Engineering Conference Series, Materials Science and
  Engineering Conference Series, vol 562, p 012076,
  \doi{10.1088/1757-899X/562/1/012076}

\bibitem[{{Yatsu} et~al.(2014){Yatsu}, {Ito}, {Kurita}, {Arimoto}, {Kawai},
  {Matsushita}, {Kawajiri}, {Kitamura}, {Matunaga}, {Kimura}, {Kataoka},
  {Nakamori}, and {Kubo}}]{Yatsu2014}
{Yatsu} Y, {Ito} K, {Kurita} S, {Arimoto} M, {Kawai} N, {Matsushita} M,
  {Kawajiri} S, {Kitamura} S, {Matunaga} S, {Kimura} S, {Kataoka} J, {Nakamori}
  T, {Kubo} S (2014) {Pre-flight performance of a micro-satellite TSUBAME for
  X-ray polarimetry of gamma-ray bursts}. In: {Takahashi} T, {den Herder} JWA,
  {Bautz} M (eds) Space Telescopes and Instrumentation 2014: Ultraviolet to
  Gamma Ray, Society of Photo-Optical Instrumentation Engineers (SPIE)
  Conference Series, vol 9144, p 91440L, \doi{10.1117/12.2056275}

\bibitem[{{Yonetoku} and {et al.}(2010)}]{Yonetoku2010}
{Yonetoku} D, {et al} (2010) {GAP aboard the solar-powered sail mission}. In:
  {Bellazzini} R, {Costa} E, {Matt} G, {Tagliaferri} G (eds) X-ray Polarimetry:
  A New Window in Astrophysics by Ronaldo Bellazzini, Enrico Costa, Giorgio
  Matt and Gianpiero Tagliaferri.~Cambridge University Press, 2010.~ ISBN:
  9780521191845, p.~339, p 339, \doi{10.1017/CBO9780511750809.051}

\bibitem[{{Yonetoku} et~al.(2011){Yonetoku}, {Murakami}, {Gunji}, {Mihara},
  {Sakashita}, {Morihara}, {Kikuchi}, {Takahashi}, {Fujimoto}, {Toukairin},
  {Kodama}, {Kubo}, and {Ikaros Demonstration Team}}]{2011PASJ...63..625Y}
{Yonetoku} D, {Murakami} T, {Gunji} S, {Mihara} T, {Sakashita} T, {Morihara} Y,
  {Kikuchi} Y, {Takahashi} T, {Fujimoto} H, {Toukairin} N, {Kodama} Y, {Kubo}
  S, {Ikaros Demonstration Team} (2011) {Gamma-Ray Burst Polarimeter (GAP)
  aboard the Small Solar Power Sail Demonstrator IKAROS}. \pasj 63:625,
  \doi{10.1093/pasj/63.3.625}, \eprint{1010.5305}

\bibitem[{Yonetoku et~al.(2011)Yonetoku, Murakami, Gunji, Mihara, Toma,
  Sakashita, Morihara, Takahashi, Toukairin, Fujimoto, Kodama, Kubo, and
  Team}]{Yonetoku.2011}
Yonetoku D, Murakami T, Gunji S, Mihara T, Toma K, Sakashita T, Morihara Y,
  Takahashi T, Toukairin N, Fujimoto H, Kodama Y, Kubo S, Team ID (2011)
  {DETECTION OF GAMMA-RAY POLARIZATION IN PROMPT EMISSION OF GRB 100826A}. The
  Astrophysical Journal Letters 743(2):L30, \doi{10.1088/2041-8205/743/2/l30},
  \urlprefix\url{http://arxiv.org/abs/1111.1779v1}, \eprint{1111.1779}

\bibitem[{{Yonetoku} et~al.(2012){Yonetoku}, {Murakami}, {Gunji}, {Mihara},
  {Toma}, {Morihara}, {Takahashi}, {Wakashima}, {Yonemochi}, {Sakashita},
  {Toukairin}, {Fujimoto}, and {Kodama}}]{Yonetoku2012}
{Yonetoku} D, {Murakami} T, {Gunji} S, {Mihara} T, {Toma} K, {Morihara} Y,
  {Takahashi} T, {Wakashima} Y, {Yonemochi} H, {Sakashita} T, {Toukairin} N,
  {Fujimoto} H, {Kodama} Y (2012) {Magnetic Structures in Gamma-Ray Burst Jets
  Probed by Gamma-Ray Polarization}. \apjl 758:L1,
  \doi{10.1088/2041-8205/758/1/L1}, \eprint{1208.5287}

\bibitem[{Zhang et~al.(2019)Zhang, Kole, Bao, Batsch, Bernasconi, Cadoux, Chai,
  Dai, Dong, Gauvin, Hajdas, Lan, Li, Li, Li, Liu, Liu, Marcinkowski, Produit,
  Orsi, Pohl, Rybka, Shi, Song, Sun, Szabelski, Tymieniecka, Wang, Wang, Wen,
  Wu, Wu, Wu, Xiao, Xiong, Zhang, Zhang, Zhang, Zhang, and
  Zwolinska}]{Zhang.2019}
Zhang SN, Kole M, Bao TW, Batsch T, Bernasconi T, Cadoux F, Chai JY, Dai ZG,
  Dong YW, Gauvin N, Hajdas W, Lan MX, Li HC, Li L, Li ZH, Liu JT, Liu X,
  Marcinkowski R, Produit N, Orsi S, Pohl M, Rybka D, Shi HL, Song LM, Sun JC,
  Szabelski J, Tymieniecka T, Wang RJ, Wang YH, Wen X, Wu BB, Wu X, Wu XF, Xiao
  HL, Xiong SL, Zhang LY, Zhang L, Zhang XF, Zhang YJ, Zwolinska A (2019)
  {Detailed polarization measurements of the prompt emission of five gamma-ray
  bursts}. Nature Astronomy 3(3):258--264, \doi{10.1038/s41550-018-0664-0},
  \eprint{1901.04207}

\bibitem[{{Zhitnik} et~al.(2006){Zhitnik}, {Logachev}, {Bogomolov}, {Denisov},
  {Kavanosyan}, {Kuznetsov}, {Morozov}, {Myagkova}, {Svertilov}, {Ignat'ev},
  {Oparin}, {Pertsov}, and {Tindo}}]{Zhitnik2006}
{Zhitnik} IA, {Logachev} YI, {Bogomolov} AV, {Denisov} YI, {Kavanosyan} SS,
  {Kuznetsov} SN, {Morozov} OV, {Myagkova} IN, {Svertilov} SI, {Ignat'ev} AP,
  {Oparin} SN, {Pertsov} AA, {Tindo} IP (2006) {Polarization, temporal, and
  spectral parameters of solar flare hard X-rays as measured by the SPR-N
  instrument onboard the CORONAS-F satellite}. Solar System Research
  40:93--103, \doi{10.1134/S003809460602002X}

\bibitem[{{Zhitnik} et~al.(2014){Zhitnik}, {Logachev}, {Bogomolov},
  {Bogomolov}, {Denisov}, {Kavanosyan}, {Kuznetsov}, {Morozov}, {Myagkova},
  {Svertilov}, {Ignatiev}, {Oparin}, and {Pertsov}}]{Zhitnik2014}
{Zhitnik} IA, {Logachev} YI, {Bogomolov} AV, {Bogomolov} VV, {Denisov} YI,
  {Kavanosyan} SS, {Kuznetsov} SN, {Morozov} OV, {Myagkova} IN, {Svertilov} SI,
  {Ignatiev} AP, {Oparin} SN, {Pertsov} AA (2014) {Experiment with the SPR-N
  Instrument Onboard the CORONAS-F Satellite: Polarization, Temporal, and
  Spectral Characteristics of the Hard X-Ray of the Solar Flares}. In:
  {Kuznetsov} V (ed) Astrophysics and Space Science Library, Astrophysics and
  Space Science Library, vol 400, p 129, \doi{10.1007/978-3-642-39268-9\_4}

\bibitem[{{Zombeck}(2006)}]{Zombeck_2006}
{Zombeck} MV (2006) {Handbook of Space Astronomy and Astrophysics}. {Cambridge
  University Press}

\end{thebibliography}

\end{document}